\documentclass[twocolumn,amsmath,amssymb,prb]{revtex4-2}
\usepackage{graphicx}
\usepackage{subcaption}
\usepackage{float}
\usepackage{dcolumn}
\usepackage{bm}
\usepackage[utf8]{inputenc}
\usepackage{newunicodechar}
\usepackage{textgreek}
\usepackage{natbib}
\usepackage{xcolor}
\begin{document}

\title{DFT modelling of stacking faults in hexagonal and cubic GaN}
\author{Zijie Wang}
\email{zijie.wang.23@ucl.ac.uk}
\affiliation{Dept. Physics \& Astronomy, University College London, Gower Street, London, WC1E 6BT, United Kingdom.}
\affiliation{London Centre for Nanotechnology, University College London, 17-19 Gordon Street, London, WC1H 0AH, United Kingdom.}

\author{Mazharul M. Islam}
\email{ucanmmi@ucl.ac.uk}
\affiliation{London Centre for Nanotechnology, University College London, 17-19 Gordon Street, London, WC1H 0AH, United Kingdom.}

\author{David R. Bowler}%
\email{david.bowler@ucl.ac.uk}
\affiliation{London Centre for Nanotechnology, University College London, 17-19 Gordon Street, London, WC1H 0AH, United Kingdom.}
\affiliation{Research Centre for Materials Nanoarchitectonics (WPI-MANA), National Institute for Materials Science (NIMS), 1-1 Namiki, Tsukuba, Ibaraki 305-0044, Japan}
\affiliation{Dept. Physics \& Astronomy, University College London, Gower Street, London, WC1E 6BT, United Kingdom.}

\date{\today}

\begin{abstract}
  We have performed density functional theory (DFT) calculations to characterize the energetics, and the atomic and electronic structure, of stacking faults in GaN, both in the stable hexagonal wurtzite (wz) phase and in the metastable cubic zincblende (zb) phase.
  In wz GaN, SFs on the $\{0001\}$ planes can be divided into three different intrinsic stacking faults (I$_{1}$, I$_{2}$, and I$_{3}$) and one extrinsic stacking fault (E). Based on the calculated formation energy, I$_{1}$ is the most stable SF of wz GaN in agreement with experiment.  In zb GaN, SFs form on $\{111\}$ planes, giving one type each of intrinsic, extrinsic and twin SFs.  In our calculations, the three types of SFs have similar formation energy.
  To characterize the effect of the stacking faults on the electronic structure of the material, we examined the band density.  We found that the bands near the valence band maximum in wz GaN are localised on the Ga-polar side of the stacking fault (i.e. on the Ga side of the Ga-N bonds perpendicular to the SF), with the bands near the conduction band minimum more on the N-polar side, though somewhat delocalised.  We found the opposite trend in zb GaN; this behaviour is caused by a redistribution of charge near the interface.  We also show the band offsets for the stacking faults, finding that they are very sensitive to local conditions, but can all be described as type II interfaces, with the presence of a stacking fault \emph{reducing} the gap locally in both wz and zb.
\end{abstract}

\maketitle

\section{Introduction}
\label{sec:introduction}

III-V nitrides are increasingly important in electronic and optoelectronic devices \cite{ding2018improving,ito2007simple}. These applications are affected by a number of material problems, mostly due to the lack of cost-effective suitable substrates on which the materials are grown. 
Growth on mismatched substrates causes the epilayers to contain a very high concentration of extended defects, principally dislocations in c-type wz GaN \cite{lester1995high,moram2009origin} and stacking faults in non-polar oriented wz \cite{Haberlen:2010jw,zakharov2005structural}. 
Most commercial GaN devices are based on the stable $c$-oriented wurtzite (wz) GaN, which shows a spontaneous polarisation; moreover, in the strained heterostructures required for devices, piezoelectric polarisation is induced \cite{tawfik2014stress}. As a result of this phenomenon, band bending occurs through the quantum confined Stark effect (QCSE) \cite{bernardini1997spontaneous, morkocc1999material}.

Through epitaxial growth of GaN films on cubic substrates, a metastable cubic zincblende (zb) structure can be stabilized  \cite{yang1996initial,trampert1997direct,ito2007simple}, which is not polar like the wz phase, providing a potential improvement in issues associated with the QCSE that occur in wz optoelectronic devices. Recent studies show that, due to the absence of polarization fields and the lower band gap compared to wz GaN, zb GaN may show higher efficiency in longer wavelength emission (green, amber, red) \cite{Church:2021hk}.  The stacking faults in zb GaN are often considered as wz insertions in a zb matrix \cite{binks2022cubic} which modify the optical properties of zb GaN films and propagate into active layers (though we show here that this model is not correct). In quantum wells and electron blocking layers, the segregation of alloying elements has been observed at stacking faults, leading to the formation of quantum wires and polarized emission \cite{ding2020alloy}.

Many questions have yet to be answered concerning the high resolution characterization of SFs' structures, alloy segregation to SFs and their influence on the optoelectronic properties of GaN-based devices.
Recently, a combination of high-resolution scanning transmission electron microscopy (STEM) and energy dispersive x-ray spectrometry was used to investigate the effects of alloy segregation around stacking faults in a zb GaN light-emitting structure \cite{ding2020alloy}.
A key question remaining is to identify the change in atomic spacing at the SFs, and whether this plays a significant role in segregation phenomena, which has been suggested as a possible mechanism for the different behaviour of Al and In \cite{ding2020alloy}: here, atomistic modelling can complement experiments, and provide insights beyond experimental resolution.  DFT modelling has previously focused on the energetics and electronic structure of basal stacking faults of wz GaN \cite{stampfl1999density,batyrev2011dislocations,benbedra2025energetics} with forcefield modelling considering energetics in zb GaN \cite{antovs2020intersections}.   
However, there have been no DFT studies on SFs in zb GaN, and no examination of the effects of stacking faults on the local electronic structure and band edges in either wz or zb GaN.

In this paper, our aim is to study the atomic structure and energetics of SFs, alongside their electronic structure and differences to bulk material in both wz and zb phases of GaN, and to compare to recent experimental findings on those SFs (e.g. wz GaN \cite{barchuk2011diffuse,zakharov2005structural,moram2009understanding} and zb GaN \cite{yucelen2018phase,xiu2023polarity}).
The rest of the paper is laid out as follows: in Sec. II, the computational approach is presented; the results for the structural and energetic properties are compared in Sec. III; the electronic structure of the SFs is discussed in Sec. IV; Sec. V gives a brief summary and concludes the paper.

\section{Computational Methods and Models}
\label{sec:comp-meth-models}

In this work, calculations were performed using Density Functional Theory (DFT) as implemented in the \textsc{Conquest} code which is designed for large-scale, massively parallel simulations on thousands to millions of atoms\cite{Bowler2002pt,Miyazaki2004,Nakata:2020dn}. The  DFT calculations used the generalized gradient approximation (GGA) based Perdew–Burke–Ernzerhof (PBE) exchange-correlation functional \cite{PhysRevLett.77.3865}. Core electrons were modeled using Hamman's optimised norm-conserving pseudopotentials \cite{PhysRevB.88.085117} from the PseudoDojo database\cite{Setten:2018xv}, while valence electrons were represented by real-space local orbital basis functions, specifically pseudo atomic orbitals (PAOs) \cite{Bowler:2019fv} at the double-zeta plus polarization (DZP) level. The PAOs were generated using the equal radii approach, with details given in the Supporting Information (SI).  The real-space integration grid spacing was set to be equivalent to a 400~Ha kinetic energy, or 0.1~Bohr, which converges total energies to better than 1~$\mu$Ha and stresses to better than 0.01~GPa.  

The structural relaxation was performed in two steps. First, the simulation cell was optimized to a tolerance of 0.1~GPa using the conjugate gradient algorithm, with fixed fractional ionic coordinates; to mimic the effects of a fixed substrate, only the stress along $c$ was minimised, with values of $a$ and $b$ fixed to bulk parameters. The stabilized quasi-Newton method (SQNM)\cite{Schaefer:2015zk} was then used to optimize the ionic positions with a threshold of 0.02~eV\AA$^{-1}$.

The wurtzite structure, which is hexagonal, is characterized by a stacking sequence of repeating layers AB along the [0001] direction\footnote{The stacking sequence of the layers in bulk GaN are often notated AaBb (wz) and AaBbCc (zb) with Aa indicating the alternation of planes of Ga and N along the [0001] direction (wz) or [111] direction (zb).  We will shorten this to AB and ABC respectively for simplicity of notation.}.  However, as \textsc{Conquest} supports only orthorhombic cells, we must use a cell where we constrain the $b/a$ ratio to maintain the correct shape (in the simplest cell, the ratio is $\sqrt{3}$).
The orthorhombic wz GaN simulation cell contains eight atoms, which we modelled using a $9 \times 6 \times 6$ gamma-centred Monkhorst–Pack k-point mesh; this gives convergence of total energy to around 10~$\mu$Ha and stress to around 0.01~GPa.
The zincblende phase is cubic and can be characterized by a stacking sequence of repeating layers ABC along the [111] direction. For the calculations of stacking faults and to simplify comparison with the wurtzite structure, we used a simulation cell oriented along the [111] direction of the cubic cell.  We again constrain the $b/a$ ratio to $\sqrt{3}$ to maintain the correct crystal geometry.  The resulting simulation cell contains twelve atoms, which we modelled using a $9 \times 6 \times 4$ gamma-centred Monkhorst–Pack k-point mesh.  The optimised lattice parameters agree well with experimental \cite{ding2021study,schulz1977crystal,frentrup2017x} and previous DFT \cite{gao2019point} studies, and are given in the SI, along with lattice parameters for  InN and AlN.

The formation energy per unit area of the simulation cell for the optimized SFs is calculated as:

\begin{equation}
\label{DE}
        E_{form} = \frac{E_{SF cell} - n E_{Ref}}{A} 
\end{equation}

\noindent where $E_{SF cell}$ is the total energy of the specific SF in a simulation cell containing $n$ stacking layers, and $E_{Ref}$ is the energy \emph{per layer} of the host material.  Note that a positive energy indicates that the simulation cell with a stacking fault is less stable than the host material.

When considering band offsets between perfect bulk cells and cells with stacking faults, we need a common zero.  The semi-core $d$ electrons in Ga form a natural low-lying band which allows comparison; however there are two groups of low-lying $d$ bands separated by around 1.9~eV, with the lowest energy band showing dispersion of around 0.7~eV, while the upper group of $d$ bands show dispersion of less than 0.1~eV. This upper group forms a suitable reference point to compare the energies of different sets of bands, and we choose the value at the gamma point to set the common zero.

\section{Physical structure and energetics}
\label{sec:phys-struct-energ}

Taking the optimized lattice parameters and atomic coordinates of bulk structures, a series of simulations cells was constructed containing stacking faults, with progressively increasing separation between the stacking faults to test the convergence. For wz GaN, the three intrinsic stacking faults I$_{1}$, I$_{2}$, I$_{3}$ in isolation would be formed as \ldots AB/CB\ldots, \ldots AB/CA\ldots and \ldots AB/C/BA\ldots respectively.  When modelled with periodic boundary conditions some adjustment is required, and we need two stacking faults in each cell. For n repeats of each stacking sequence, we then have /nAB/nCB/, /nAB/nCA/C/, and /nAB/C/nBA/C/ where the stacking faults are in the middle and at the edges of the cell, and an extra C atomic layer has been added to both I$_{2}$ and I$_{3}$ to restore periodicity.  For the extrinsic stacking fault, we have \ldots AB/C/AB\ldots in isolation, and /nAB/C/ in a periodic cell (in this case, we need only one stacking fault per simulation cell).

\begin{figure}[ht]
\subfloat[]{\includegraphics[height=1.8in]{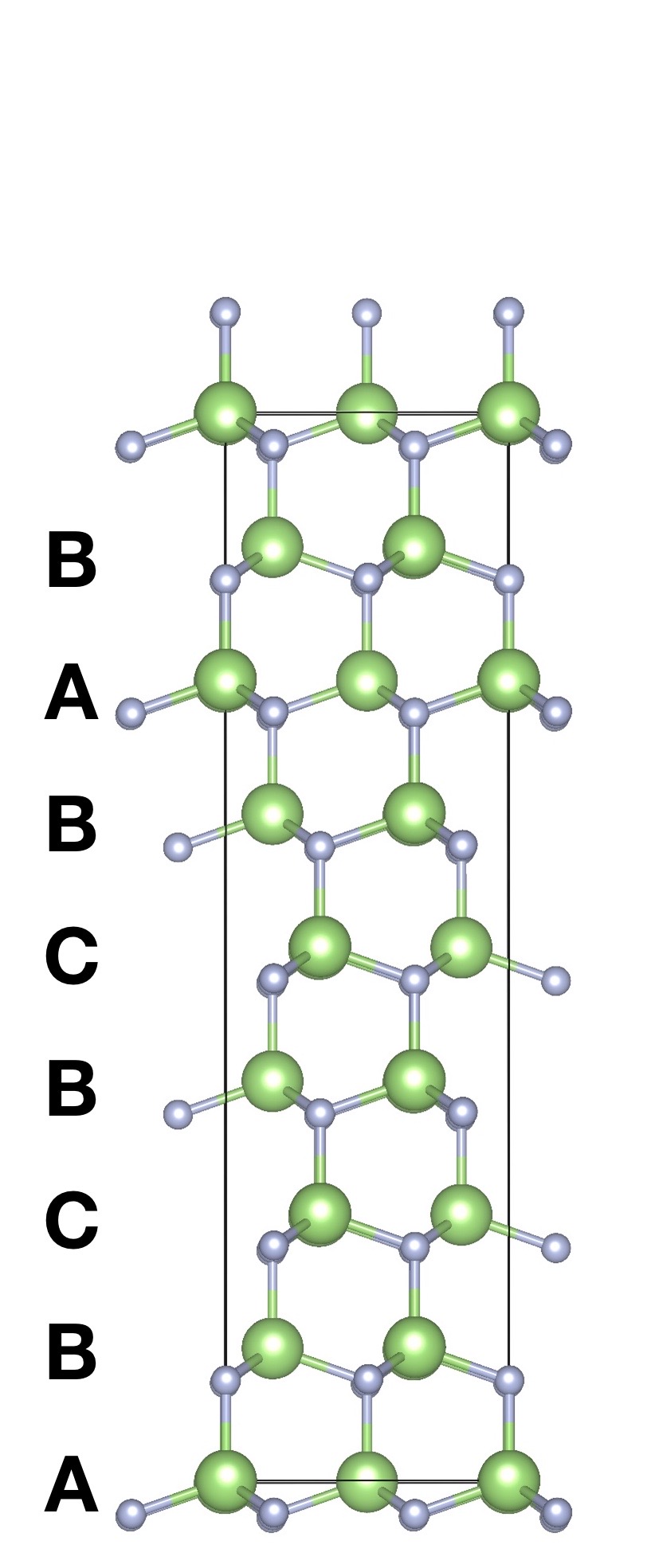}}\hspace{0.25cm}
\subfloat[]{\includegraphics[height=1.8in]{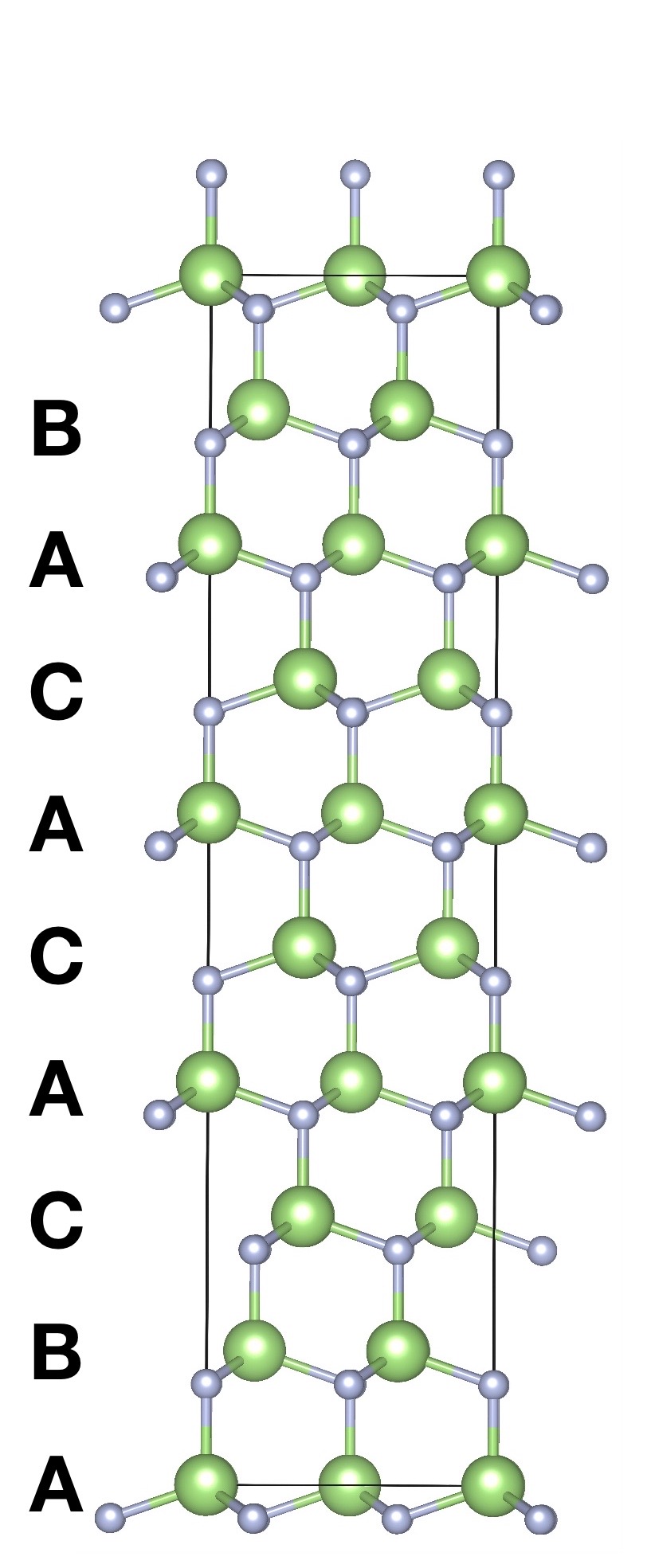}}\hspace{0.25cm}
\subfloat[]{\includegraphics[height=1.8in]{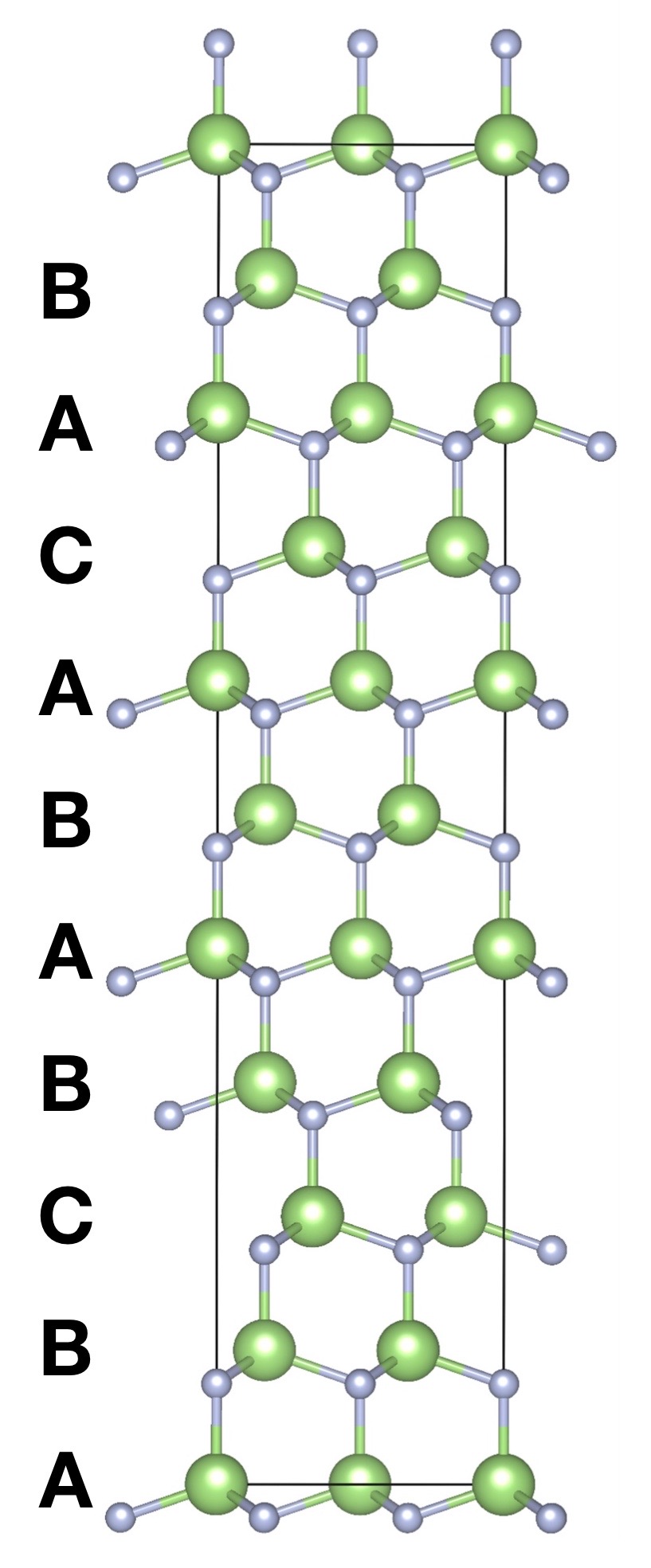}}\hspace{0.25cm}
\subfloat[]{\includegraphics[height=1.8in]{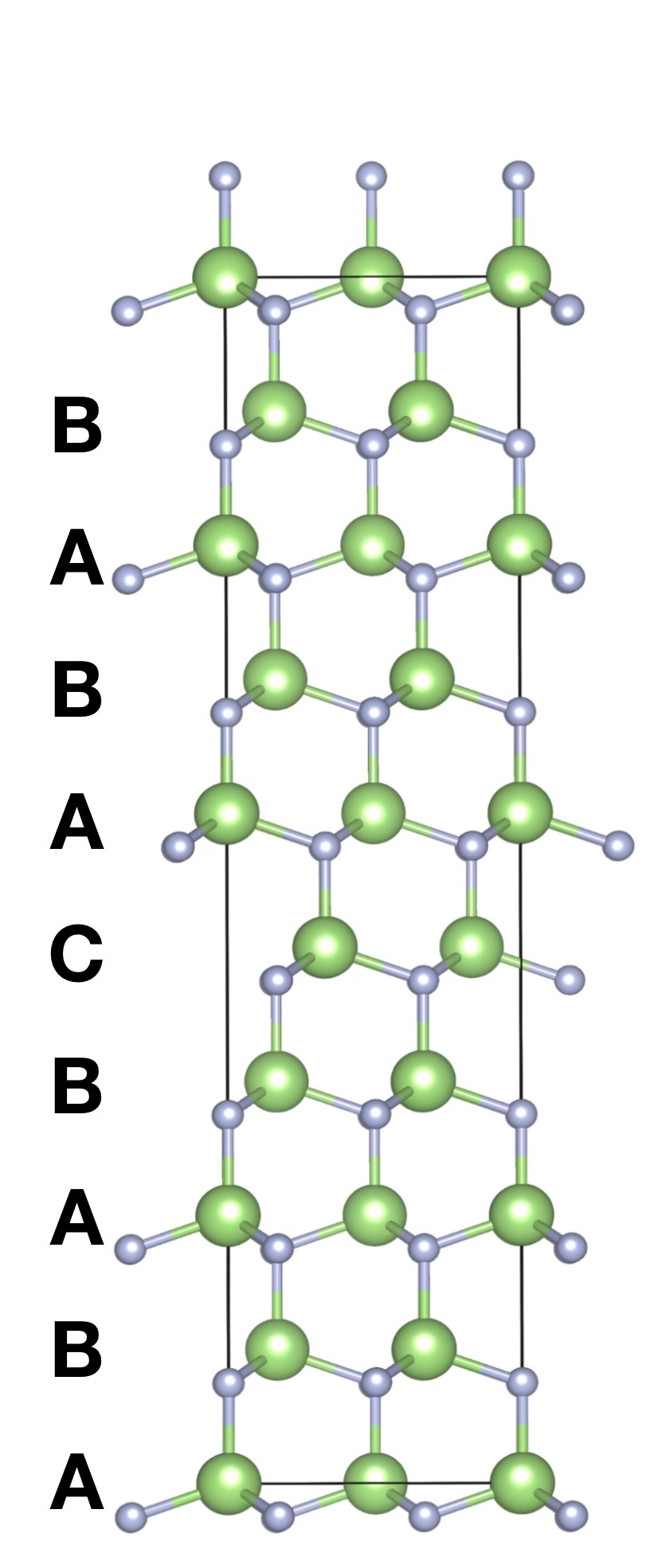}}
\caption{Stacking faults in wurtzite GaN. (a) Intrinsic-1 (I$_{1}$) /AB/CBCB/AB/, (b) Intrinsic-2 (I$_{2}$) /AB/CACA/C/AB/, (c) Intrinsic-3 (I$_{3}$) /AB/C/BABA/C/AB/ and (d) Extrinsic /ABAB/C/ABAB/. Large green and small silver spheres represent gallium and nitrogen atoms respectively. A, B and C atomic layers are marked in the figure. Note that these are highly shortened cells for illustration only.}
\label{fgr:Fig-1_New}
\end{figure}

\begin{figure}[ht]
\subfloat[]{\includegraphics[height = 1.8in]{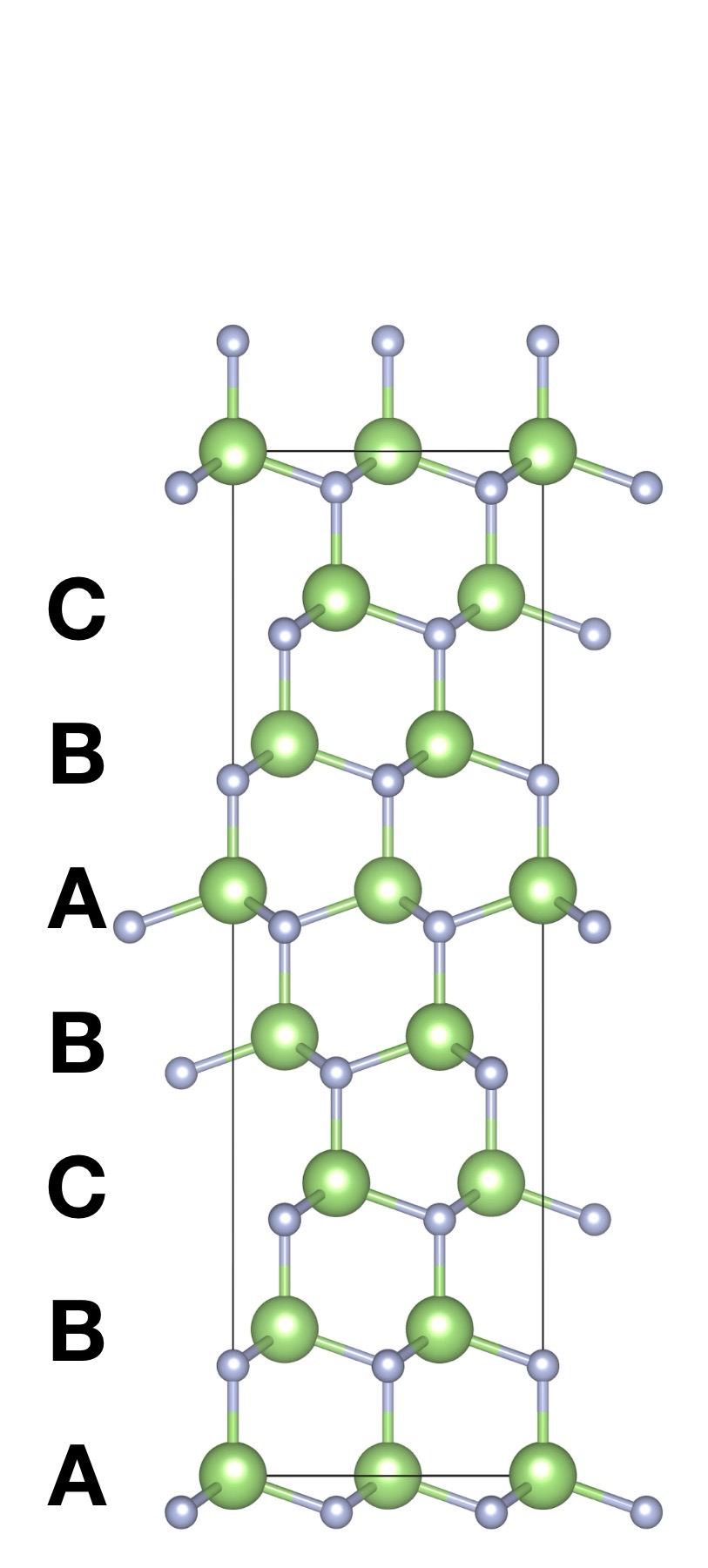}}\hspace{0.5cm} 
\subfloat[]{\includegraphics[height = 1.8in]{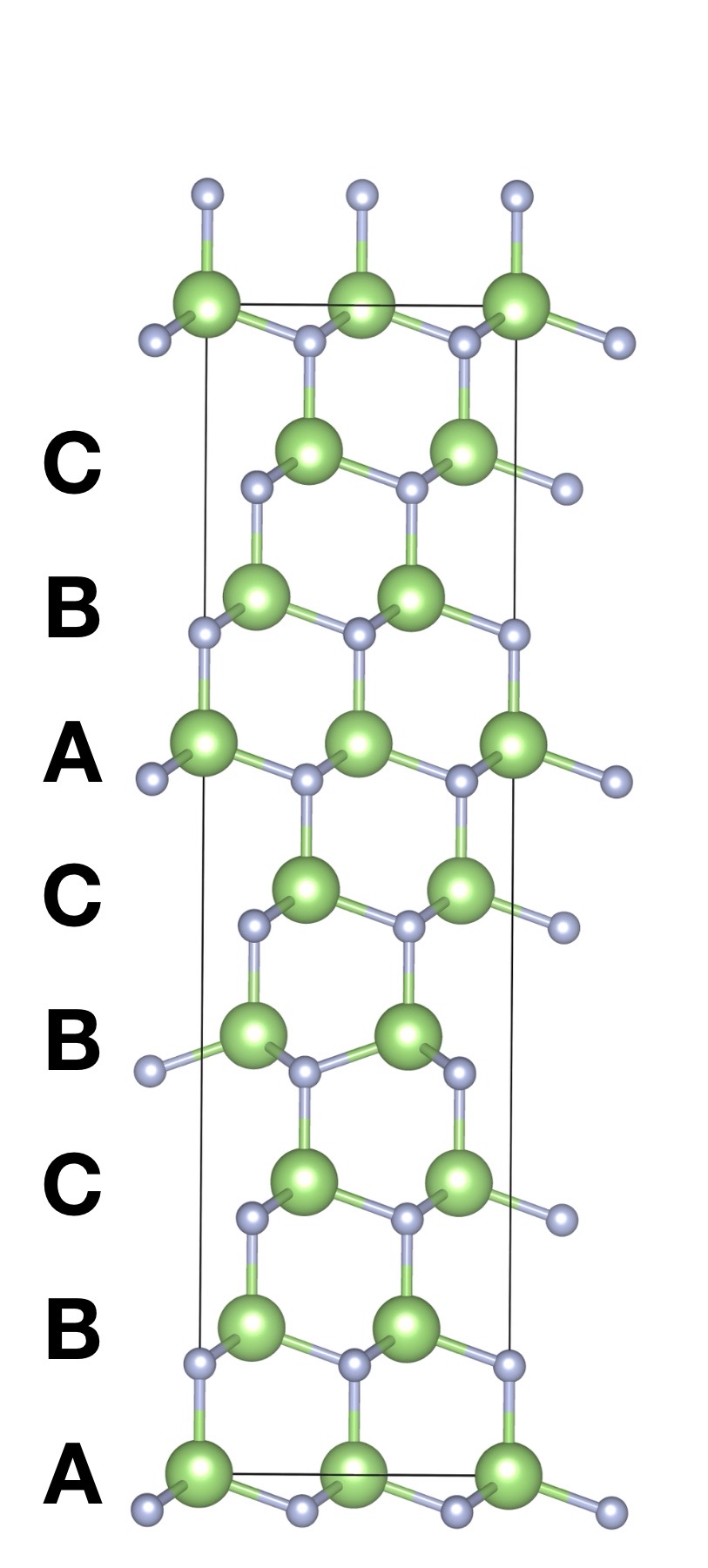}}\hspace{0.5cm}
\subfloat[]{\includegraphics[height = 1.8in]{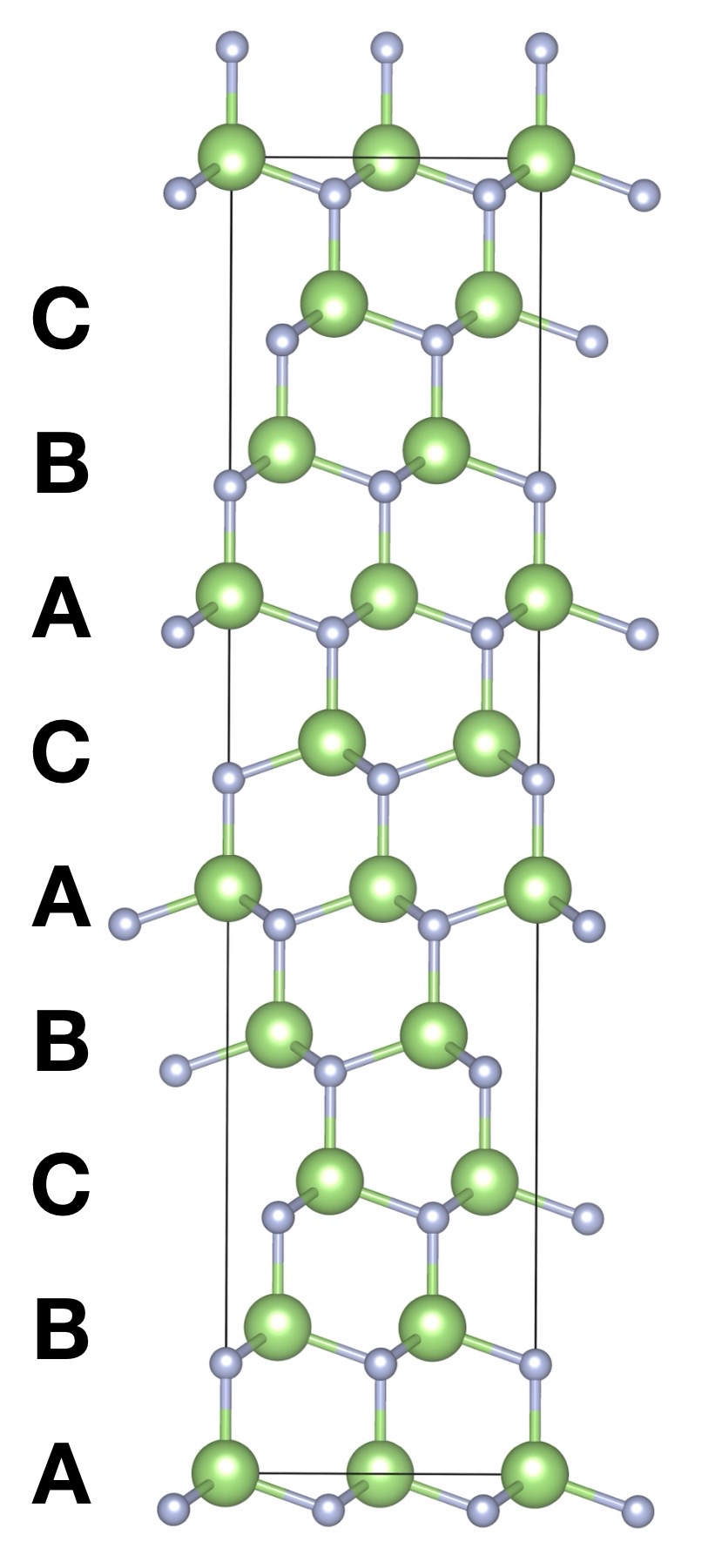}}
\caption{Stacking faults in Zincblende GaN. (a) Extrinsic /ABC/B/ABC/, (b) Intrinsic /ABC/BC/ABC/ and (c) Twin /ABC/BAC/ABC/. Large green and small silver spheres represent gallium and nitrogen atoms respectively. A, B and C atomic layers are marked in the figure.  Note that these are highly shortened cells for illustration only.}
\label{fgr:Fig-2_New}
\end{figure}

For zb GaN, in isolation, the intrinsic, extrinsic and twin stacking faults are formed as \ldots ABC/BC/ABC\ldots, \ldots ABC/B/ABC\ldots\ and \ldots ABC/BAC/ABC\ldots\ respectively.  We can accommodate a single stacking fault in a periodic cell in all three cases, which for n repeats of the stacking sequence we write as /nABC/BC/, /nABC/B/ and /nABC/BAC/ in each case.  These models are illustrated in Figs. \ref{fgr:Fig-1_New} and \ref{fgr:Fig-2_New}.

We tested the effect of increasing the number of layers of bulk material between stacking faults in both wz and zb, to find the converged value of the formation energy, which we give in mJ/m$^{2}$.  The converged values are given in Table~\ref{tab:SFenergy} along with number of layers.  Full details of the convergence of the energy with number of layers are given in the supplementary information, both as tables and a figure; based on the convergence data and the structural optimisation criteria, we estimate that the formation energies have an error of $\pm$ 1~mJ/m$^{2}$.
The formation energies show that all SFs in wz GaN have a positive energy, with I$_{1}$ being the most stable SF.  This agrees well with the experimental observation that 90\% of observed basal plane stacking faults in wz GaN films are of I$_{1}$-type \cite{zakharov2005structural,moram2009x}. The second most common SF is I$_{2}$ followed by I$_{3}$ and E; experimentally, the extrinsic type of SF is rarely observed \cite{moram2009understanding}. The observed trend is in good accord with energies calculated in other DFT investigations \cite{stampfl1999density, batyrev2011dislocations,benbedra2025energetics} though none of these studies report details of convergence tests of spacing between SFs; the I$_{1}$ energy is also in good agreement with the available experimental estimate\cite{takeuchi1999stacking} of 20 $\pm$ 3 mJ/m$^{2}$.  The positive formation energies indicate that the SFs are less stable than the perfect bulk material.

On the other hand, in zb GaN, all stacking faults have a negative formation energy, indicating that these cells are more stable than pure zb.  As zb is a metastable phase, and the stacking faults change the layer ordering towards that of the more stable wz phase, this is not surprising, and it fits well with the experimental observations of many stacking faults in zb GaN material.
All of the SFs have almost the same energy (certainly the difference between them is within the accuracy limit of these calculations) though the experimental observation is that the intrinsic stacking fault (SF) is the most abundant planar defect in epitaxial layers \cite{ding2020alloy}.
All stacking faults have a very low formation energy, implying that these defects may form very easily in both wz and zb GaN, though they are much more likely to form in the zb phase. We note that the rotation of the Ga-N bonds visible in ABF (annular bright field) STEM images of the intrinsic stacking fault\cite{xiu2023polarity} is reproduced in the calculated charge density of the modelling.

The low SFE in GaN can be understood in terms of the close structural and bonding similarity between the wurtzite and zincblende phases, and the small energy difference between the two (we calculate it to be 0.02eV per Ga-N pair, in line with other DFT results). In both polymorphs, the local bonding is tetrahedral with very similar bond lengths and bond angles. Importantly, the c/a ratio we calculate in wz GaN is 1.631, which is very close to the ideal value (1.633).  As a result, the introduction of a stacking fault—--corresponding to a local change in stacking sequence from AB (wz) to ABC (zb) or vice versa--—preserves the local tetrahedral bonding network, requiring only minimal structural rearrangement. Consequently, the energetic penalty associated with forming such faults is very small.  The key factor governing the low SFE is not the nature of the bonding, but rather the preservation of bond topology across the stacking fault.

\begin{table} [ht]
  \caption{Calculated values of formation energy (mJ/m$^{2}$) of stacking faults ($\Delta$E) and converged number of atomic layers (N) for wz (upper) and zb (lower) phases of GaN.\label{tab:SFenergy}}
\begin{center}
  \begin{tabular}{ccccc}
    \hline
    \hline
    \multicolumn{5}{c}{Wurtzite GaN}     \\
    \hline
    \hline
     & I$_{1}$  & I$_{2}$ & I$_{3}$ & E  \\
     $\Delta$E & 17 & 28 & 36 & 55 \\
    N & 24 & 33 & 34 & 25 \\
    \hline
    \hline
    \multicolumn{5}{c}{Zincblende GaN}     \\
    \hline
    \hline
     & I          & E           & T         &  \\
    $\Delta$E & -33   &  -34   &  -34 & \\
    N & 26 & 25 & 27 & \\
    \hline
    \hline
  \end{tabular}
\end{center}
\end{table}

Accurate determination of the stacking fault thickness ($l$) is essential for calculating the band profile of stacking faults functioning as quantum wells (QWs) \cite{binks2022cubic}. The calculated stacking-fault thickness $l$ for the optimized models of all the considered types of SFs are provided in Table \ref{tbl:Str}.  Here we use the inter-planar spacing as a measure of stacking-fault thickness, which is consistent with a recent experimental investigation based on high-resolution scanning transmission electron microscopy \cite{ding2020alloy}. 
At 300 K, the $c$ lattice parameter of bulk wz GaN and the lattice parameter of bulk zb GaN are measured to be 5.186 \AA\ and 4.506 \AA, respectively, corresponding to inter-planar spacings of 2.593 \AA\ and 2.602 \AA\ in wz-GaN and zb-GaN \cite{frentrup2017x,ding2021study,schulz1977crystal}. 
Bulk zb-GaN exhibits a larger inter-planar spacing between close-packed planes compared to bulk wz-GaN, which may be attributed to stronger Coulombic interactions between third-nearest-neighbor Ga and N atoms in the wurtzite phase. Our calculated lattice parameters for the bulk wz GaN are $a$ = 3.239 \AA\ and $c$ = 5.284 \AA\ and for bulk zb GaN, it is $a$ = 4.578 \AA. The resulting inter-planar spacing d$_{0001}$ in bulk wz-GaN is 2.642 \AA\ and d$_{111}$ in bulk zb-GaN is 2.644 \AA. These show good agreement with the experimental measurements\cite{frentrup2017x,ding2021study,schulz1977crystal} though perhaps the difference between the phases is a little small.
There is some question over the spacings near SFs, and whether the presence of the SF changes the inter-planar spacing relative to the non-defective materials. In Table \ref{tbl:Str}, the calculated largest ($l_{max}$) and smallest ($l_{min}$) layer spacings of the SFs are presented. Our calculated stacking faults layer spacings range from 2.639 to 2.643 \AA\ for wz GaN and 2.643 to 2.644 \AA\ for zb GaN, showing a very small variation, and good agreement with the bulk value; plots of the evolution of the layer spacing along the $c$ axis of the simulation cell are given in the supporting information.
(The alternative approach to define SF thickness using multiples of the Ga-N bond length \cite{benbedra2025energetics,lahnemann2012direct} gives similar insight but with less detail, so we do not present it.)

\begin{table}[ht]
\begin{center}
  \caption{Calculated range of layer spacings in stacking faults with respect to ideal crystals of wz and zb GaN.}
  \label{tbl:Str}
  \begin{tabular}{*{5}{c}}
    \hline
    \hline
    \multicolumn{5}{c}{Stacking faults based on Wurtzite GaN}     \\
    \hline
    \hline
    Bulk & I$_{1}$  & I$_{2}$ & I$_{3}$ & E    \\
    2.642  &    2.640--2.643  &  2.640--2.643  &   2.639--2.643     & 2.640--2.642 \\
    \hline
    \hline
    \multicolumn{5}{c}{Stacking faults based on Zincblende GaN}    \\
    \hline
    \hline
    Bulk & I & E  & T &     \\
   2.644  &   2.643--2.644   &  2.643--2.644    & 2.643--2.644    &      \\
    \hline
     \hline
  \end{tabular}
\end{center}
\end{table}

\section{Electronic structure}
\label{sec:electronic-structure}

It is important to characterise the effect of the stacking faults on the electronic structure of the materials (both wz and zb GaN).  While SFs can be seen as simply a change in the layer ordering, it is often suggested that they act as a small inclusion of the other crystal (i.e. zb in wz and vice versa) which will involve a change from polar to non-polar (or vice versa) material, with potentially significant effects on the electronic structure; below, we present evidence which shows that this picture is not correct.

\begin{figure}[ht]
\subfloat[]{\includegraphics[height = 3.2 in]{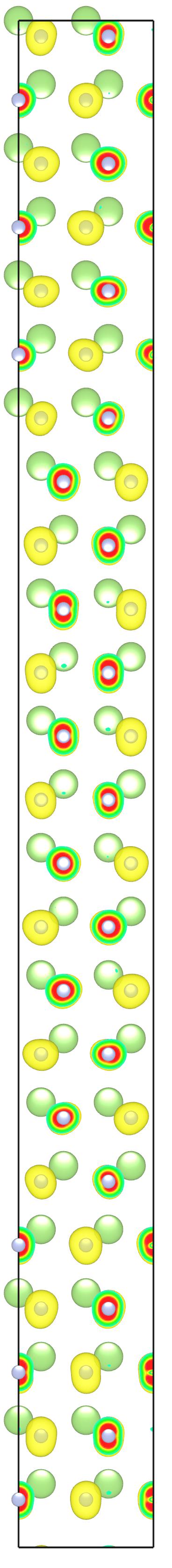}} 
\subfloat[]{\includegraphics[height = 3.2 in]{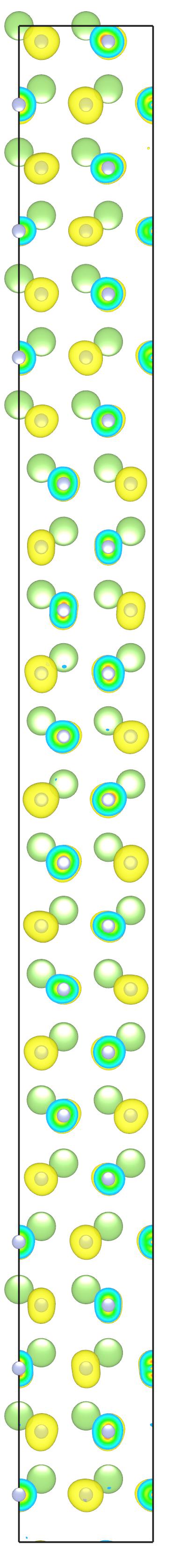}}
\subfloat[]{\includegraphics[height = 3.2 in]{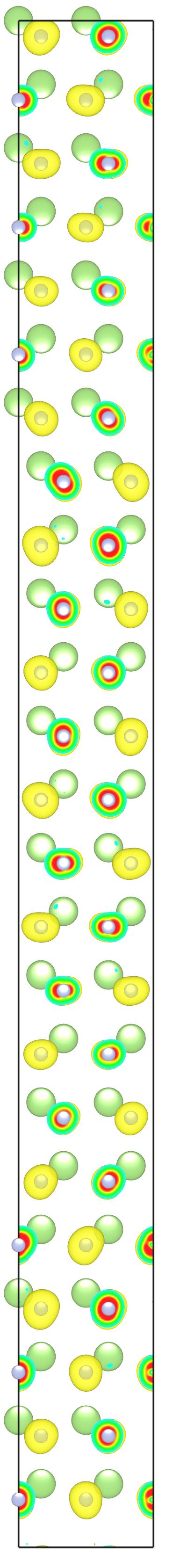}}
\subfloat[]{\includegraphics[height = 3.2 in]{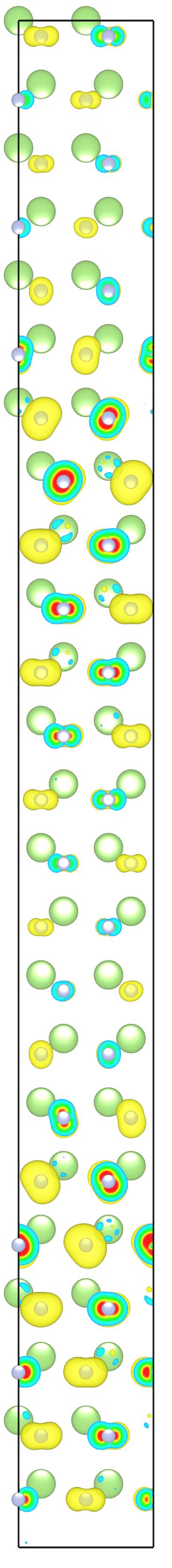}}
\subfloat[]{\includegraphics[height = 3.2 in]{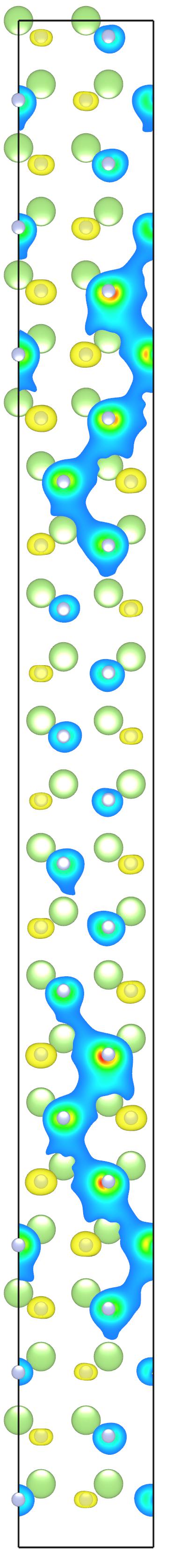}}
\subfloat[]{\includegraphics[height = 3.2 in]{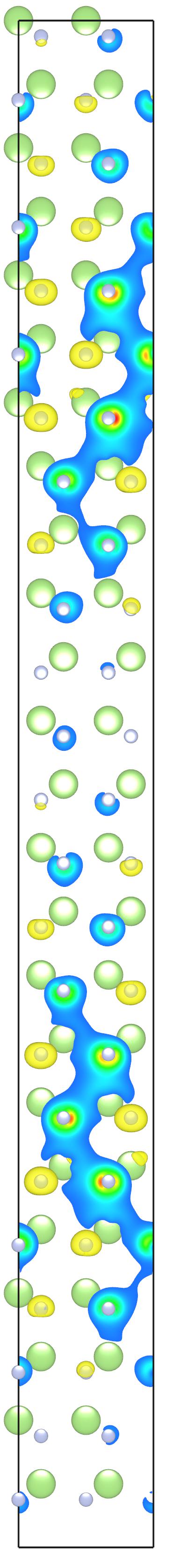}}
\subfloat[]{\includegraphics[height = 3.2 in]{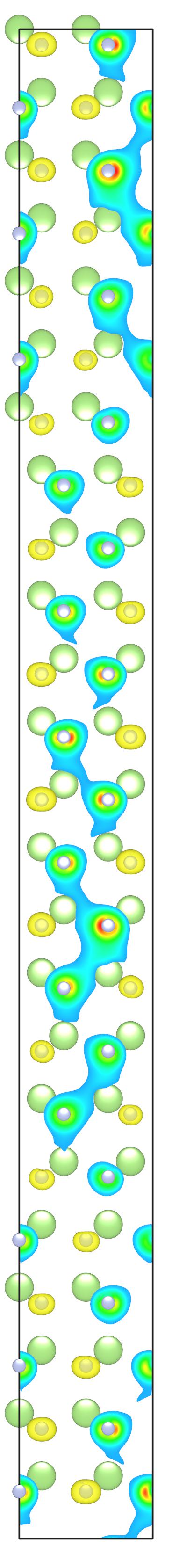}}
\subfloat[]{\includegraphics[height = 3.2 in]{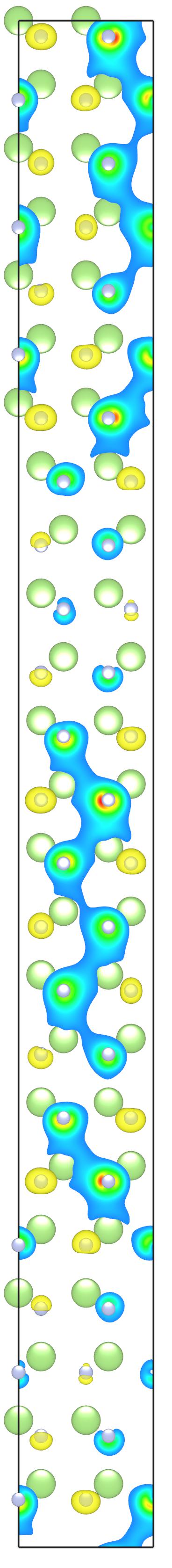}}
\caption{Band-resolved densities of Intrinsic-1 stacking fault of wz GaN, here shown with six repeats of the stacking sequence (or twelve layers) between stacking faults (/3AB/6CB/3AB/). The bands below the Fermi energy are shown in panels (a-d), and the bands above the Fermi energy  are shown in panels (e-h). The isosurfaces were plotted at a density of 0.0005 electrons per Bohr$^{3}$.  Note that there are two stacking faults in the simulation cell, 25\% and 75\% of the way along the cell.}
\label{fgr:Band_Density_I1}
\end{figure}

\begin{figure}[ht]
\subfloat[]{\includegraphics[height = 3.2 in]{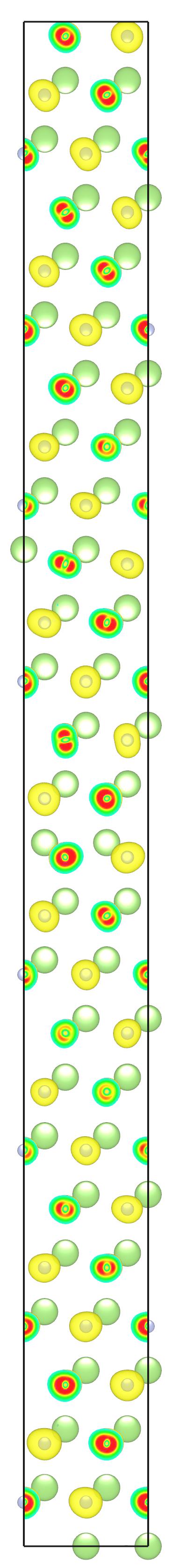}}
\subfloat[]{\includegraphics[height = 3.2 in]{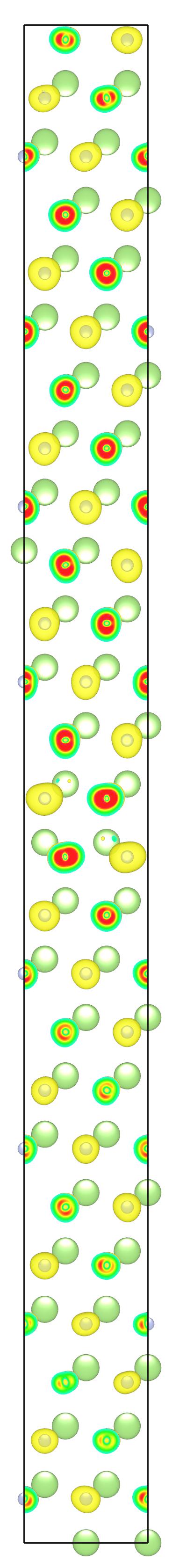}}
\subfloat[]{\includegraphics[height = 3.2 in]{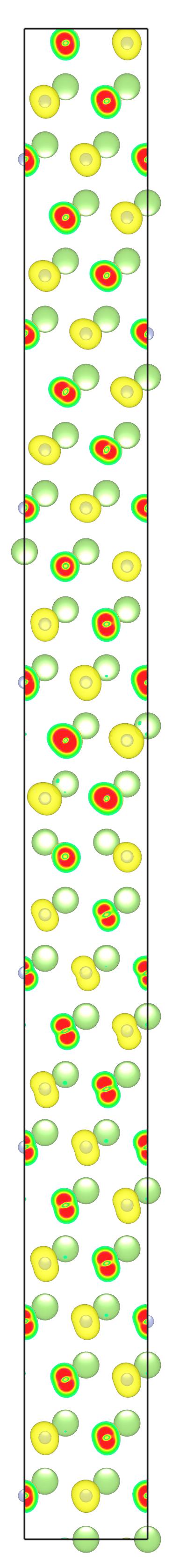}}
\subfloat[]{\includegraphics[height = 3.2 in]{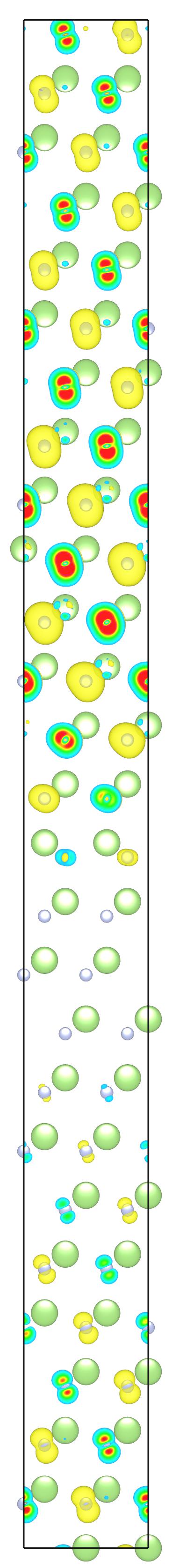}}
\subfloat[]{\includegraphics[height = 3.2 in]{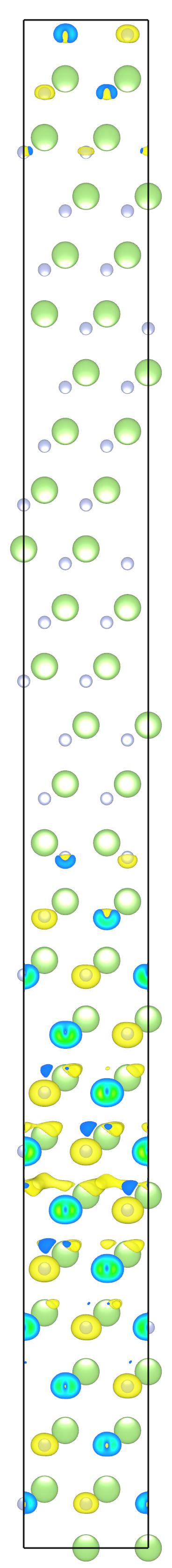}}
\subfloat[]{\includegraphics[height = 3.2 in]{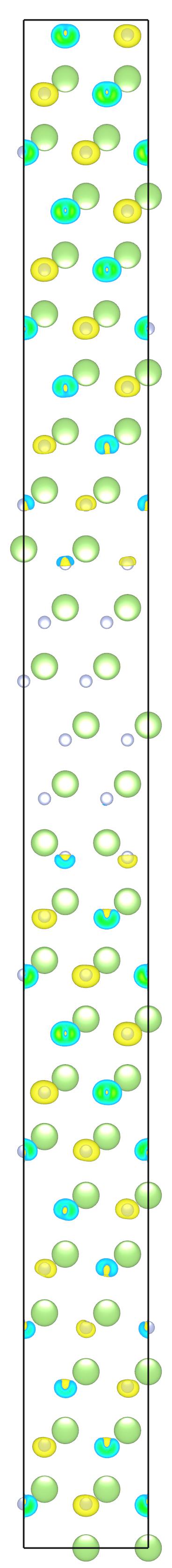}}
\subfloat[]{\includegraphics[height = 3.2 in]{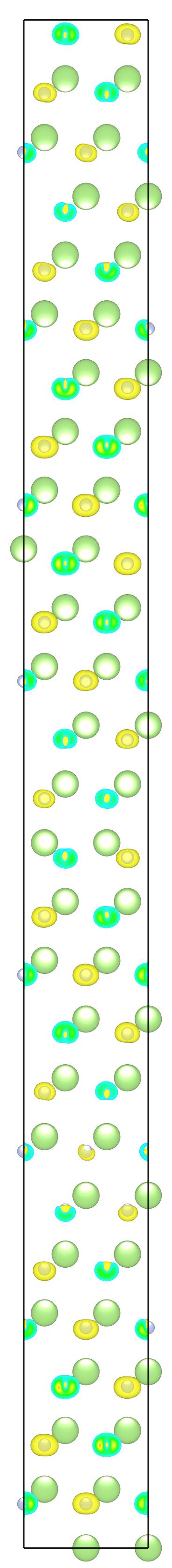}}
\subfloat[]{\includegraphics[height = 3.2 in]{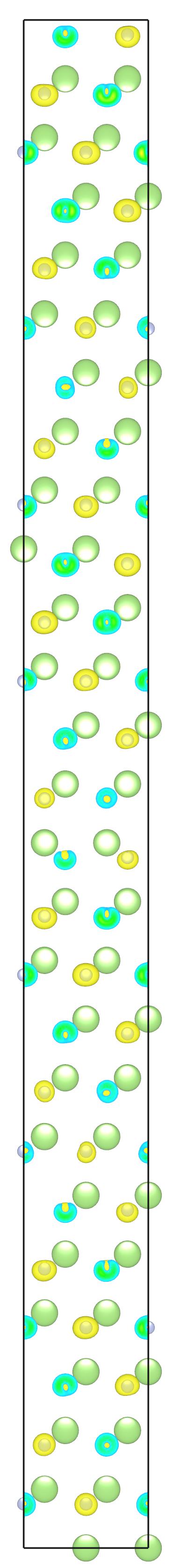}}
\caption{Band-resolved densities of Intrinsic stacking fault for zb GaN along [111], here shown with eight repeats of the stacking sequence (or twenty four layers) between stacking faults (/4ABC/BC/4ABC/). The bands below the Fermi energy are shown in panels (a-d), and the bands above the Fermi energy are shown in panels (e-h). The isosurfaces were plotted at a density of 0.0005 electrons per Bohr$^{3}$.  There is one stacking fault, halfway along the simulation cell.}
\label{fgr:Band_Density_ZB_In}
\end{figure}

In the first instance, we plot band-resolved charge densities to visualise the changes around the Fermi energy in real-space.  In the usual way, we define the density for band $n$ as $\rho_{n}(\mathbf{r})=\sum_{\mathbf{k}} w_{\mathbf{k}}|\psi_{n\mathbf{k}}(\mathbf{r})|^2 $ where $w_{\mathbf{k}}$ is the weight for each point in the Brillouin zone; the resulting plots are a solid state equivalent of a molecular orbital plot.  These band densities show similar trends for all SFs in each material, so we show here the bands for the most common SF in wz (I$_{1}$) in Fig.~\ref{fgr:Band_Density_I1} and in zb (I) in Fig.~\ref{fgr:Band_Density_ZB_In}.  Band densities for all other SFs are given in the supporting information.

In both wz and zb the bands near the top of the valence band (Fig.~\ref{fgr:Band_Density_I1}(a)-(d) in wz and Fig.~\ref{fgr:Band_Density_ZB_In}(a)-(d) in zb) are localised on one side of the SF, and the bands near the bottom of the conduction band (Fig.~\ref{fgr:Band_Density_I1}(e)-(h) in wz and Fig.~\ref{fgr:Band_Density_ZB_In}(e)-(h) in zb) are localised on the other side (though this is perhaps a little clearer in zb).  Similar behaviour is seen in all other stacking faults in both materials.

However, the localisation occurs on different sides of the SF in wz and in zb: in wz, the valence bands are localised on the Ga-polar side of the stacking fault (i.e. on the Ga side of the Ga-N bonds perpendicular to the SF, or below the stacking fault as plotted) with the conduction bands localised on the N-polar side.  By contrast, in zb the opposite trend is seen, with the valence bands on the N-polar side of the SF (above the stacking fault as plotted) and conduction bands on the Ga-polar side (below the stacking fault as plotted).

We seek to identify the mechanism underlying this localisation of the bands on one side of the SF, and the origin of the difference between the different behaviour in the two materials.  We characterise the problem by comparing the electrostatic potential in the faulted material with perfect material, and consider the potential along the $c$-axis of the simulation cell, taking a planar average in the $a-b$ plane.  The resulting potential trace is shown for the most stable SFs in the top panels of Fig.~\ref{fig:potentialtrace}(a) and (b) for wz and zb respectively (we again show plots for all other SFs in the supplementary information).

We see that the potential changes rapidly across the SFs in both materials (noting that there are \emph{two} SFs in the wz simulation cell) with periodicity leading to a slow variation of the potential across the rest of the simulation cell.  We can get insight into the extent of charge localisation by plotting the planar average of the density difference between the faulted and perfect cells (middle panel) and by integrating this along the c axis (bottom panel).  We can see the expected consistency between charge difference and potential difference.

\begin{figure}[ht]
\subfloat[]{\includegraphics[width = 0.95\columnwidth]{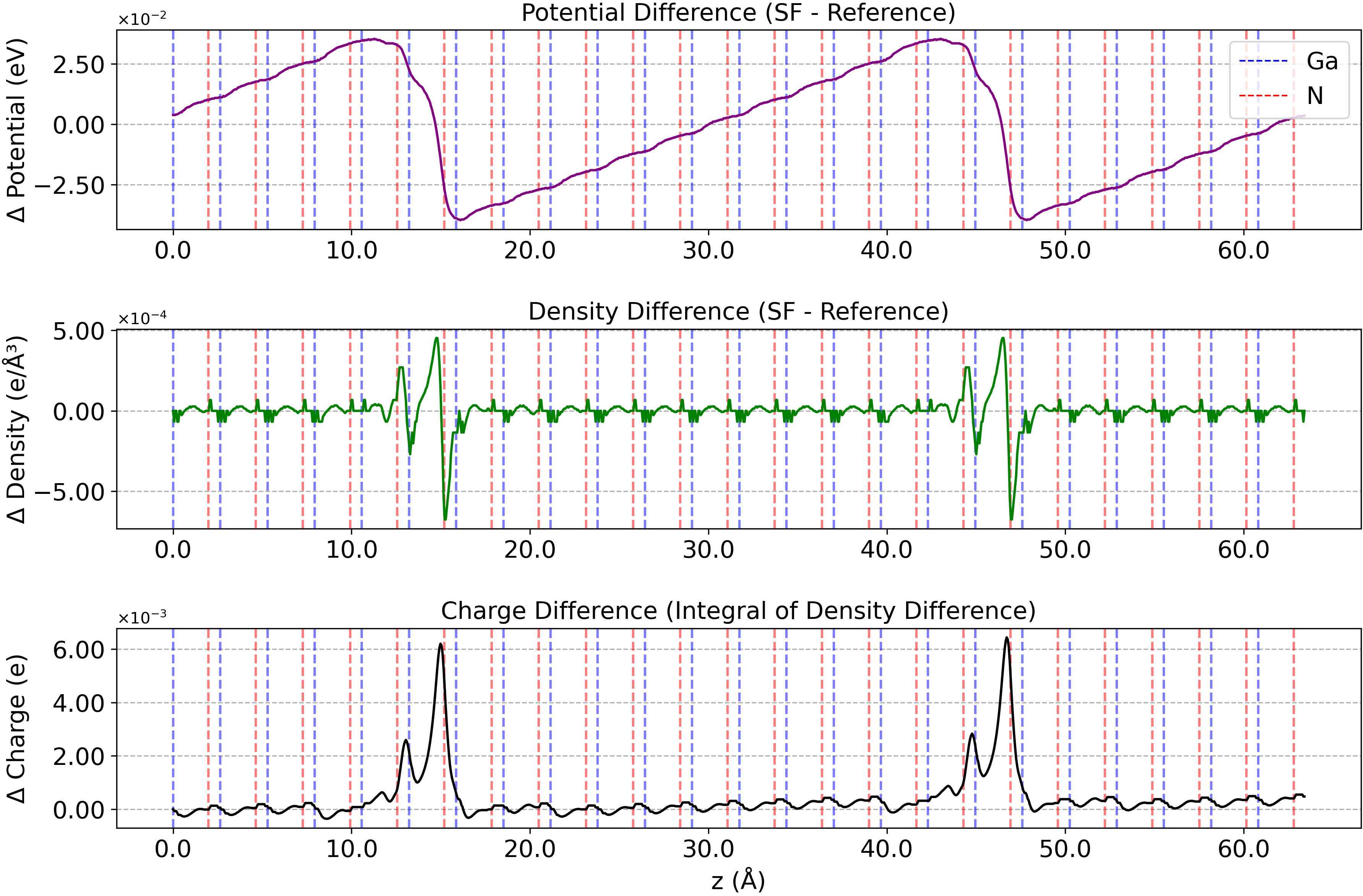}}\\ 
\subfloat[]{\includegraphics[width = 0.95\columnwidth]{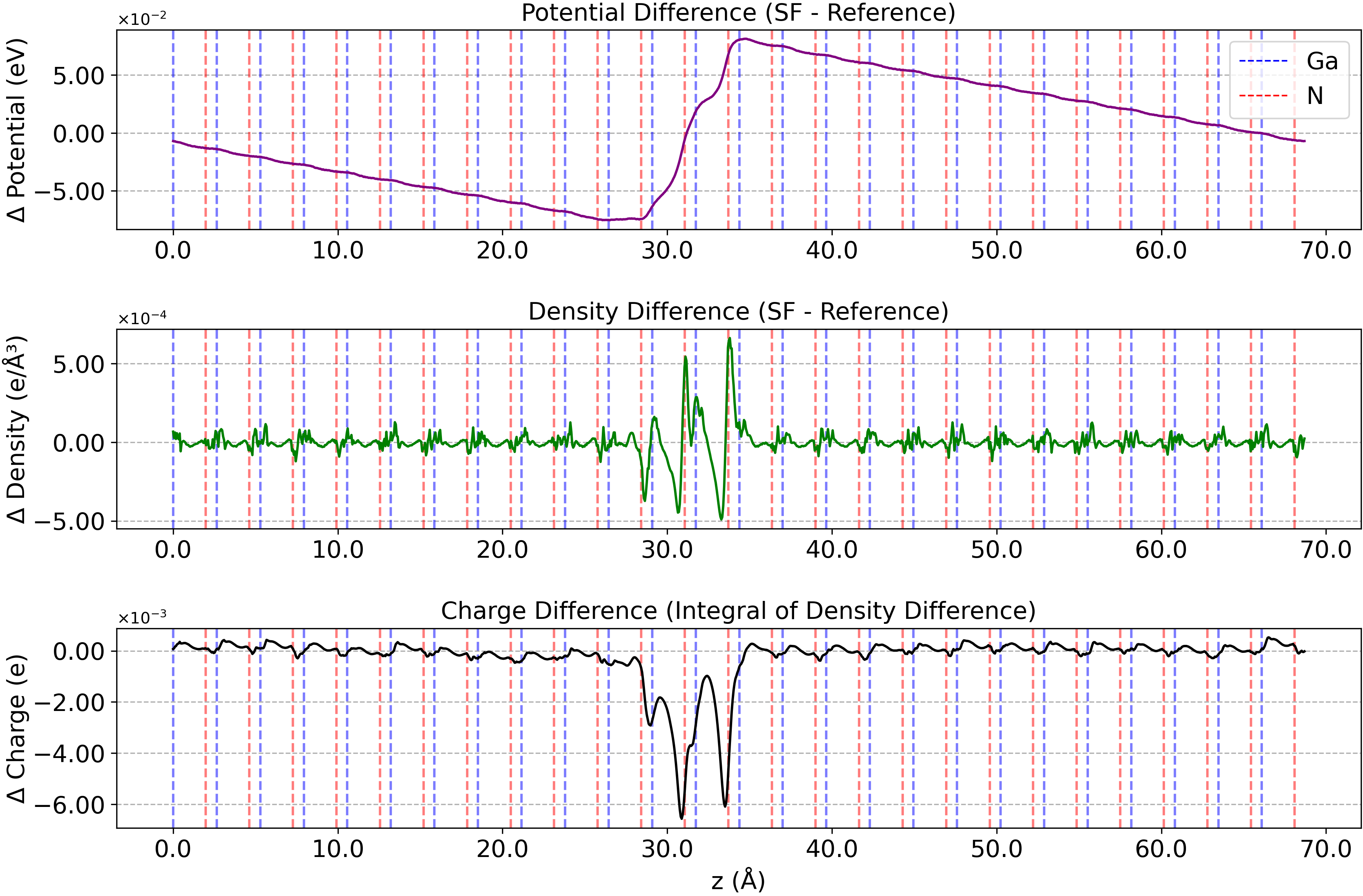}}
\caption{Potential difference between faulted and perfect material, averaged in the x-y plane (top), density difference found in the same way (middle) and density difference integrated along z (bottom) for (a) I$_{1}$ SF in wz and (b) Intrinsic SF in zb.}
\label{fig:potentialtrace}
\end{figure}

It is clear from Fig.~\ref{fig:potentialtrace}(a) that, relative to perfect material, charge accumulates on the Ga side of the SF in wz in agreement with Fig.~\ref{fgr:Band_Density_I1}.  In Fig.~\ref{fig:potentialtrace}(b) we see that the opposite is true in zb, with charge reducing on the Ga side of the SF and accumulating on the N side, consistent with both Fig.~\ref{fgr:Band_Density_ZB_In} and the change in the potential.  The sign difference between the two cases occurs because the change in the stacking sequence in wz (from AB to ABC) is the exact opposite of the stacking sequence in zb (from ABC to AB).  (In terms of Fig.~\ref{fig:potentialtrace}(a), for the potential we plot $V_{Faulted} - V_{wz}$ which in the area around the SF will be $V_{ABC} - V_{AB}$; in Fig.~\ref{fig:potentialtrace}(b) in the area around the SF we will effectively be plotting $V_{BC} - V_{ABC}$.)  The localisation of the band densities is then easily understood in terms of the potential change across the stacking faults.

The change in potential arises from the break in the stacking sequence, or equivalently the break in translational symmetry.  The nature of the bonding does not change at the stacking fault: bond angles and lengths are essentially the same, and the layer spacing does not change (plots for layer spacings are shown in the SI).  As seen in Fig.~\ref{fig:potentialtrace}, the potential change across the stacking faults is no more than 0.1\,eV, which is a small perturbation on the variation in electrostatic potential from layer to layer (as is standard in covalent systems this is of the order of 10\,eV).  Only the bands within a few tenths of an electron volt of the Fermi energy are affected by this change.

It would be useful to consider the effect of the stacking faults on the polarisation of the material, but this is challenging: while polarisation can be calculated using the modern theory of polarisation\cite{Resta:2007gd}, it is only polarisation differences between two systems which can be adiabatically transformed from one to the other that are meaningful.  For the stacking faults in GaN, there are three cases where this is possible because the simulation cell has the same number of layers as an equivalent perfect bulk cell: the twin SF in zb; and the intrinsic 1 and intrinsic 3 SFs in wz.  Using Resta's formulation\cite{Resta:1999us} we find polarisation differences of 0.002~C/m$^{2}$ for the twin SF in zb, and -0.003 and -0.007~C/m$^{2}$ for the I$_{1}$ and I$_{3}$ SFs in wz (note that there are two SFs in each wz cell).  These numbers suggest that there is only a small effect on the polarisation of the underlying material.

We can consider this further through consideration of the mechanism that gives rise to polarisation in wz GaN: the deviation of the internal u parameter and the c/a ratio from their ideal values (from 0.375 to 0.376, and 1.633 to 1.631 respectively).  It is not possible to study these values directly in a simulation cell containing a SF which has undergone structural optimisation, but we can identify a proxy: the ratio of the distance from a Ga layer to a N layer along the c-axis of the simulation cell to the distance from the N layer to the next Ga layer along the c-axis.  In zb GaN, this ratio is 3.00 (reflecting the non-polarity of the material) while we calculate it to be 3.03 in bulk wz GaN.  We plot this ratio for all stacking faults in the supplementary information, finding that in all cases there is only a small deviation from the ideal value as we pass through the stacking fault: in zb simulation cells it remains around 3.00, while in wz simulation cells it is close to 3.03.  This underlines our earlier conclusion that the SF produces only a small perturbation on the underlying material, and that the standard picture of a SF acting as a small inclusion of one material in the other is not accurate.

\begin{figure}[ht]
\subfloat[WZ CBM]{\includegraphics[width = 0.47\columnwidth]{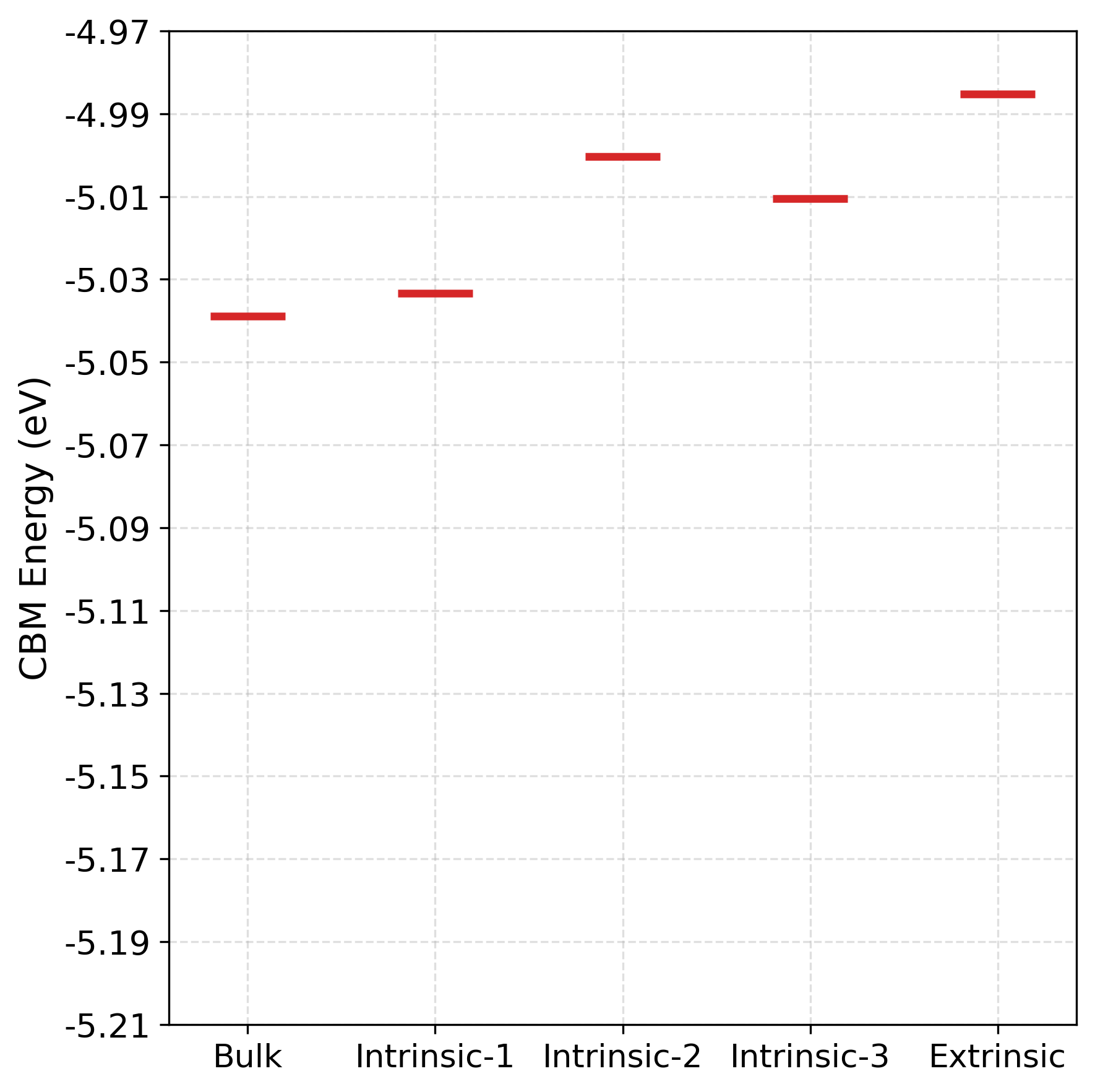}}
\subfloat[ZB CBM]{\includegraphics[width = 0.47\columnwidth]{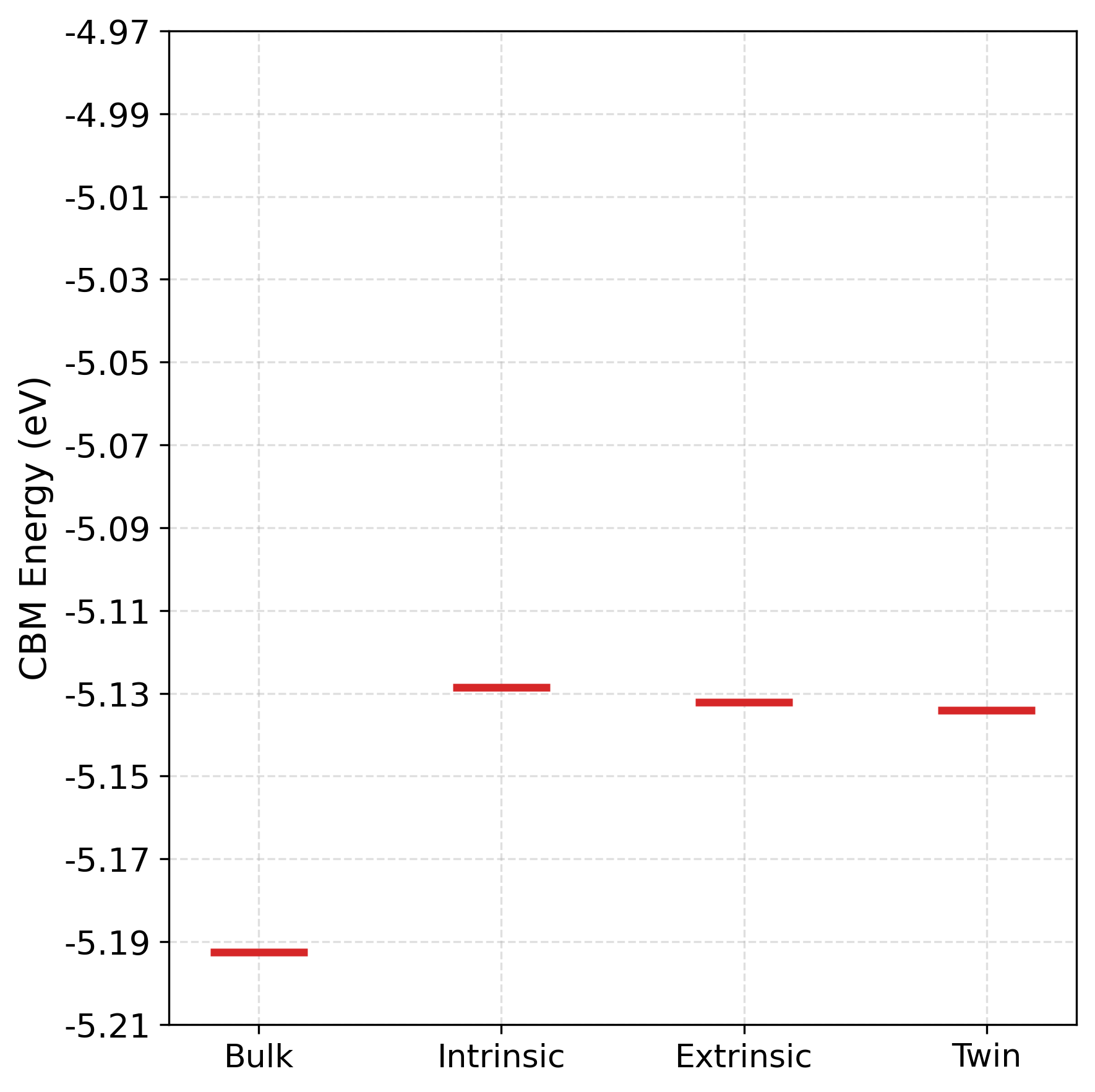}}\\ 
\subfloat[WZ VBM]{\includegraphics[width = 0.47\columnwidth]{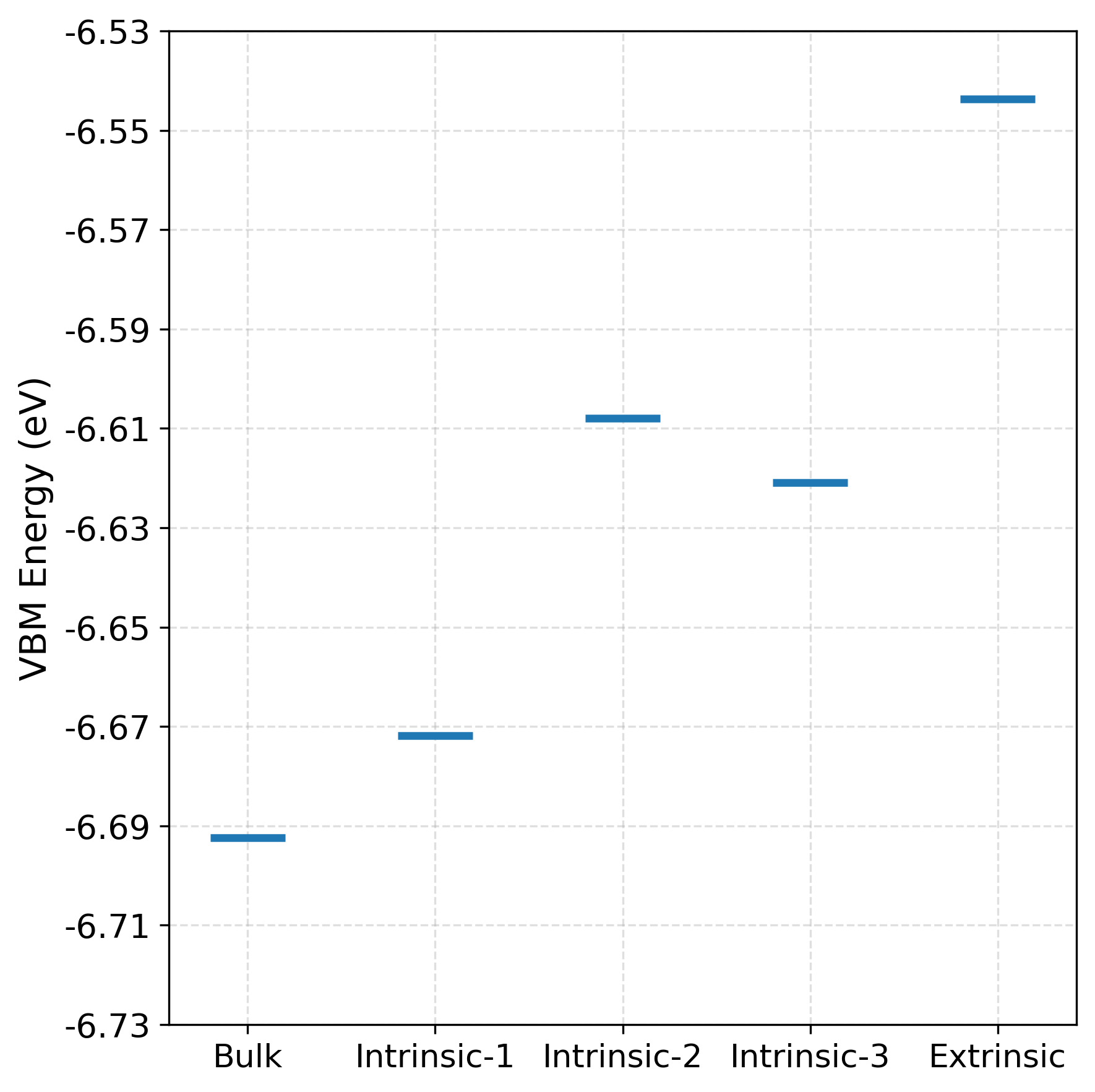}} 
\subfloat[ZB VBM]{\includegraphics[width = 0.47\columnwidth]{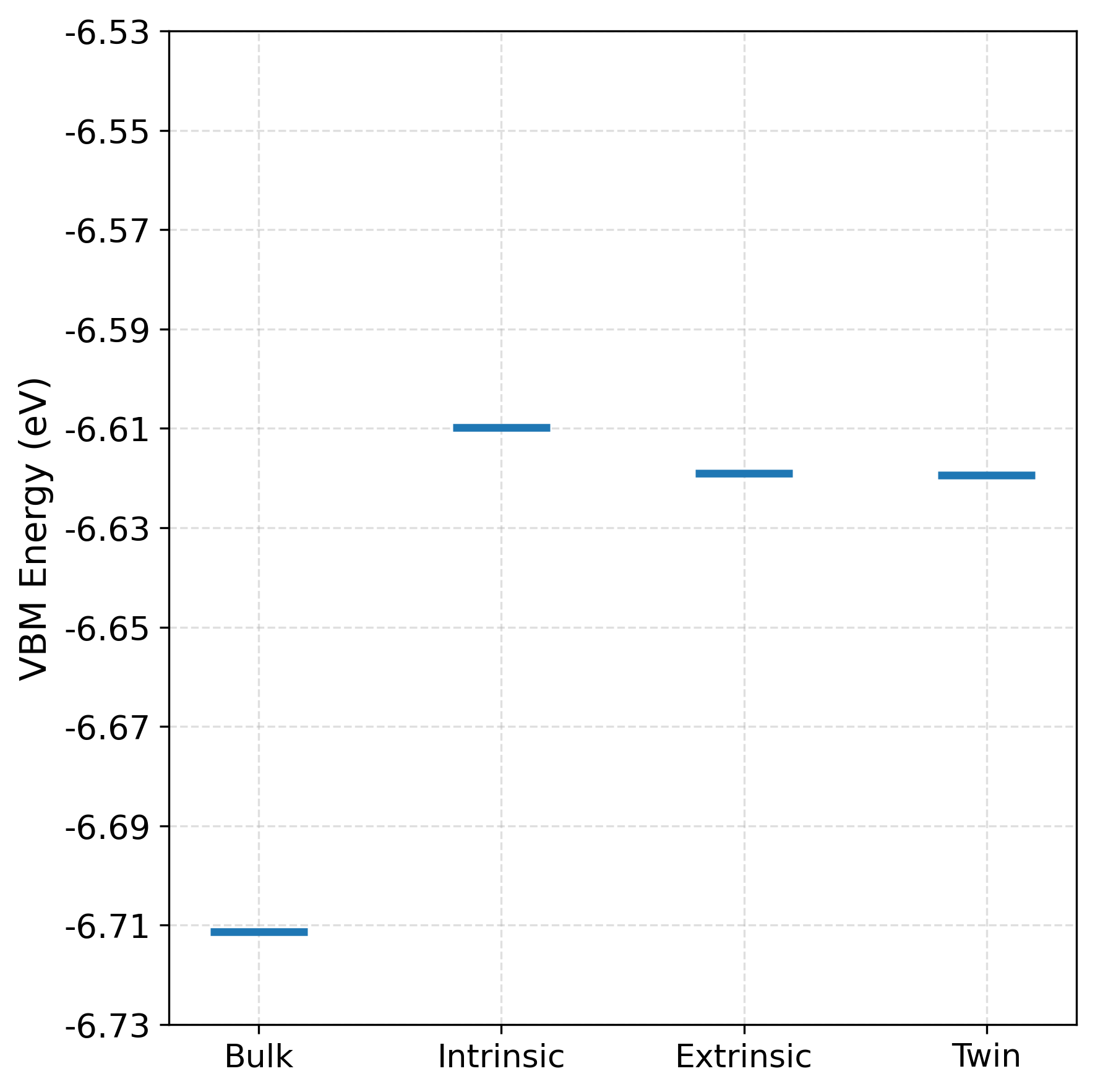}}\\ 
\caption{Band edges: (a) WZ CBM; (b) ZB CBM; (c) WZ VBM; (d) ZB VBM.}
\label{fig:bandedges}
\end{figure}

It is important to understand the nature of the band alignment around the SFs (whether a type I or type II interface) as it can affect the trapping of electrons and holes and hence the optoelectronic properties\cite{caro2013theory}.
We compare the band edges in simulation cells with stacking faults to cells of perfect bulk, ensuring that the deep-lying non-dispersive $d$-bands are aligned as explained in Sec.~\ref{sec:comp-meth-models}.  The band edges are shown in Fig.~\ref{fig:bandedges} for both wz and zb.
It should be noted that, as is standard for most DFT calculations, our calculated band gaps 1.65 eV (bulk wz GaN) and 1.52 eV (bulk zb GaN) are significantly underestimated compared to the experimental values (3.51 eV and 3.30 eV for bulk wz and zb GaN respectively\cite{Vurgaftman:2003dm}).
This underestimate is a well-known feature of standard DFT functionals, and our calculations are in good agreement with other DFT values\cite{benbedra2025energetics}.

We see that, for all SFs in zb, these defects are type II interfaces, with both valence band maximum and conduction band minimum shifted higher than in the bulk.  For wz the same is true, though for the I$_{1}$ SF (the most common) the conduction band minimum is extremely close to the energy in the bulk; the behavior is likely to be very sensitive to local fields and perturbations of the electronic structure, so that hints of type I behavior might be seen.  This is in broad agreement with existing simulation results\cite{lahnemann2012direct,stampfl1999density,Lu:2003so,Sun:2002vi}, though we are not aware of systematic testing of the size of simulation cell in other theory work.  We also see that the gap is reduced by $\sim$0.01~eV in the wz I1 SF and by $\sim$0.04~eV in zb I SF.  This result is in contradiction to simple models of SFs in zb, assumed to be wz insertions into the zb matrix, which would lead to an increase of the gap.  This highlights the importance of detailed electronic structure modelling of realistic models of materials.  Experimental CL and PL results show bands associated with SFs in wz samples which are at lower energies than the band edge\cite{Liu:2005nt,Haberlen:2010jw}, with shifts of $\sim$0.06--0.18~eV.
We note that the question of calculating band offsets with a GGA such as PBE used here can be challenging, particularly between different materials.  However, here we are considering offsets between systems with only a small structural change between them (from perfect bulk to the stacking faults) where these challenges with band alignment will be very small.

\begin{figure}[h]
\includegraphics[width = 0.95\columnwidth]{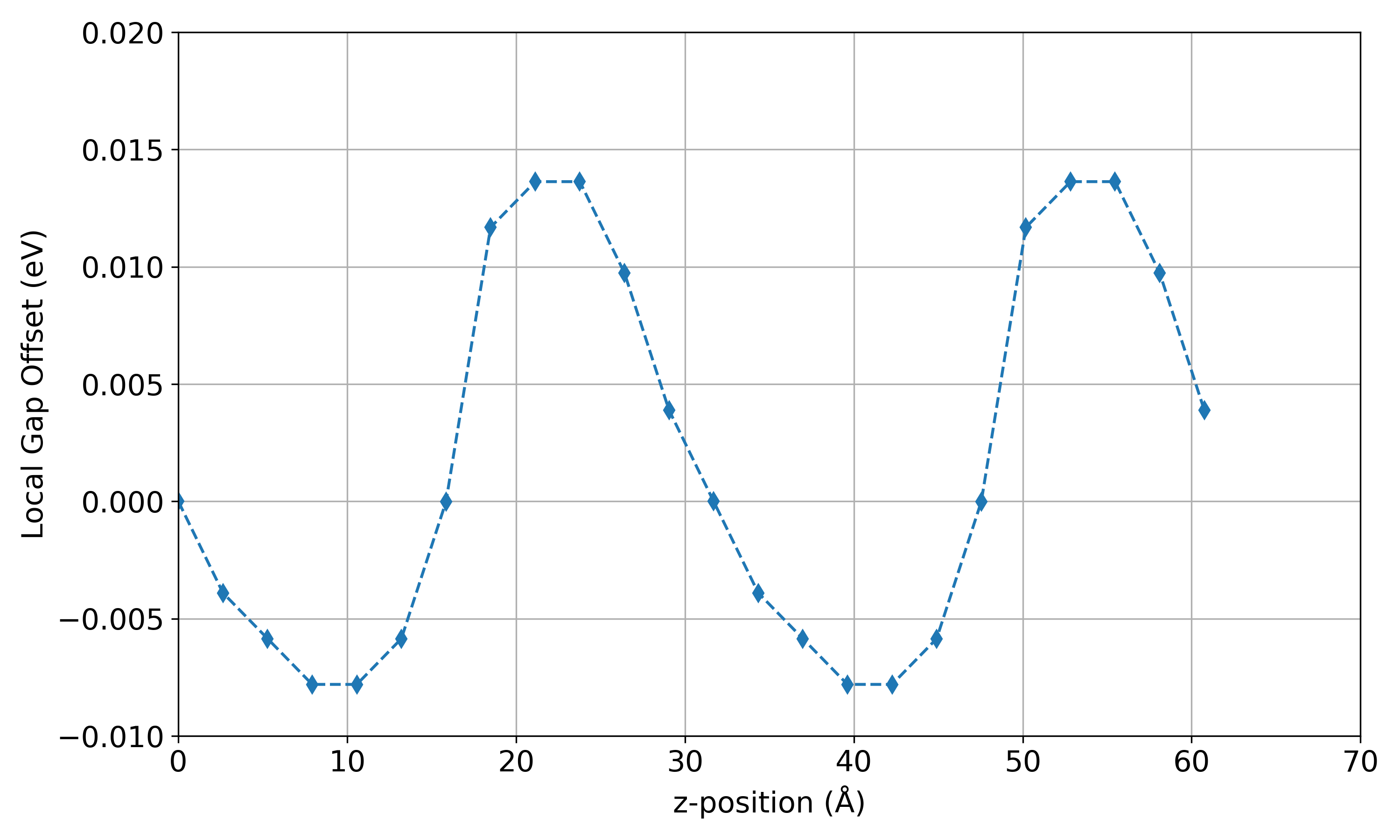}\\
\includegraphics[width = 0.95\columnwidth]{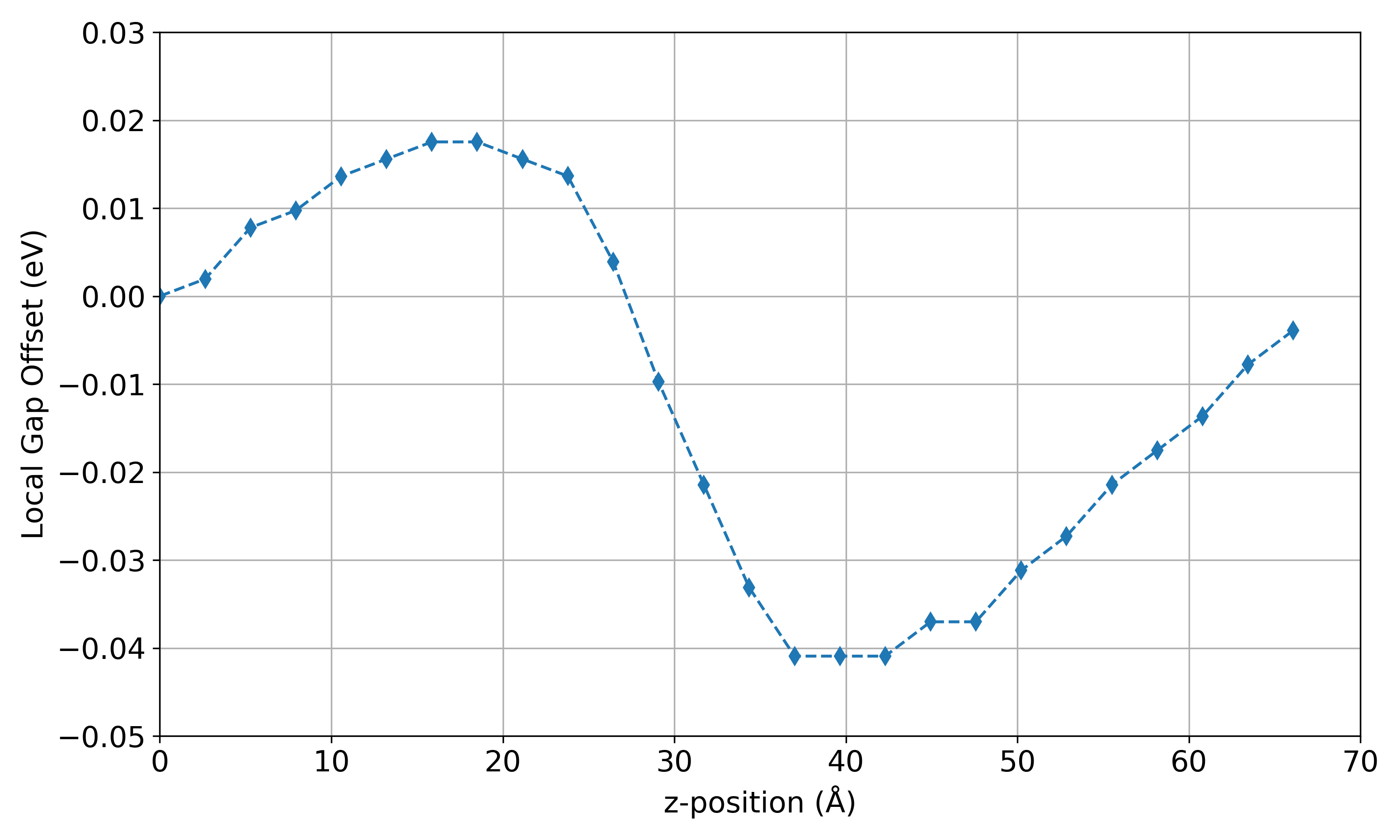}
\caption{Local band gap plotted across simulation cell for I$_{1}$ SF in wz GaN (a) and  I SF in zb GaN (b).}
\label{fgr:Band_Gap}
\end{figure}

We can further investigate the electronic properties of the stacking faults by resolving them spatially via the projected density of states (pDOS), and using this data to track the local positions of the VBM and CBM within the simulation cell.  Since our focus is on the band edges, the pDOS is not shown in the main text, but is provided for all the SFs in both wz and zb in the supplementary information.  In general the valence band maximum is more strongly affected than the conduction band minimum.
The variation of the local band gap through the simulation cell is shown in Fig.~\ref{fgr:Band_Gap} for the two most common stacking faults in wz and zb, with data for all the other SFs given in the supplementary information.  We see that the change in the gap is anti-correlated with the potential differences in Fig.~\ref{fig:potentialtrace}, with the gap widening where the potential is negative.  This also coincides with the location of the depletion in the valence bands, consistent with the shift of the VBM to lower energies.
Calculated band gap changes $\Delta E_g$ (eV), band offsets $\Delta E_C$ (eV) and $\Delta E_V$ (eV) for the stacking faults with respect to respective bulk crystals of wz and zb GaN are presented in supporting information.

\section{Conclusions}
\label{sec:conclusions}

We have performed a systematic study of stacking faults in wurtzite and zincblende phases of GaN using periodic DFT calculations. We have investigated all possible types of stacking faults structures, considering their stability and electronic properties to understand the structure properties relationship. This investigation is of central importance since stacking faults are very common defects in zb III-nitride materials, and non-polar oriented wz materials. From the structural relaxation, we can observe that the stacking faults have a rather small influence on lattice parameters and atomic positions; the inter-planar distance varies very slightly around the stacking faults. 

The formation energies are quite low. In wz, the I$_{1}$ intrinsic stacking fault has the lowest formation energy, while all stacking faults in zb have similar energies.  The formation energy in zb is much smaller than that in wz, so stacking faults are more likely to be formed in zb phase compared to the wz phase.

The analysis of the band resolved density reveals that near the Fermi energy the valence bands are concentrated on the opposite side of the stacking fault to the conduction bands.  However, the concentration happens on opposite sides in wz and zb, which is understood in light of the potential distribution around the SFs which take opposite signs in wz and zb as the change in stacking sequence is opposite.

Finally, a detailed analysis of the electronic structure shows that, in contrast to standard models, the band gap of stacking faults is \emph{smaller} than that of the ideal wurtzite and zincblende phases, with a positive offset. This change in the band edges is equivalent to a type II interface in all cases, though the most common wz stacking fault is highly sensitive to local conditions.  Overall, we conclude that the common picture of a stacking fault as an inclusion of a small amount of one material in the other is not accurate.

The data that support the findings of this article are openly available\cite{Wang:2026nl}.

\acknowledgements{This research was funded through UKRI/EPSRC grant number EP/Y00423X/1.  We gratefully acknowledge many useful discussions and insights from Professor Rachel Oliver, Dr Martin Frentrup, Dr Petr Vacek and Professor David Wallis of Cambridge University.  This research was partly carried out using the high performance computing 
cluster in the London Centre for Nanotechnology. The authors acknowledge the use of the UCL Kathleen High Performance Computing Facility (Kathleen\@UCL), and associated support services, in the completion of this work.  We are grateful for computational support from the UK Materials and Molecular Modelling Hub, which is partially funded by EPSRC (EP/T022213/1, EP/W032260/1 and EP/P020194/1), for which access was obtained via the UKCP consortium and funded by EPSRC grant ref EP/P022561/1.}

\appendix
\section{Basis sets}
\label{sec:basis-sets}

The pseudo atomic orbital (PAO) basis sets\cite{Bowler:2019fv} used for Ga and N were double-zeta plus polarization (DZP) basis sets, generated with the equal radius setting in \textsc{Conquest}.  The gallium pseudopotential includes the $3d$ electrons as semi-core states, with the radii of the PAOs in Bohr as follows: $3d$: 3.58\,$a_{0}$; $4s$ and $4p$: 8.06 and 4.33\,$a_{0}$; 4d (polarization): 8.06\,$a_{0}$.  For N the radii are: $2s$ and $2p$: 5.48 and 2.81\,$a_{0}$; 3d (polarization): 5.48\,$a_{0}$.  For indium, the pseudopotential includes the $4d$ electrons as semi-core states, with the radii of the PAOs in Bohr as follows: $4d$: 4.11\,$a_{0}$; $5s$ and $5p$: 8.41 and 4.63\,$a_{0}$; 5d (polarization): 8.41\,$a_{0}$.  For aluminium, the radii of the PAOs in Bohr are as follows: $3s$ and $3p$: 8.24 and 4.48\,$a_{0}$; 3d (polarization): 8.24\,$a_{0}$.

\section{Physical parameters}
\label{sec:physical-parameters}

\begin{figure*}[ht]
  \includegraphics[width = 0.5\columnwidth]{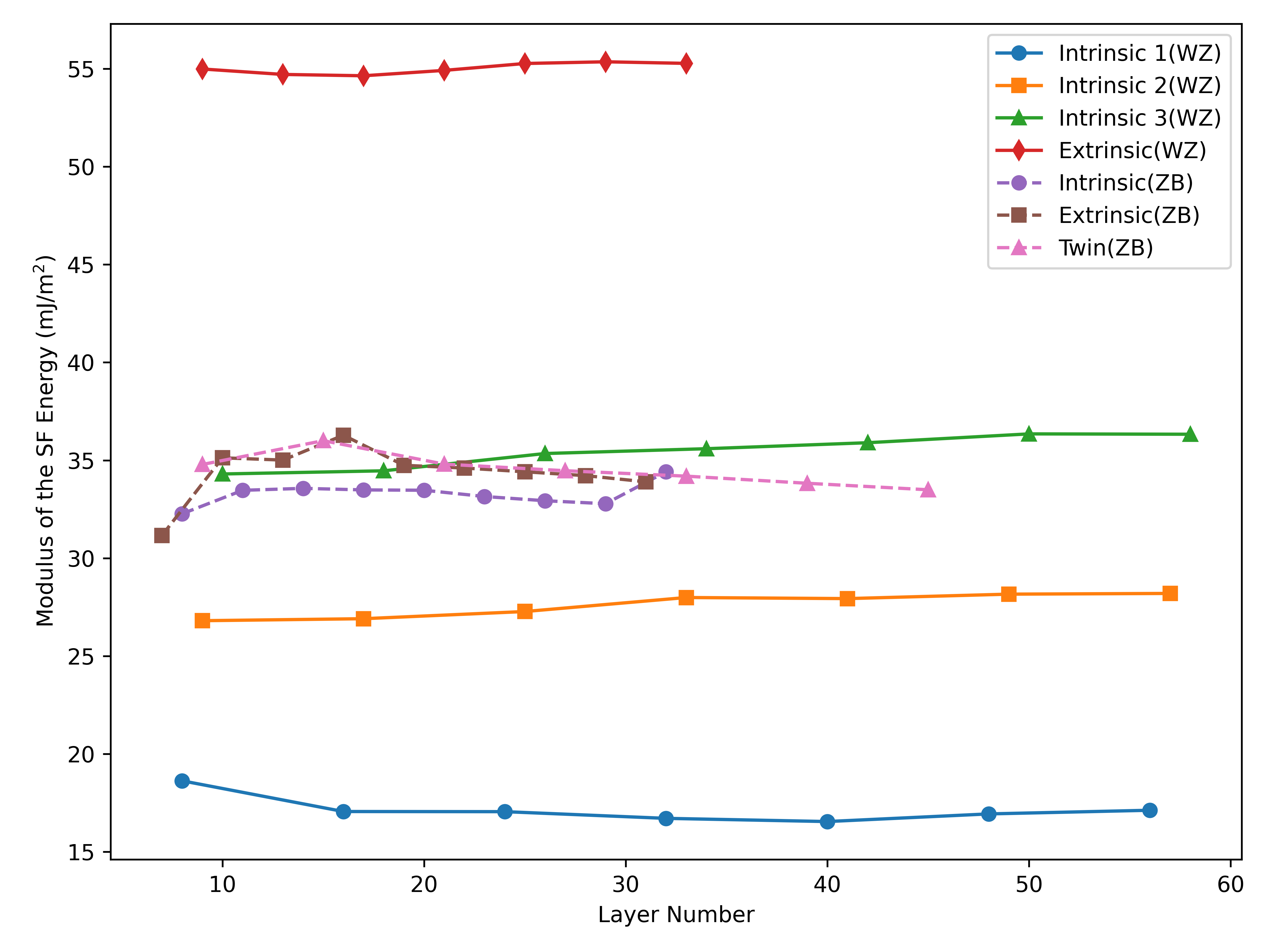}
  \caption{SF formation energies with number of layers for different types of SFs in WZ and ZB GaN.}
  \label{fig:SF_energy_conv}
\end{figure*}

In Table~\ref{tbl:Str} we give bulk lattice constants for GaN (along with AlN and InN for comparison).  In Table~\ref{tbl:WZ} we show the physical parameters for the stacking fault models, and in Table~\ref{tbl:WZ-ZB-All} we show the SF formation energy.  The convergence of SF formation energy is shown graphically in Figure~\ref{fig:SF_energy_conv}.

\onecolumngrid
\begin{table*}[ht]
\begin{center}
  \caption{Comparison of calculated hexagonal lattice parameters a, c (\AA ) of wz and cubic lattice parameter a (\AA ) in zb phases of GaN, InN and AlN with available literature data. }
  \label{tbl:Str}
  \begin{tabular}{*{8}{c}}
    \hline
    \hline
    \multicolumn{8}{c}{WZ}     \\
    \hline
& & $a$ & & &$c$ & & c/a \\
     \hline
     & This Study &  Previous Study \cite{gao2019point}  & Exp. \cite{ding2021study, schulz1977crystal} & This Study & Previous Study \cite{gao2019point}  & Exp. \cite{frentrup2017x, ding2021study, schulz1977crystal} & This Study \\
    \hline
    GaN & 3.239 & 3.24 & 3.19  & 5.284 & 5.23 &  5.19 & 1.631\\
    InN & 3.595 & 3.61 & 3.56  & 5.854 & 5.83 &  5.70 & 1.628\\
    AlN & 3.133 & 3.13 & 3.11  & 5.078 & 5.02 &  4.98 & 1.621\\
    \hline
    \multicolumn{5}{c}{ZB}     \\
    \hline
        &  &  $a$ &     \\
     \hline
        & This Study &  Previous Study \cite{de2011influence}  & Exp. \cite{frentrup2017x, ding2021study}   \\
    \hline
    GaN &  4.578   & 4.55 & 4.49 \\
    InN &  5.080   & 5.06 & 4.98 \\
    AlN &  4.421   & 4.40 & 4.37 \\
    \hline
    \hline
     \end{tabular}
\end{center}
\end{table*}

\begin{table*} [ht]
\begin{center}
  \caption{Considered models for stacking faults, number of atoms and c parameter (\AA)}
  \label{tbl:WZ}
  \begin{tabular}{*{12}{c}}
    \hline
    \hline
    \multicolumn{12}{c}{Stacking faults based on Wurtzite GaN}     \\
    \hline
    \hline
       I$_{1}$       & Atoms &  $c$  & I$_{2}$             & Atoms & $c$       &     I$_{3}$         &  Atoms & $c$  & Extrinsic & Atoms & $c$  \\
    AB-2CB-AB   & 32    & 21.154    & AB-2CA-C-AB    & 36        & 23.798    & AB-C-2BA-C-AB  &  40   & 26.441 & 2AB-C-2AB & 36  &  23.798  \\
   2AB-4CB-2AB  & 64    & 42.311    & 2AB-4CA-C-2AB  & 68        & 44.945    & 2AB-C-4BA-C-2AB & 72   & 47.598 & 3AB-C-3AB & 52  &  34.370   \\
   3AB-6CB-3AB  & 96   &  63.469    & 3AB-6CA-C-3AB  & 100       & 66.096    & 3AB-C-6BA-C-3AB & 104  & 68.741 & 4AB-C-4AB & 68  &  84.936   \\
   4AB-8CB-4AB  & 128   & 84.612    & 4AB-8CA-C-4AB  & 132       & 87.269    & 4AB-C-8BA-C-4AB & 136  & 89.891 & 5AB-C-5AB & 84   &  55.521     \\
   5AB-10CB-5AB & 160   & 105.767   & 5AB-10CA-C-5AB & 164       & 108.399   & 5AB-C-10BA-C-5AB & 168 & 111.039 & 6AB-C-6AB & 100 & 66.102     \\
   6AB-12CB-6AB & 192  & 126.946   & 6AB-12CA-C-6AB & 196       & 129.551   & 6AB-C-12BA-C-6AB & 200 & 132.198 & 7AB-C-7AB & 116 & 76.677       \\
   7AB-14CB-7AB &  224 & 279.823 & 7AB-14CA-C-7AB & 228   & 284.805   & 7AB-C-14BA-C-7AB & 232 & 289.773 & 8AB-C-8AB & 132  & 164.729        \\
    \hline
    \hline
    \multicolumn{9}{c}{Stacking faults based on Zincblende GaN}     \\
    \hline
    \hline
    Intrinsic   & Atoms  & $c$   &   Extrinsic   & Atoms  & $c$  & Twin    & Atoms &  $c$ \\
    ABC-BC-ABC  &   32     &   39.971       &  ABC-B-ABC   &  28    &  34.975  & ABC-BA-CABC       &  36 &  44.968 \\
    2ABC-BC-ABC   &   44    &   54.961      & 2ABC-B-ABC   &  40   &  49.964  & 2ABC-BA-C2ABC  &  60  &  74.946  \\
    2ABC-BC-2ABC   &   56     &   69.950    & 2ABC-B-2ABC  &   52   &  64.953  & 3ABC-BA-C3ABC     & 84   &  104.925   \\
    3ABC-BC-2ABC      &   68     &  84.939    & 3ABC-B-2ABC  &  64   &  79.943    & 4ABC-BA-C4ABC     &  108  &  134.903  \\
    3ABC-BC-3ABC  &    80     &   99.928      & 3ABC-B-3ABC  &   76    &  94.932   &  5ABC-BA-C5ABC     &  132  &  164.882\\
    4ABC-BC-3ABC      &    92     &     114.917    & 4ABC-B-3ABC       &   88    &   109.921                & 6ABC-BA-C6ABC & 156 &  194.859     \\
    4ABC-BC-4ABC      &   104    &     129.906       & 4ABC-B-4ABC       &    100   &    124.910               & 7ABC-BA-C7ABC  & 180  &    224.838      \\
    5ABC-BC-4ABC      &   116      &    144.896     & 5ABC-B-4ABC       &   112    &     139.899              &    &    &         \\
    5ABC-BC-5ABC      &  128     &    159.885       & 5ABC-B-5ABC       &   124    &     154.889              &    &    &         \\
    \hline
  \end{tabular}
\end{center}
\end{table*}

\begin{table*} [ht]
\begin{center}
  \caption{Calculated values of formation energy (mJ/m$^{2}$) of stacking faults with respect to increasing number of layers}
  \label{tbl:WZ-ZB-All}
  \begin{tabular}{*{9}{c}}
    \hline
    \hline
    \multicolumn{8}{c}{Stacking faults based on Wurtzite GaN}     \\
    \hline
    \hline
       Intrinsic 1 (I$_{1}$)& $\Delta$E  & Intrinsic 2 (I$_{2}$)& $\Delta$E  & Intrinsic 3 (I$_{3}$)& $\Delta$E & Extrinsic & $\Delta$E \\
    AB-2CB-AB   & 18.627 & AB-2CA-C-AB    & 26.806 & AB-C-2BA-C-AB    & 34.301 & 2AB-C-2AB & 54.989 \\
   2AB-4CB-2AB  & 17.058 & 2AB-4CA-C-2AB  & 26.908 & 2AB-C-4BA-C-2AB  & 34.463 & 3AB-C-3AB & 54.711 \\
   3AB-6CB-3AB  & 17.051 & 3AB-6CA-C-3AB  & 27.275 & 3AB-C-6BA-C-3AB  & 35.346 & 4AB-C-4AB & 54.643 \\
   4AB-8CB-4AB  & 16.709 & 4AB-8CA-C-4AB  & 27.990 & 4AB-C-8BA-C-4AB  & 35.592 & 5AB-C-5AB & 54.918 \\
   5AB-10CB-5AB & 16.546 & 5AB-10CA-C-5AB & 27.935 & 5AB-C-10BA-C-5AB & 35.898 & 6AB-C-6AB & 55.270 \\
   6AB-12CB-6AB & 16.936 & 6AB-12CA-C-6AB & 28.162 & 6AB-C-12BA-C-6AB & 36.349 & 7AB-C-7AB & 55.357 \\
   7AB-14CB-7AB & 17.123 & 7AB-14CA-C-7AB & 28.199 & 7AB-C-14BA-C-7AB & 36.331 & 8AB-C-8AB & 55.426
 \\
    \hline
    \hline
    \multicolumn{8}{c}{Stacking faults based on Zincblende GaN}     \\
    \hline
    \hline
    Intrinsic           & $\Delta$E  & Extrinsic       & $\Delta$E & Twin              &  $\Delta$E \\
    ABC-BC-ABC          & -32.267   & ABC-B-ABC          & -31.159   & ABC-BA-CABC       & -34.786  \\
    2ABC-BC-ABC         & -33.466   & 2ABC-B-ABC          & -35.127   & 2ABC-BA-C2ABC    & -36.002  \\
    2ABC-BC-2ABC        & -33.567   & 2ABC-B-2ABC          & -35.010   & 3ABC-BA-C3ABC     & -34.804  \\
    3ABC-BC-2ABC        & -33.487  & 3ABC-B-2ABC          & -36.281 & 4ABC-BA-C4ABC    & -34.467  \\
    3ABC-BC-3ABC      & -33.470   & 3ABC-B-3ABC           & -34.744   & 5ABC-BA-C5ABC     & -34.190  \\
    4ABC-BC-3ABC      & -33.150   & 4ABC-B-3ABC           & -34.604   & 6ABC-BA-C6ABC      & -33.826  \\
    4ABC-BC-4ABC       & -32.934   & 4ABC-B-4ABC           & -34.413   & 7ABC-BA-C7ABC      & -33.494  \\
    5ABC-BC-4ABC      & -32.783   & 5ABC-B-4ABC         & -34.217   &                   &            &           \\
    5ABC-BC-5ABC       & -34.413   & 5ABC-B-5ABC         & -33.901   &                   &            &           \\
    \hline
  \end{tabular}
\end{center}
\end{table*}

\clearpage

\subsection{Layer spacings}
\label{sec:layer-spacings}

\begin{figure*}[ht]
\subfloat[]{\includegraphics[width = 0.5\columnwidth]{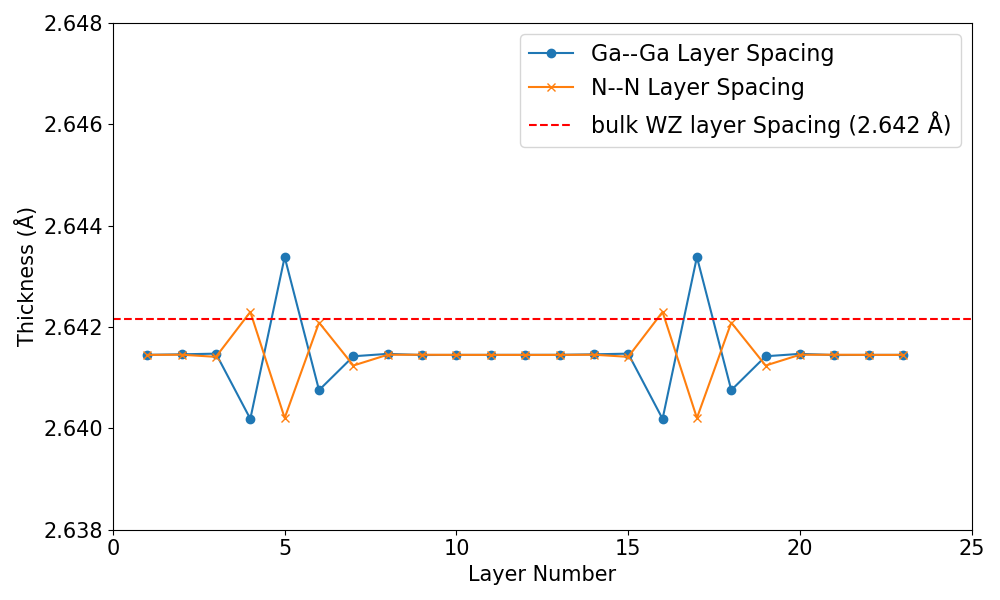}}
\subfloat[]{\includegraphics[width = 0.5\columnwidth]{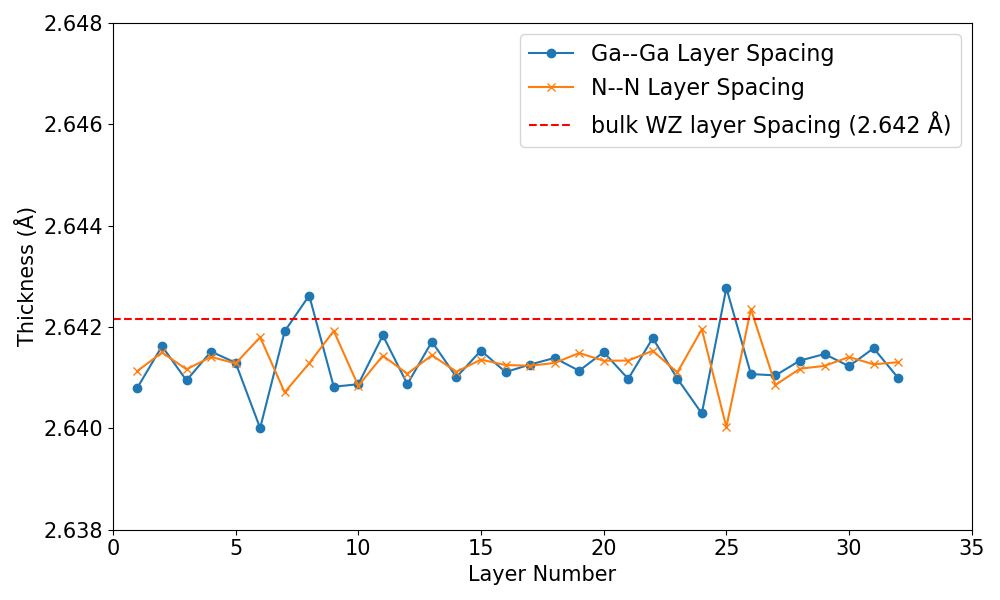}}\\
\subfloat[]{\includegraphics[width = 0.5\columnwidth]{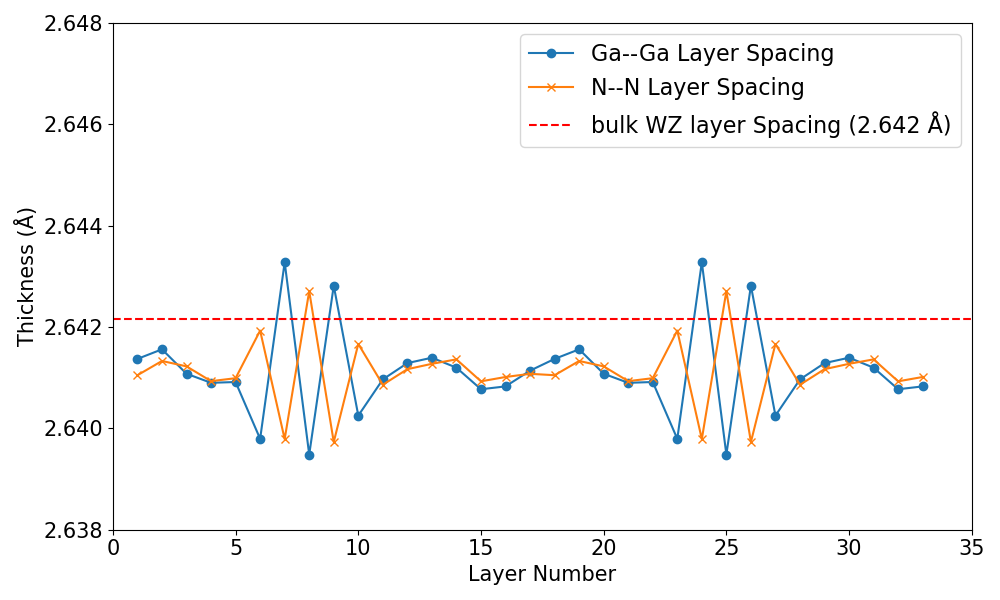}}
\subfloat[]{\includegraphics[width = 0.5\columnwidth]{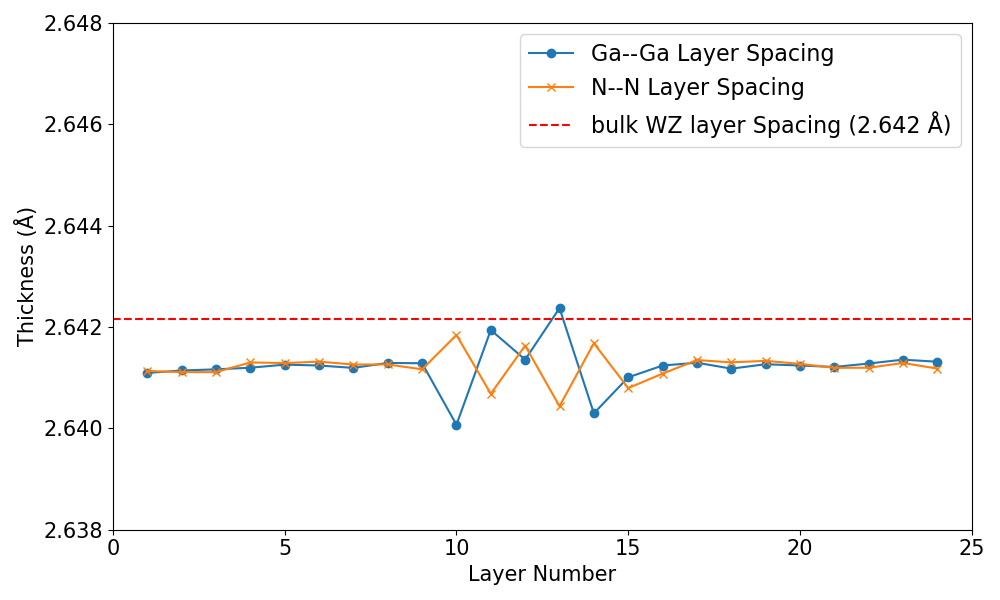}}
\caption{Evolution of layer spacing along the c-axis of the simulation cell of the wz GaN SFs with respect to calculated bulk wz GaN. I$_{1}$ (a), I$_{2}$ (b), I$_{3}$ (c) and Extrinsic (d) SFs}
\label{fgr:thickness_WZ}
\end{figure*}

\begin{figure*}[ht]
\subfloat[]{\includegraphics[width = 0.5\columnwidth]{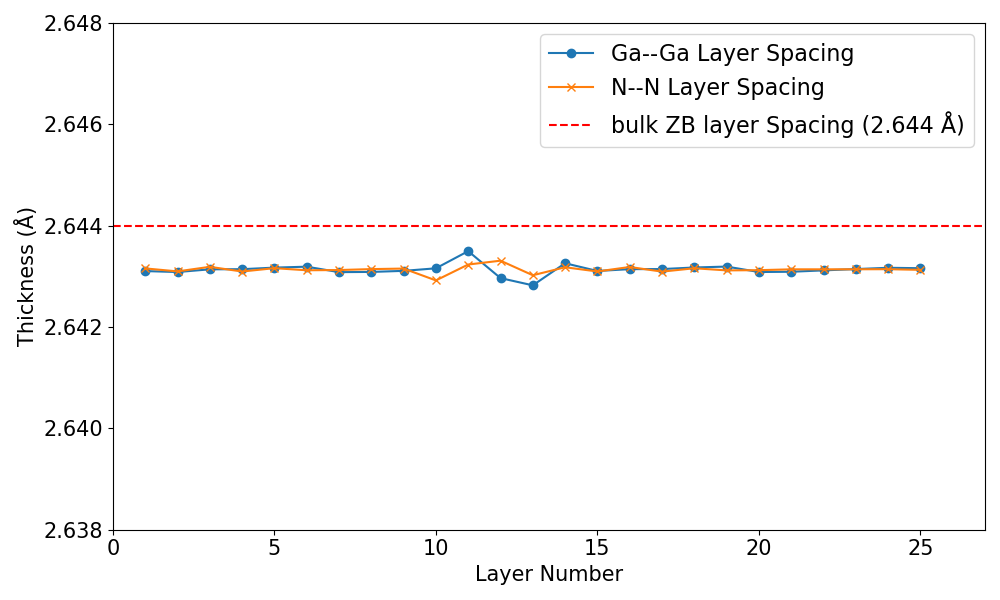}}
\subfloat[]{\includegraphics[width = 0.5\columnwidth]{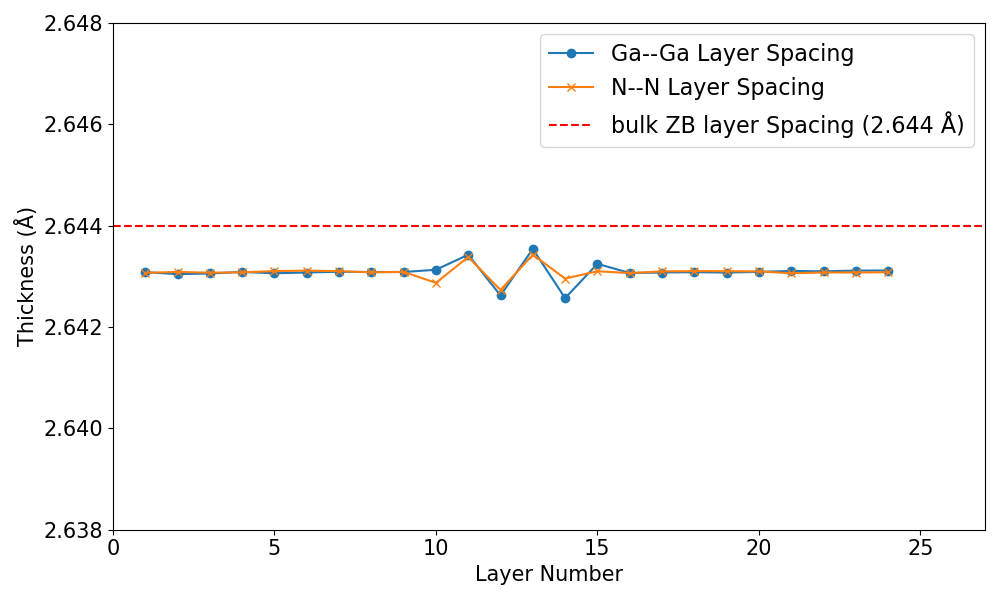}}\\
\subfloat[]{\includegraphics[width = 0.5\columnwidth]{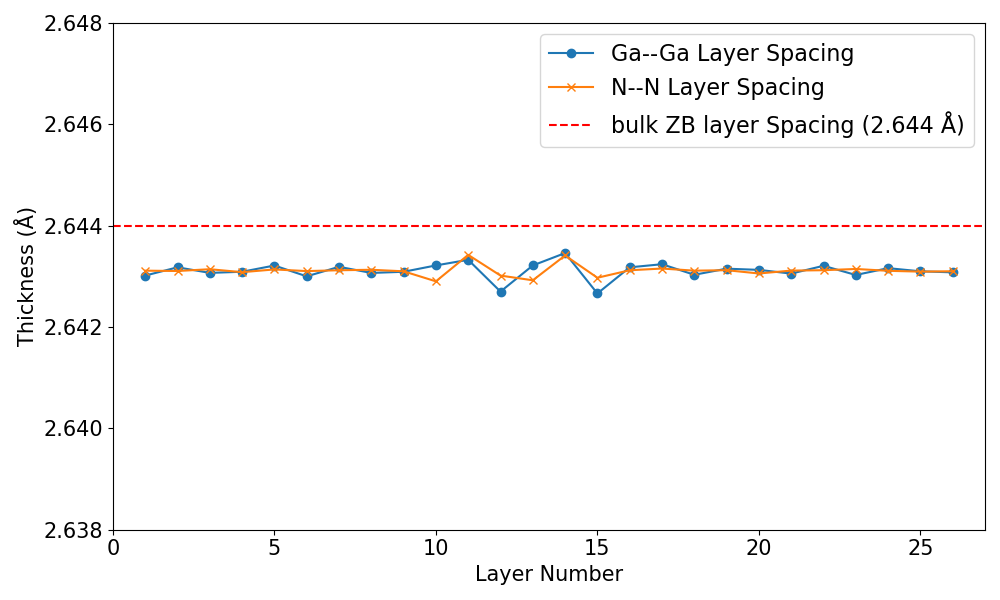}}
\caption{Evolution of layer spacing along the c-axis of the simulation cell  of the ZB GaN SFs with respect to calculated bulk zb GaN. Intrinsic (a), Extrinsic (b) and Twin (c) }
\label{fgr:thickness_zb}
\end{figure*}

\begin{figure*}[ht]
\includegraphics[width = \columnwidth]{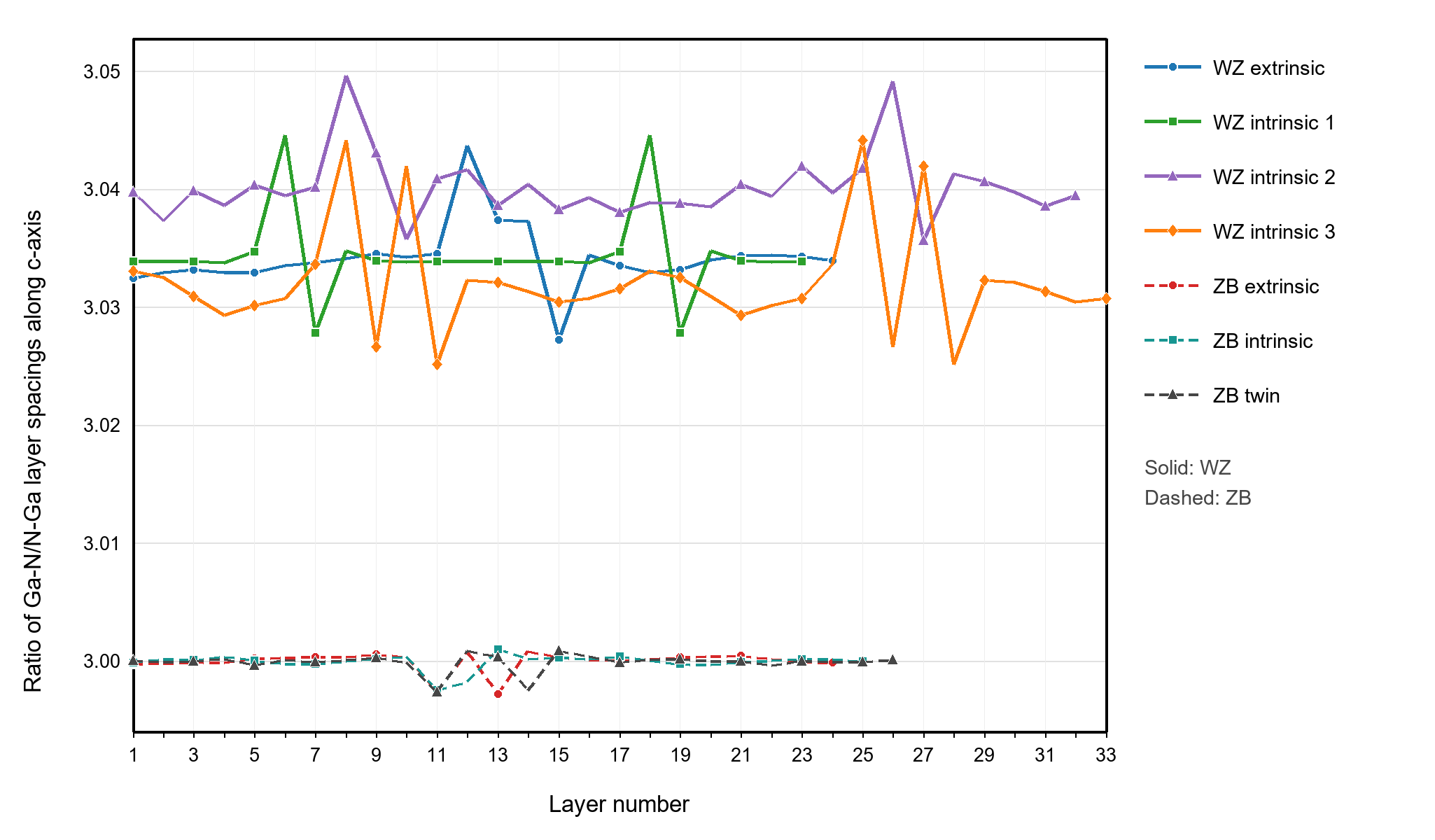}
\caption{Ratio of the adjacent interplanar Ga–N to N–Ga layer spacings along the c-axis  for different types of SFs in WZ and ZB GaN.}
\label{fig:c_axis_ratio}
\end{figure*}

We show here the spacing between layers in the SF models for wurtzite in Figure~\ref{fgr:thickness_WZ} and for zincblende in Figure~\ref{fgr:thickness_zb}, plotting both Ga-Ga and N-N distances; we see almost no change in spacing through the stacking faults.  In Fig.~\ref{fig:c_axis_ratio}, we then plot the ratio between successive Ga-N and N-Ga spacings along the c-axis to offer insight into the deviation of the stacking fault material from the ideal crystal (where the value would be 3.00 for zincblende GaN and 3.03 for wurtzite GaN).  Again, we see only a very small deviation from the ideal crystal values.

\clearpage

\section{Band Densities}
\label{sec:band-densities}

\begin{figure*}[ht]
\subfloat[]{\includegraphics[width = 0.8in]{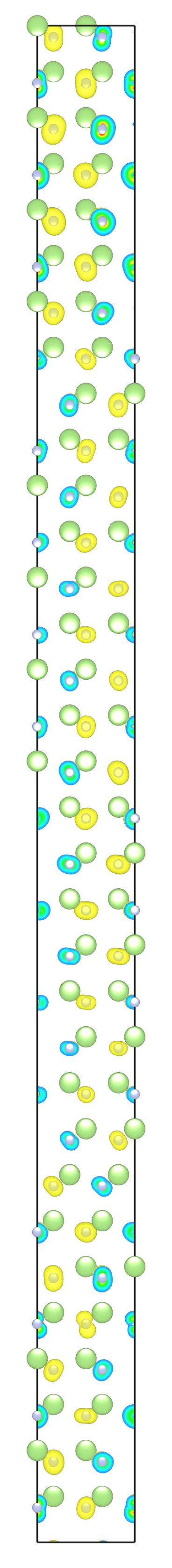}} 
\subfloat[]{\includegraphics[width = 0.8in]{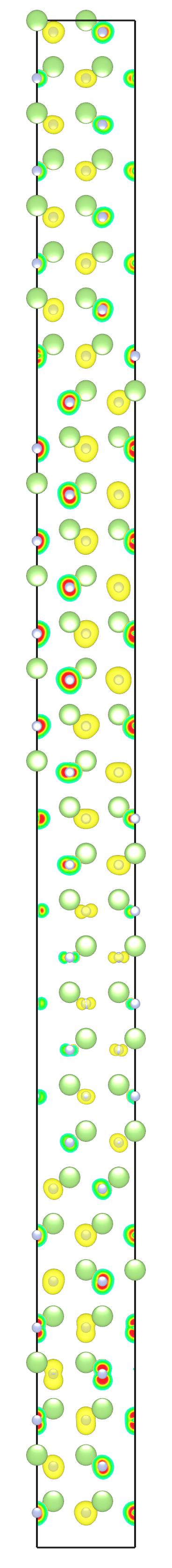}}
\subfloat[]{\includegraphics[width = 0.8in]{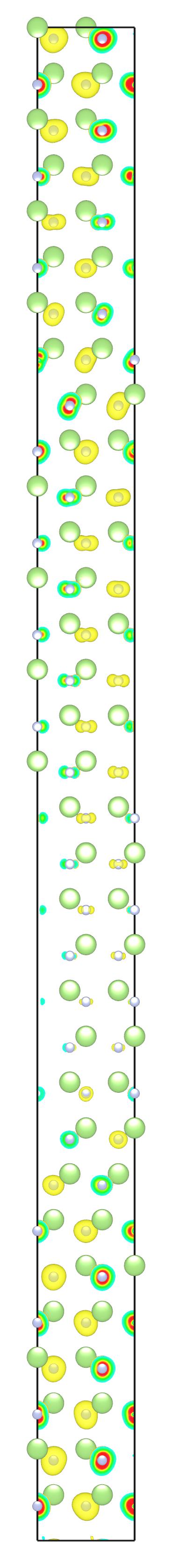}}
\subfloat[]{\includegraphics[width = 0.8in]{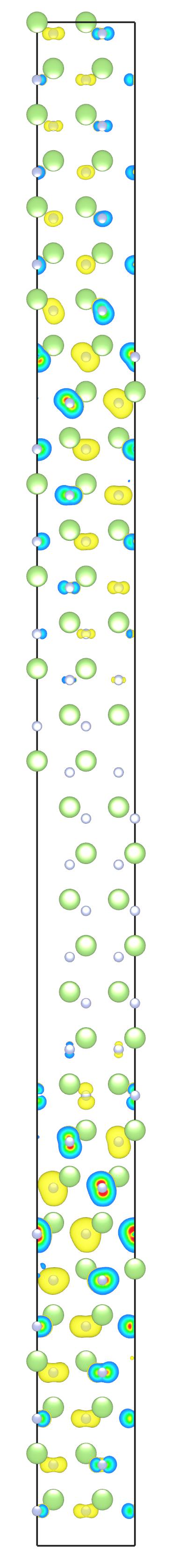}}
\subfloat[]{\includegraphics[width = 0.8in]{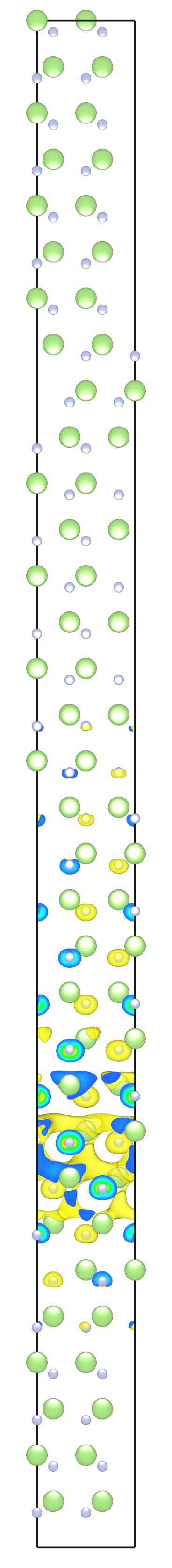}}
\subfloat[]{\includegraphics[width = 0.8in]{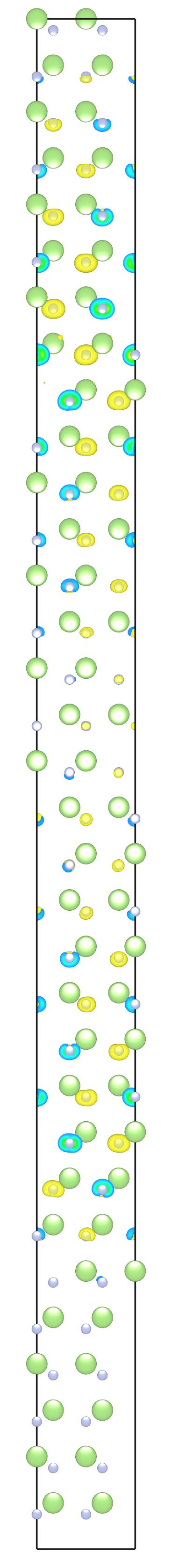}}
\subfloat[]{\includegraphics[width = 0.8in]{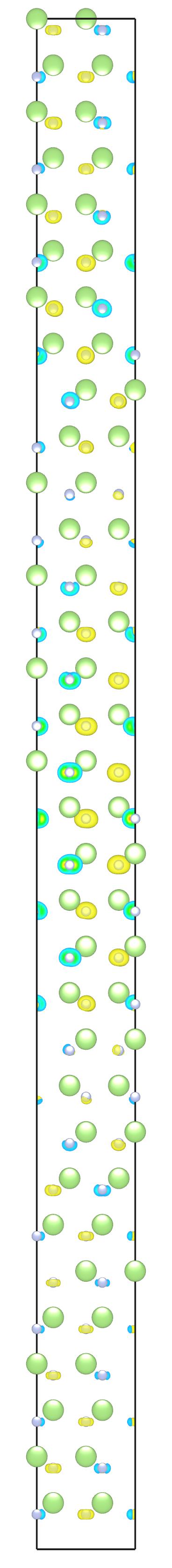}}
\subfloat[]{\includegraphics[width = 0.8in]{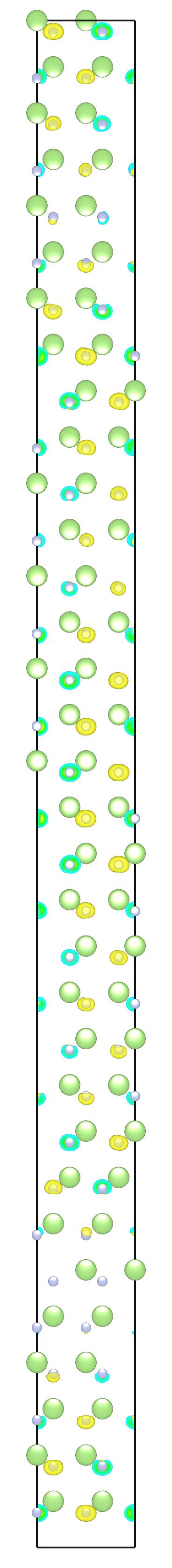}}
\caption{ Band Density of Intrinsic-2 (4AB-8CA-C-4AB)  stacking fault of wz GaN. (a) Band 591, (b) Band 592, (c) Band 593, (d) Band 594, (e) Band 595, (f) Band 596, (g) Band 597, (h) Band 598. Fermi energy (Ef: -5.86 eV) lies in between band 594 and band 595. }
\label{fgr:Fig-8}
\end{figure*}

\begin{figure*}[ht]
\subfloat[]{\includegraphics[width = 0.8in]{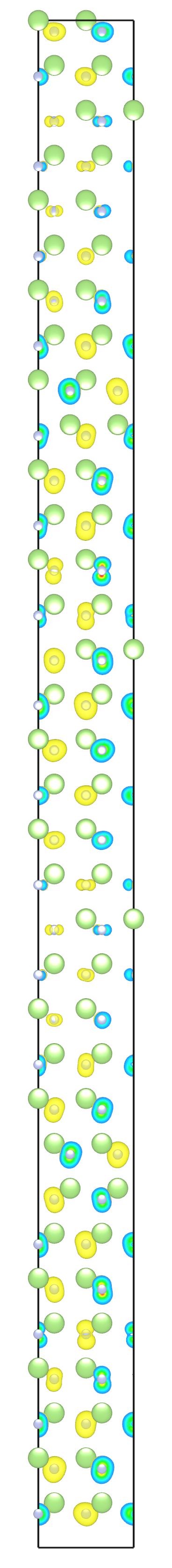}} 
\subfloat[]{\includegraphics[width = 0.8in]{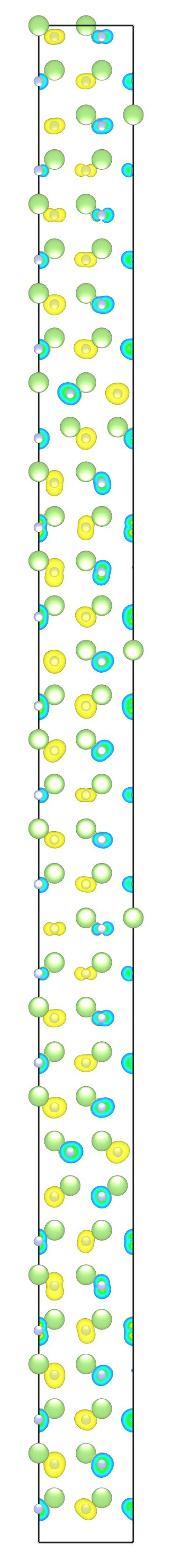}}
\subfloat[]{\includegraphics[width = 0.8in]{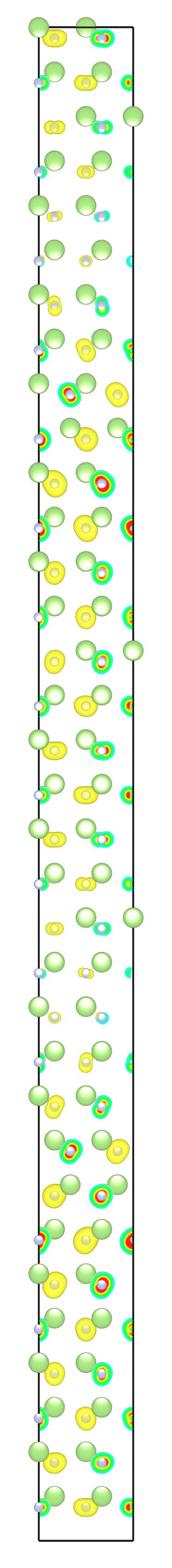}}
\subfloat[]{\includegraphics[width = 0.8in]{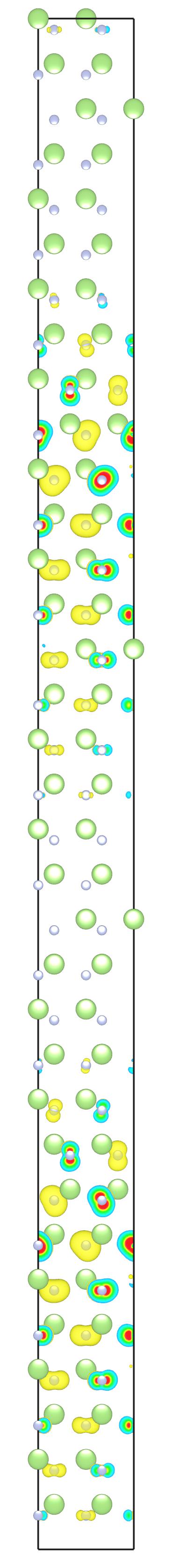}}
\subfloat[]{\includegraphics[width = 0.8in]{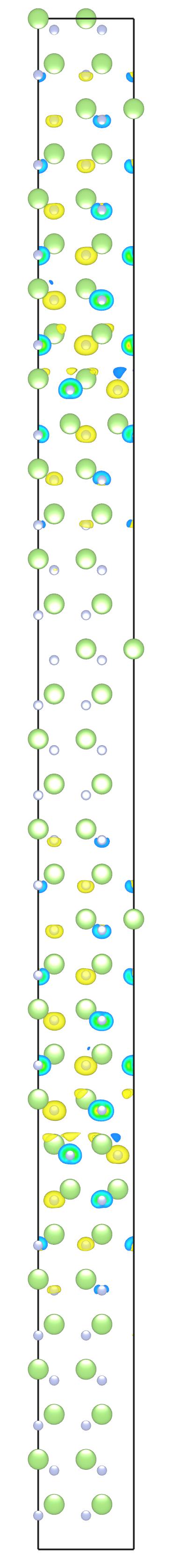}}
\subfloat[]{\includegraphics[width = 0.8in]{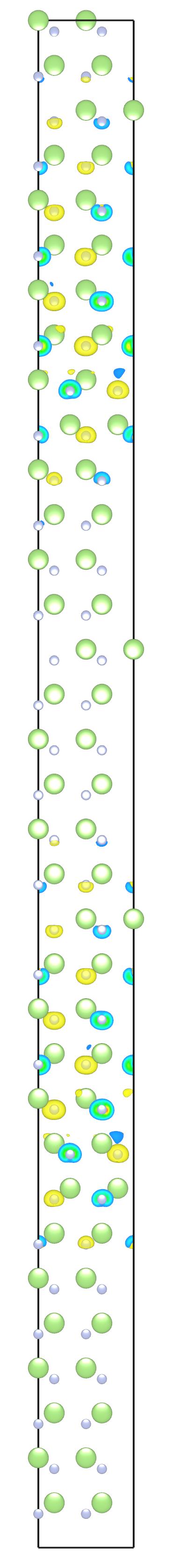}}
\subfloat[]{\includegraphics[width = 0.8in]{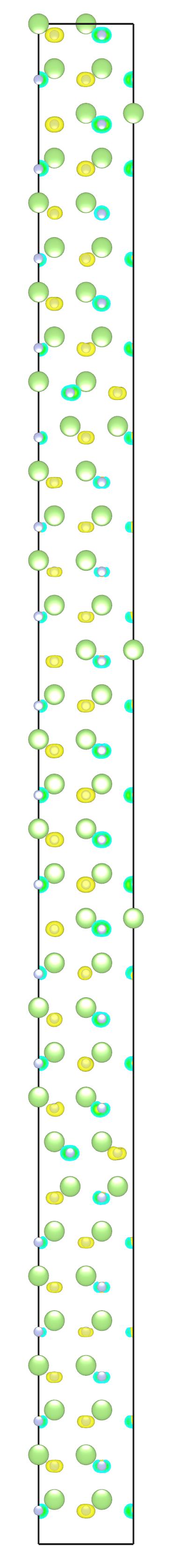}}
\subfloat[]{\includegraphics[width = 0.8in]{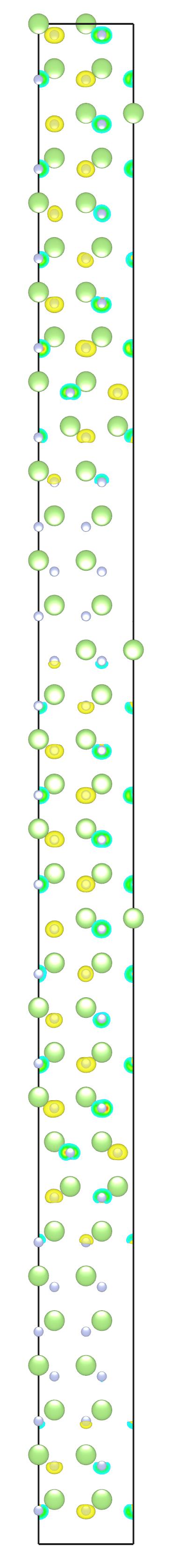}}
\caption{ Band Density of Intrinsic-3 (4AB-C-8BA-C-4AB)  stacking fault of wz GaN. (a) Band 609, (b) Band 610, (c) Band 611, (d) Band 612, (e) Band 613, (f) Band 614, (g) Band 615, (h) Band 616. Fermi energy (Ef: -5.86 eV) lies in between band 612 and band 613.}
\label{fgr:Fig-9}
\end{figure*}

\begin{figure*}[ht]
\subfloat[]{\includegraphics[width = 0.8in]{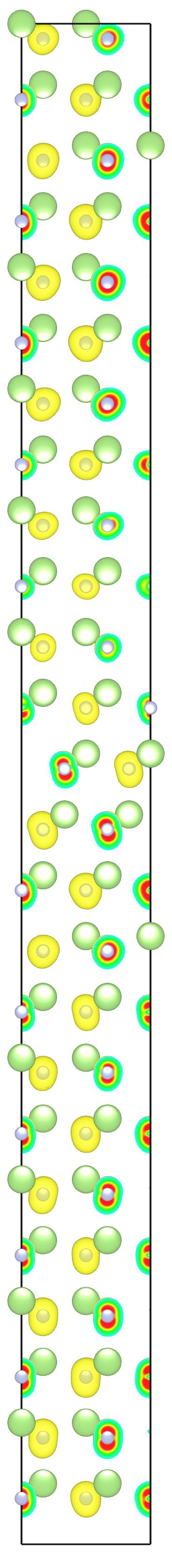}} 
\subfloat[]{\includegraphics[width = 0.8in]{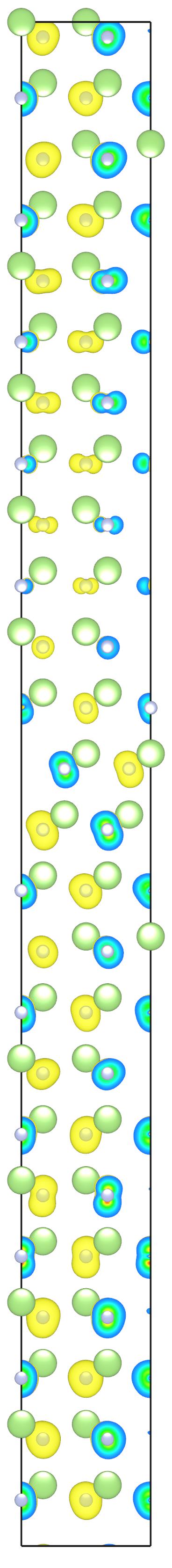}}
\subfloat[]{\includegraphics[width = 0.8in]{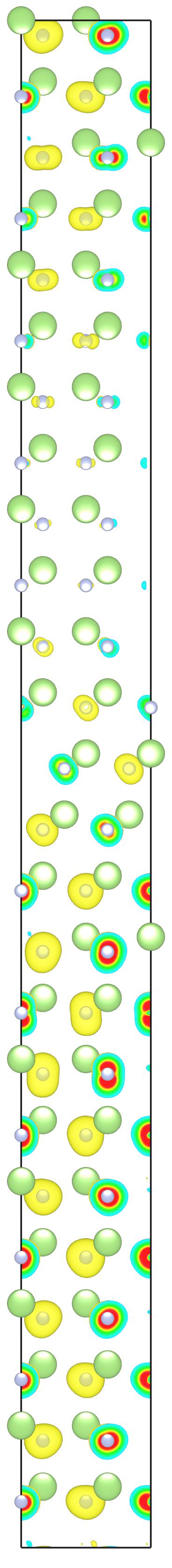}}
\subfloat[]{\includegraphics[width = 0.8in]{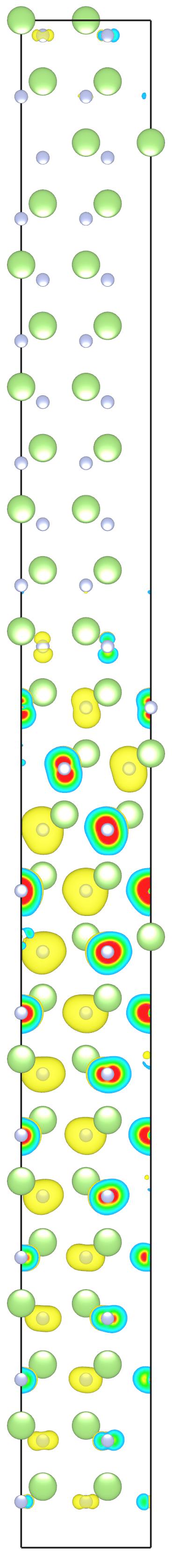}}
\subfloat[]{\includegraphics[width = 0.8in]{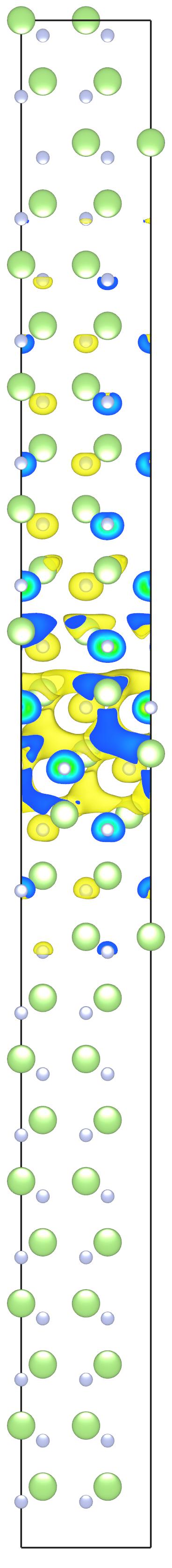}}
\subfloat[]{\includegraphics[width = 0.8in]{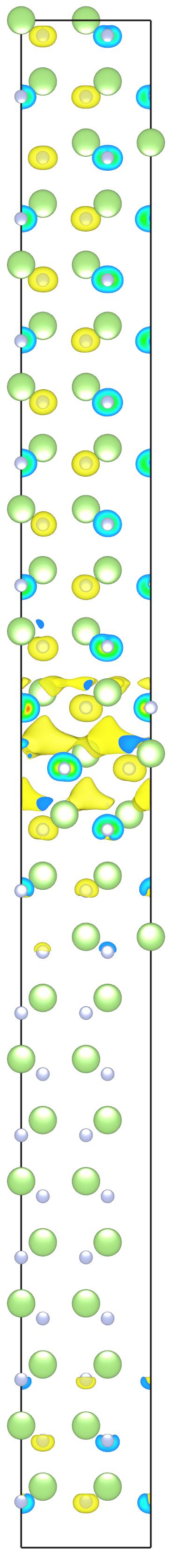}}
\subfloat[]{\includegraphics[width = 0.8in]{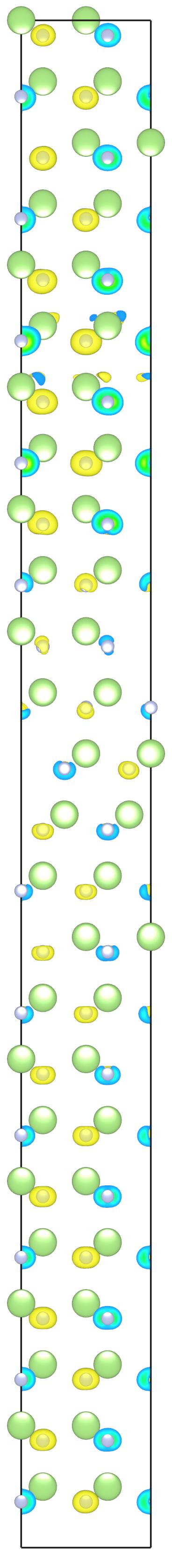}}
\subfloat[]{\includegraphics[width = 0.8in]{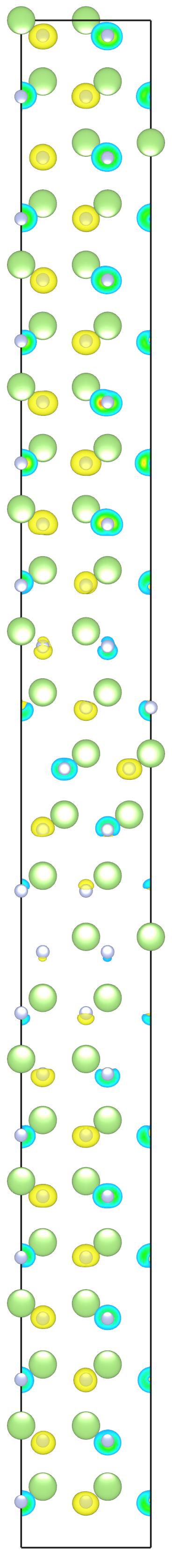}}
\caption{ Band Density of Extrinsic (6AB-C-6AB) stacking fault of wz GaN. (a) Band 447 , (b) Band 448 , (c) Band 449, (d) Band 450, (e) Band 451 , (f) Band 452 , (g) Band 453 , (h) Band 454. Fermi energy (Ef: -5.80 eV) lies in between band 450 and band 451.}
\label{fgr:Fig-10}
\end{figure*}

\begin{figure*}[ht]
\subfloat[]{\includegraphics[width = 0.8in]{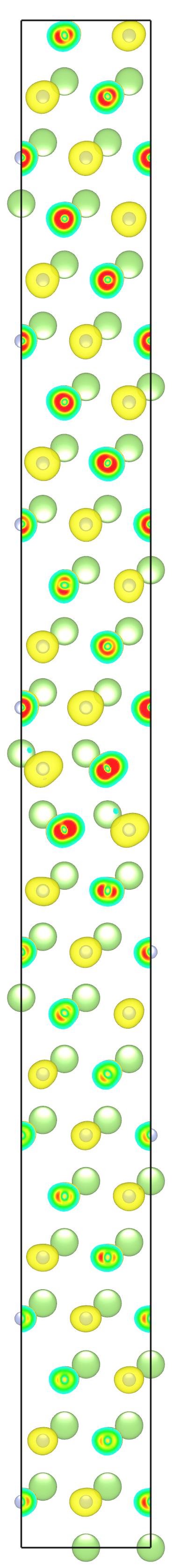}} 
\subfloat[]{\includegraphics[width = 0.8in]{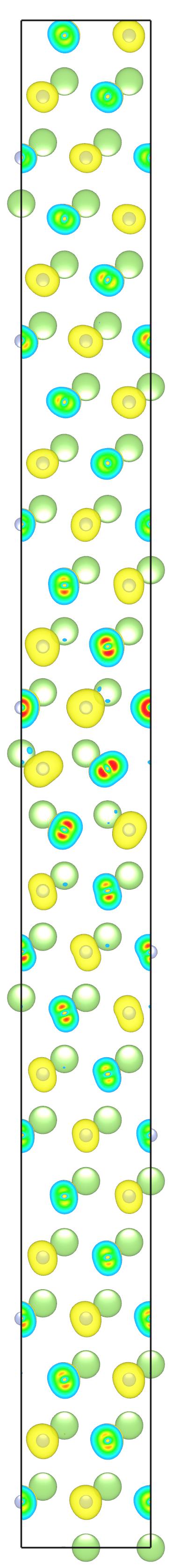}}
\subfloat[]{\includegraphics[width = 0.8in]{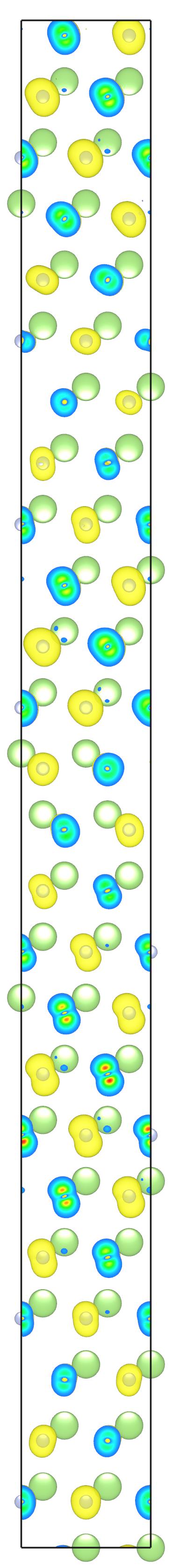}}
\subfloat[]{\includegraphics[width = 0.8in]{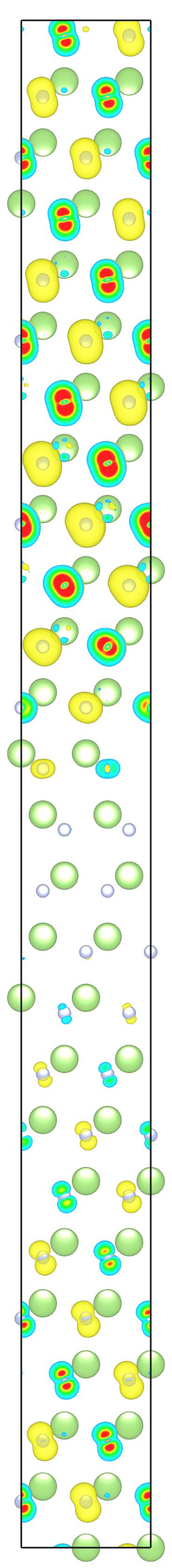}}
\subfloat[]{\includegraphics[width = 0.8in]{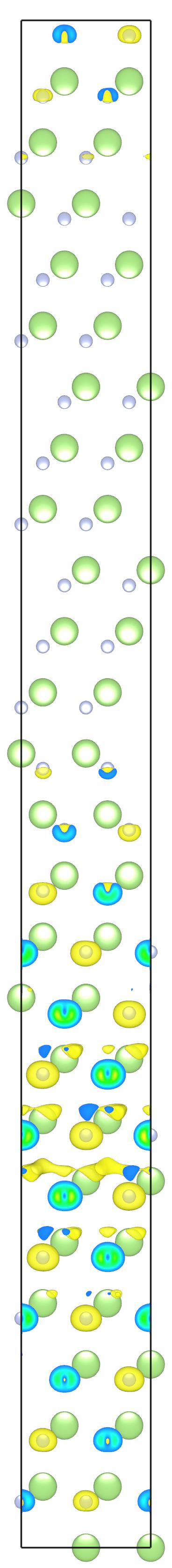}}
\subfloat[]{\includegraphics[width = 0.8in]{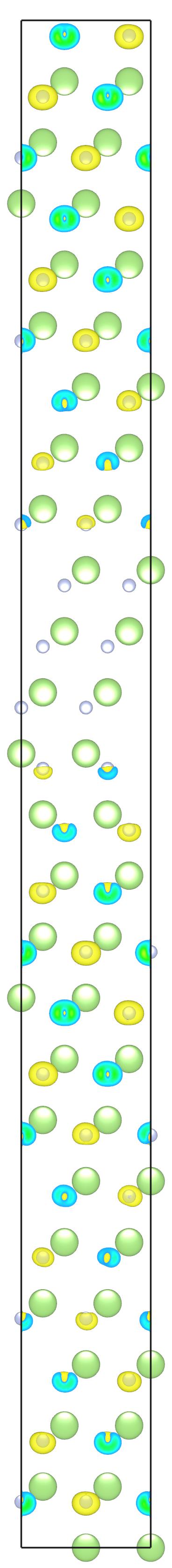}}
\subfloat[]{\includegraphics[width = 0.8in]{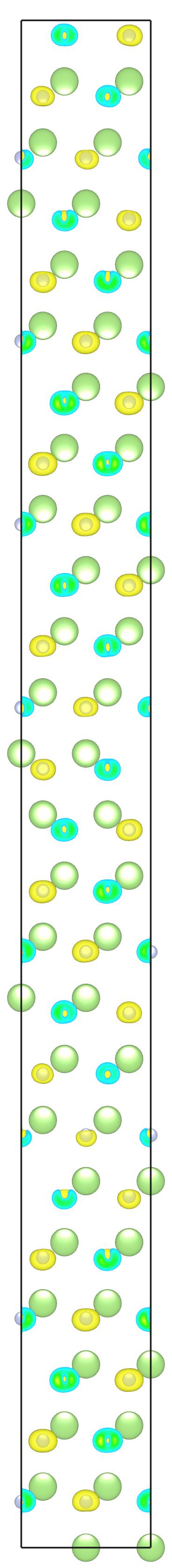}}
\subfloat[]{\includegraphics[width = 0.8in]{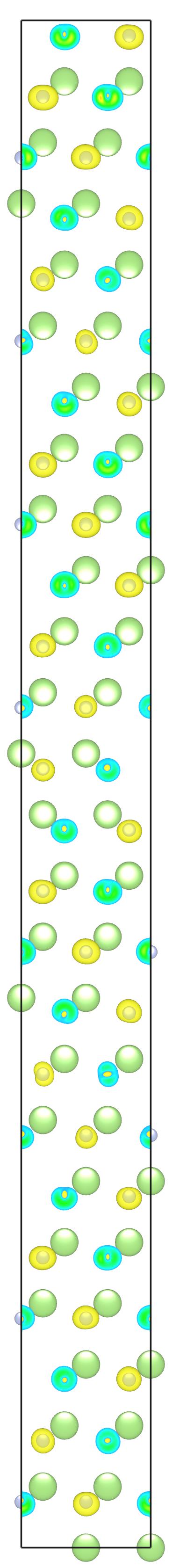}}
\caption{ Band Density of Extrinsic (4ABC-B-4ABC) stacking fault for zb GaN. (a) Band 447 , (b) Band 448 , (c) Band 449, (d) Band 450, (e) Band 451 , (f) Band 452 , (g) Band 453 , (h) Band 454. Fermi energy (Ef = -0.2146 Ha) lies in between band 450 and band 451.}
\label{fgr:Fig-12}
\end{figure*}

\begin{figure*}[ht]
\subfloat[]{\includegraphics[width = 0.8in]{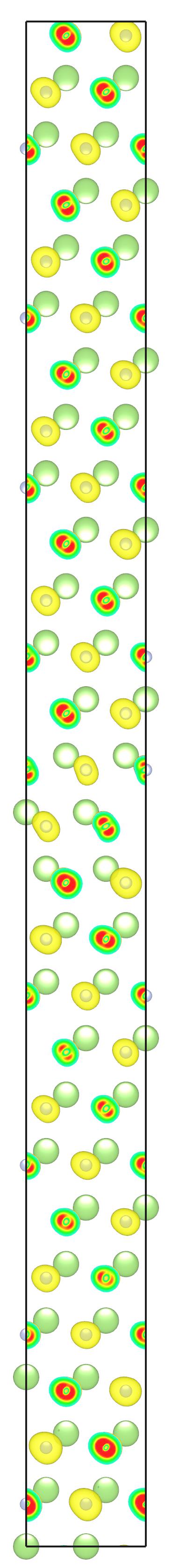}} 
\subfloat[]{\includegraphics[width = 0.8in]{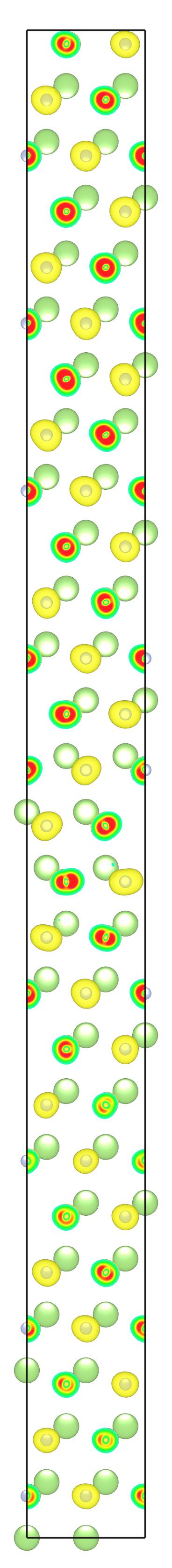}}
\subfloat[]{\includegraphics[width = 0.8in]{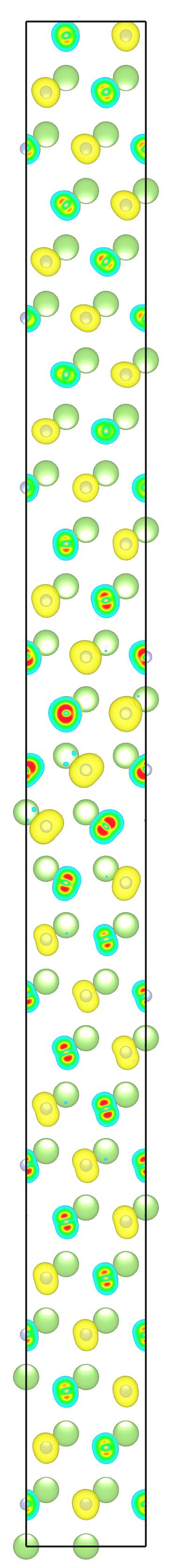}}
\subfloat[]{\includegraphics[width = 0.8in]{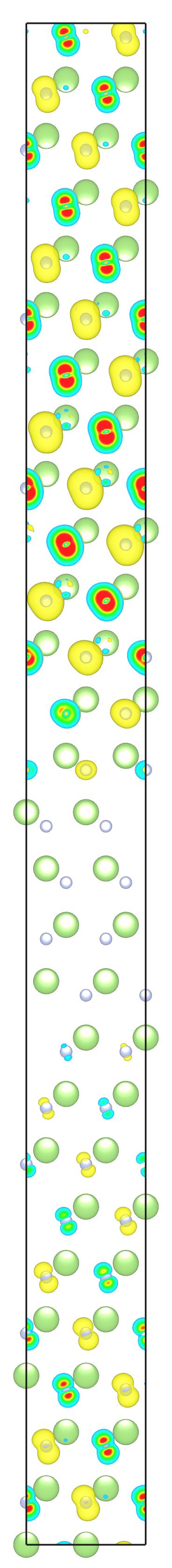}}
\subfloat[]{\includegraphics[width = 0.8in]{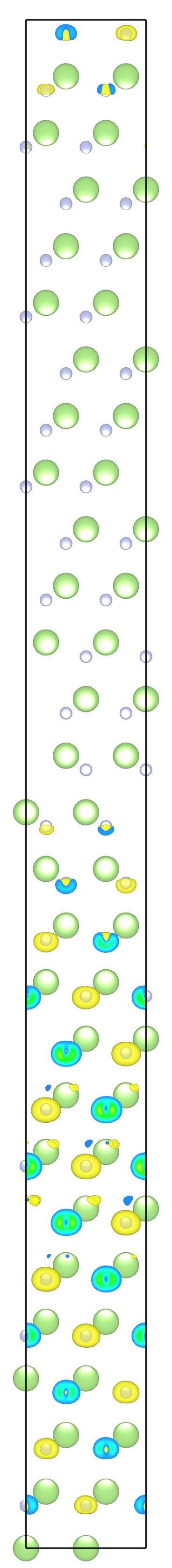}}
\subfloat[]{\includegraphics[width = 0.8in]{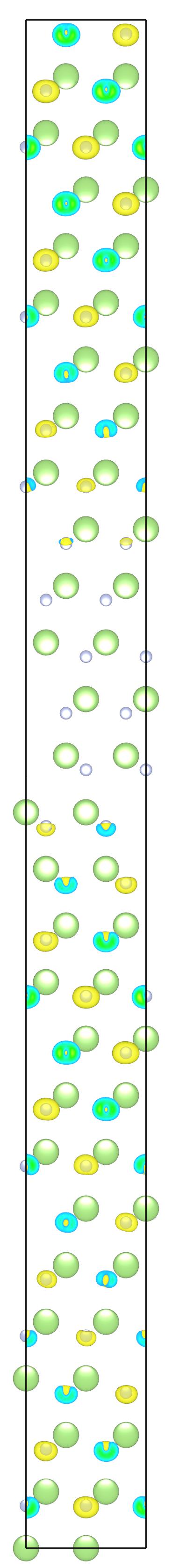}}
\subfloat[]{\includegraphics[width = 0.8in]{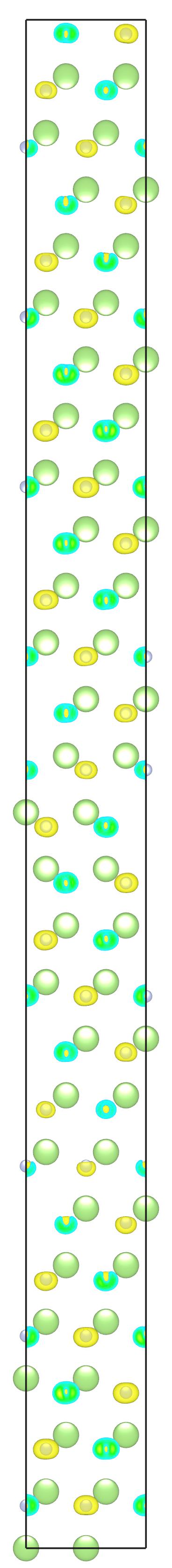}}
\subfloat[]{\includegraphics[width = 0.8in]{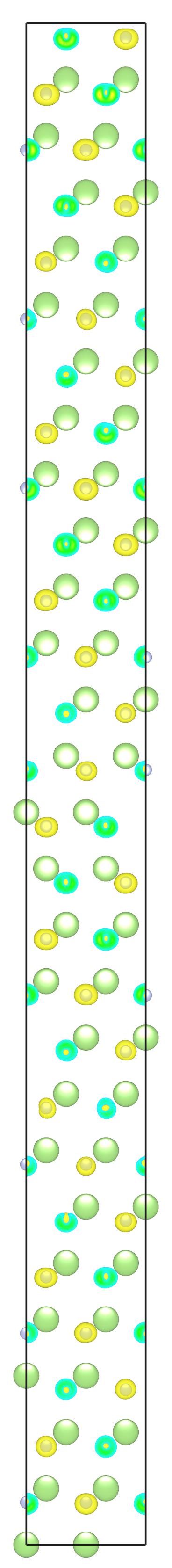}}
\caption{ Band density of Twin (4ABC-BAC-4ABC) stacking fault for zb GaN. (a) Band 483 , (b) Band 484 , (c) Band 485, (d) Band 486, (e) Band 487 , (f) Band 488 , (g) Band 489 , (h) Band 490. Fermi energy (Ef = --0.2145805742 Ha ) lies in between band 486 and band 487.}
\label{fgr:Fig-13}
\end{figure*}

The band densities for the I$_{1}$ stacking fault in wz GaN and the I stacking fault in zb GaN are shown in the main text.  Here we show the other band densities in Figs.~\ref{fgr:Fig-8} to~\ref{fgr:Fig-13}.

\clearpage

\section{Potentials}
\label{sec:potentials}

\begin{figure*}[ht]
\subfloat[]{\includegraphics[width = 0.45\columnwidth]{AP_WZ_IN1_new.png}}\hspace{0.5cm}
\subfloat[]{\includegraphics[width = 0.45\columnwidth]{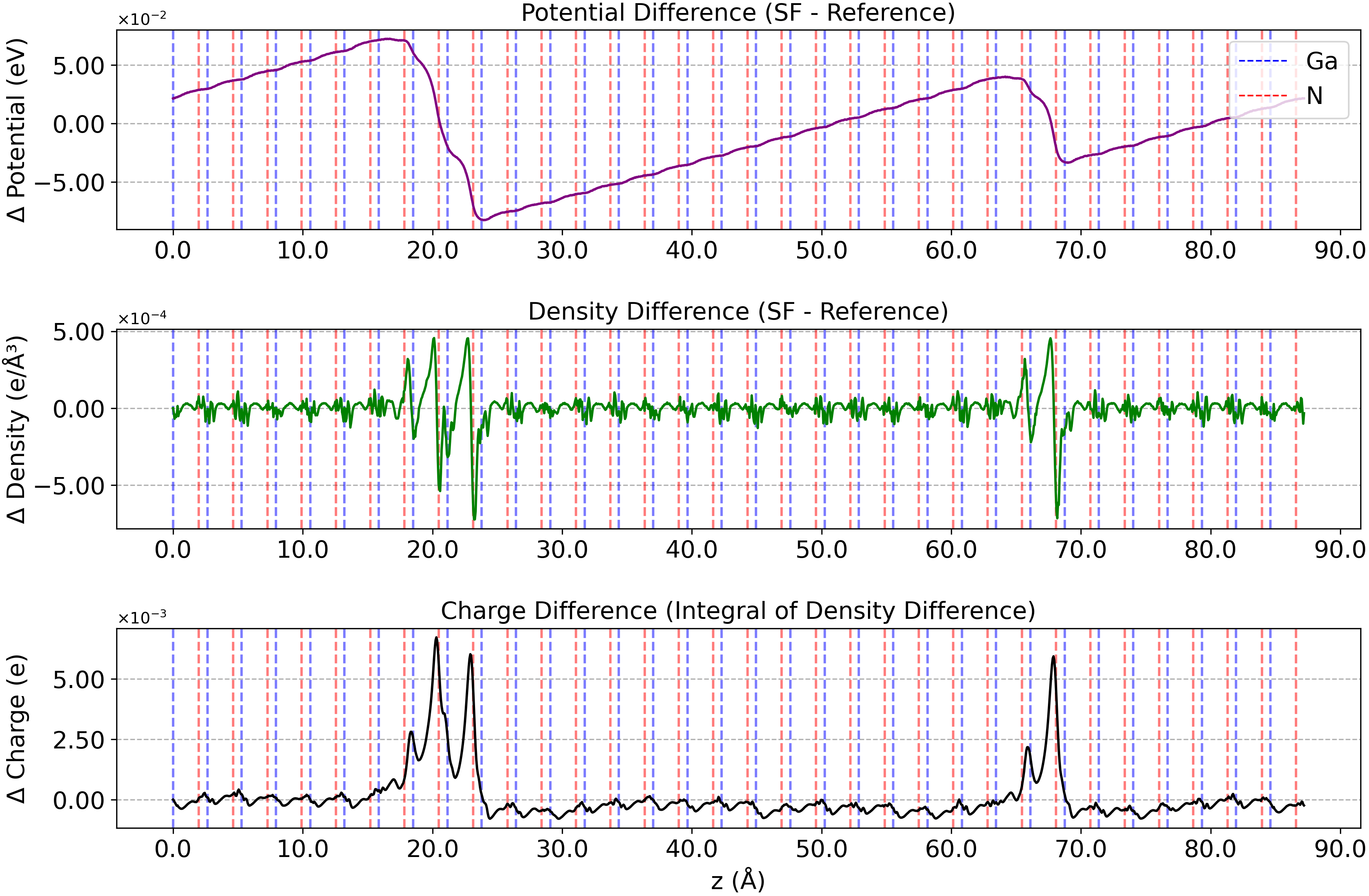}}\\
\subfloat[]{\includegraphics[width = 0.45\columnwidth]{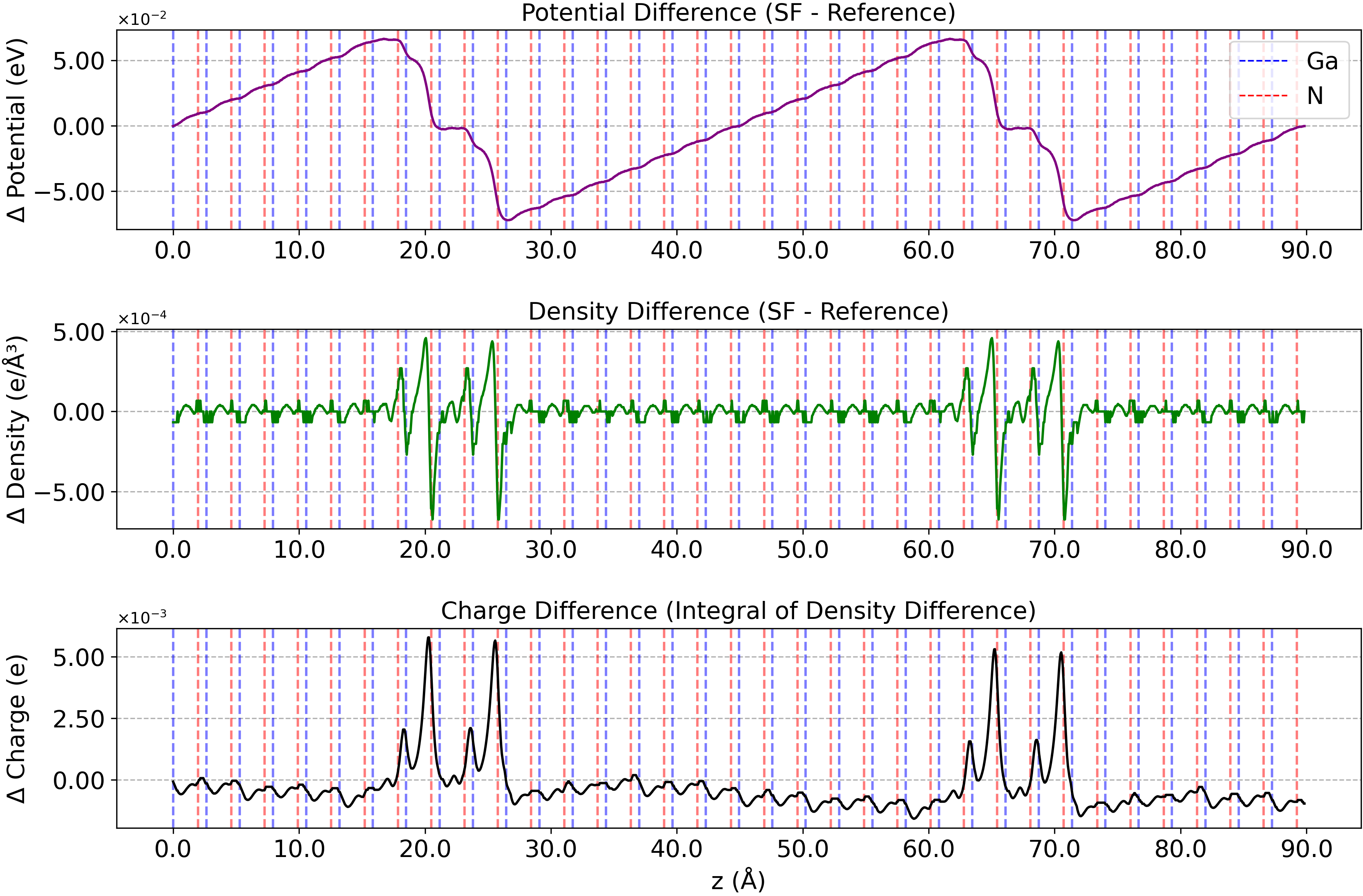}}\hspace{0.5cm}
\subfloat[]{\includegraphics[width = 0.45\columnwidth]{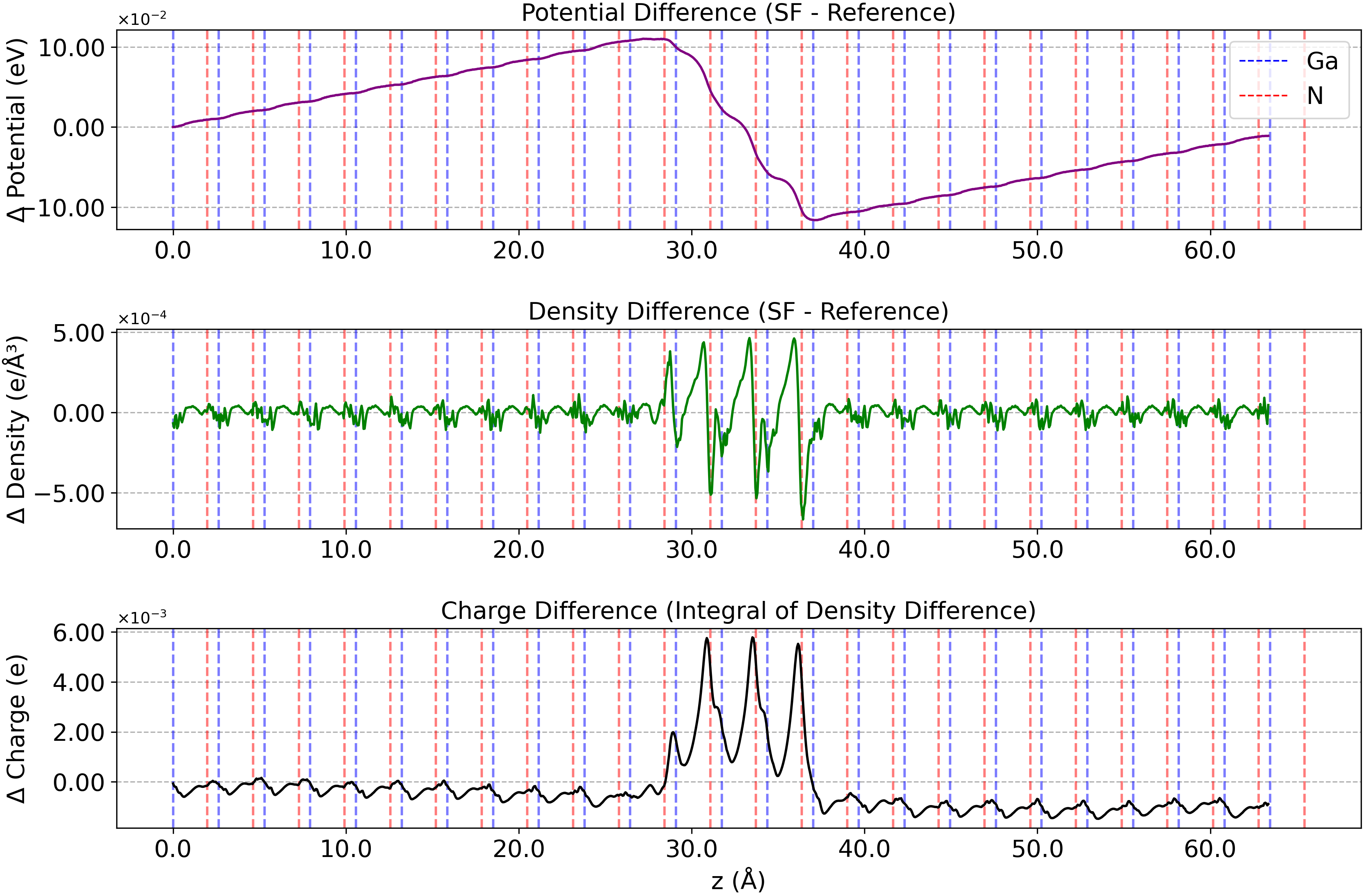}}
\caption{Average potential of stacking faults based on wz GaN (a) I$_{1}$ (3AB-6CB-3AB) (b) I$_{2}$ (4AB-8CA-C-4AB) (c) I$_{3}$ (4AB-C-8BA-C-4AB) (d) Extrinsic (6AB-C-6AB).}
\label{fgr:AP_WZ}
\end{figure*}

\begin{figure*}[ht]
\subfloat[]{\includegraphics[width = 0.45\columnwidth]{AP_ZB_In_new.png}}\\
\subfloat[]{\includegraphics[width = 0.45\columnwidth]{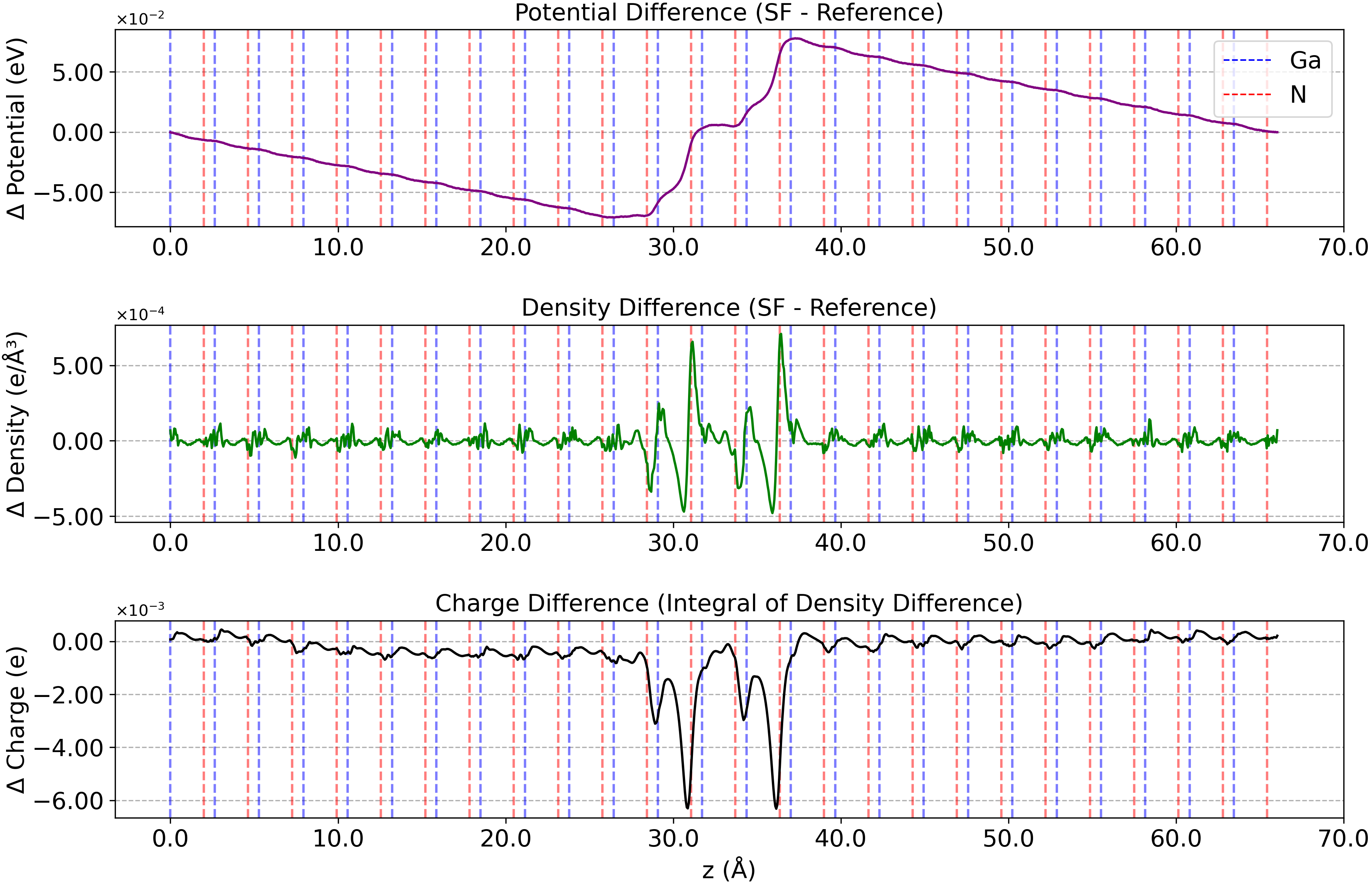}}\\
\subfloat[]{\includegraphics[width = 0.45\columnwidth]{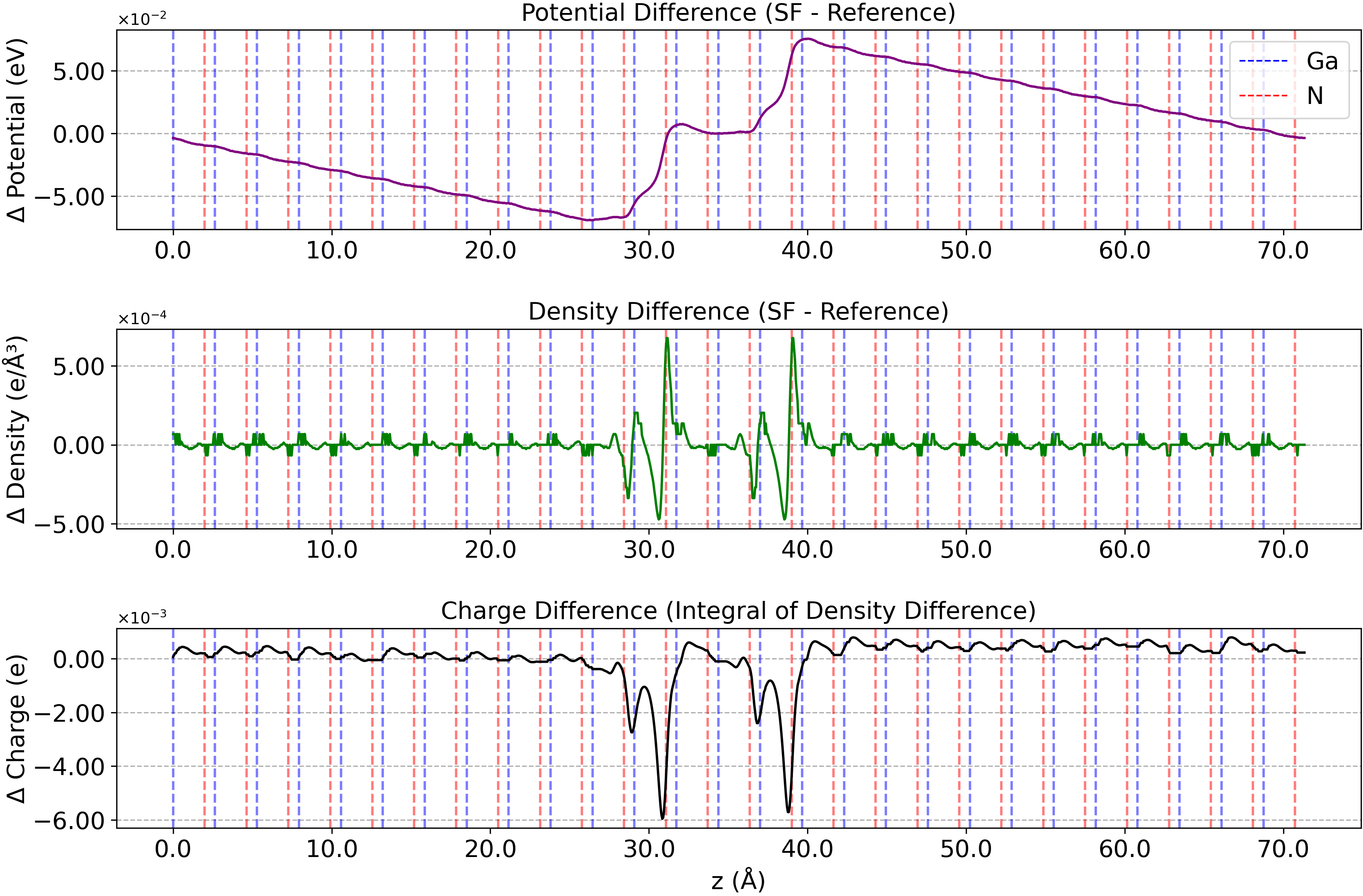}}
\caption{Average potential of stacking faults based on zb GaN (a) intrinsic (b) extrinsic and (c) twin.}
\label{fgr:AP_ZB}
\end{figure*}

\clearpage

\section{Projected DOS}
\label{sec:projected-dos-band}

\begin{figure*}[ht]
\subfloat[]{\includegraphics[width = 0.32\columnwidth]{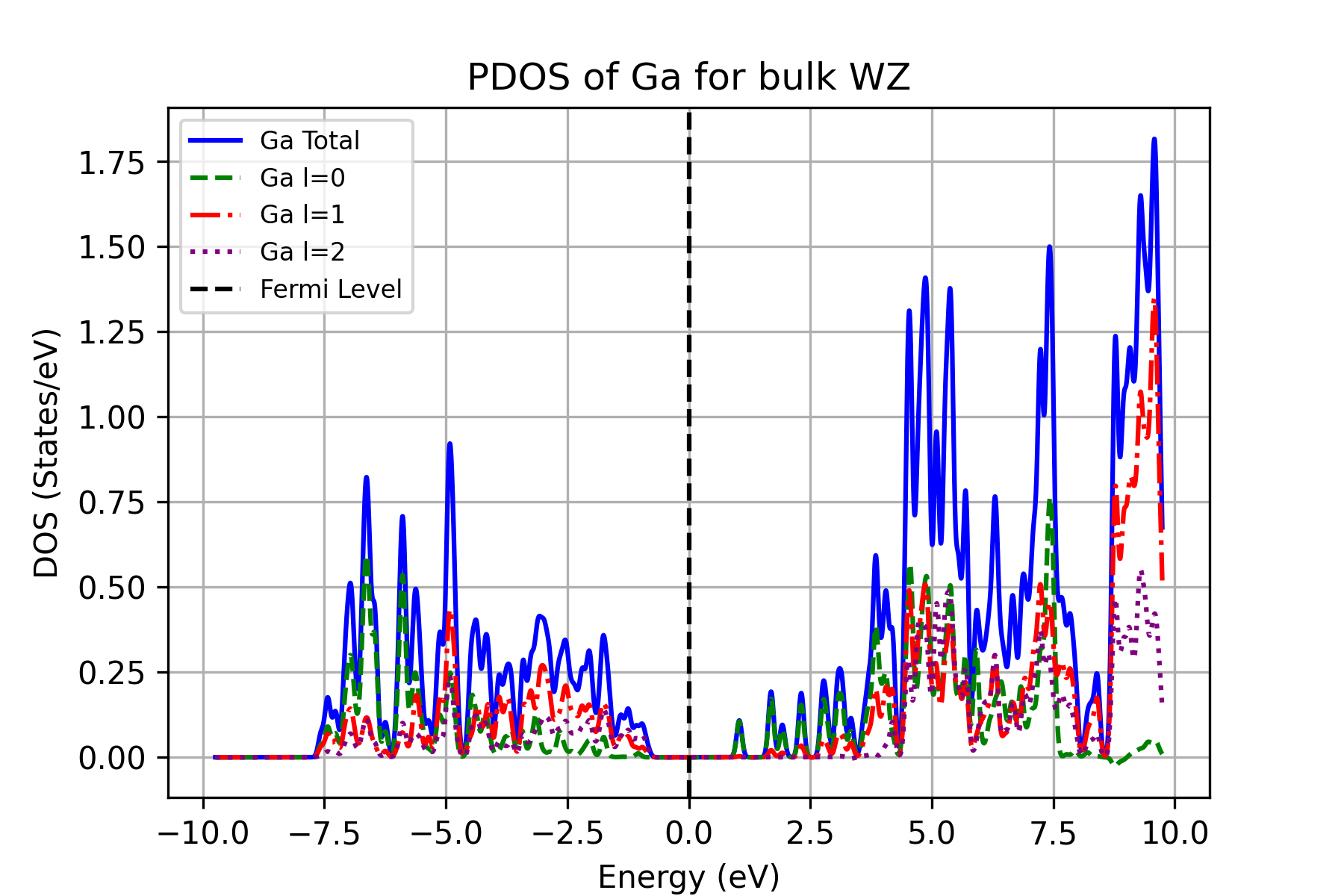}} 
\subfloat[]{\includegraphics[width = 0.32\columnwidth]{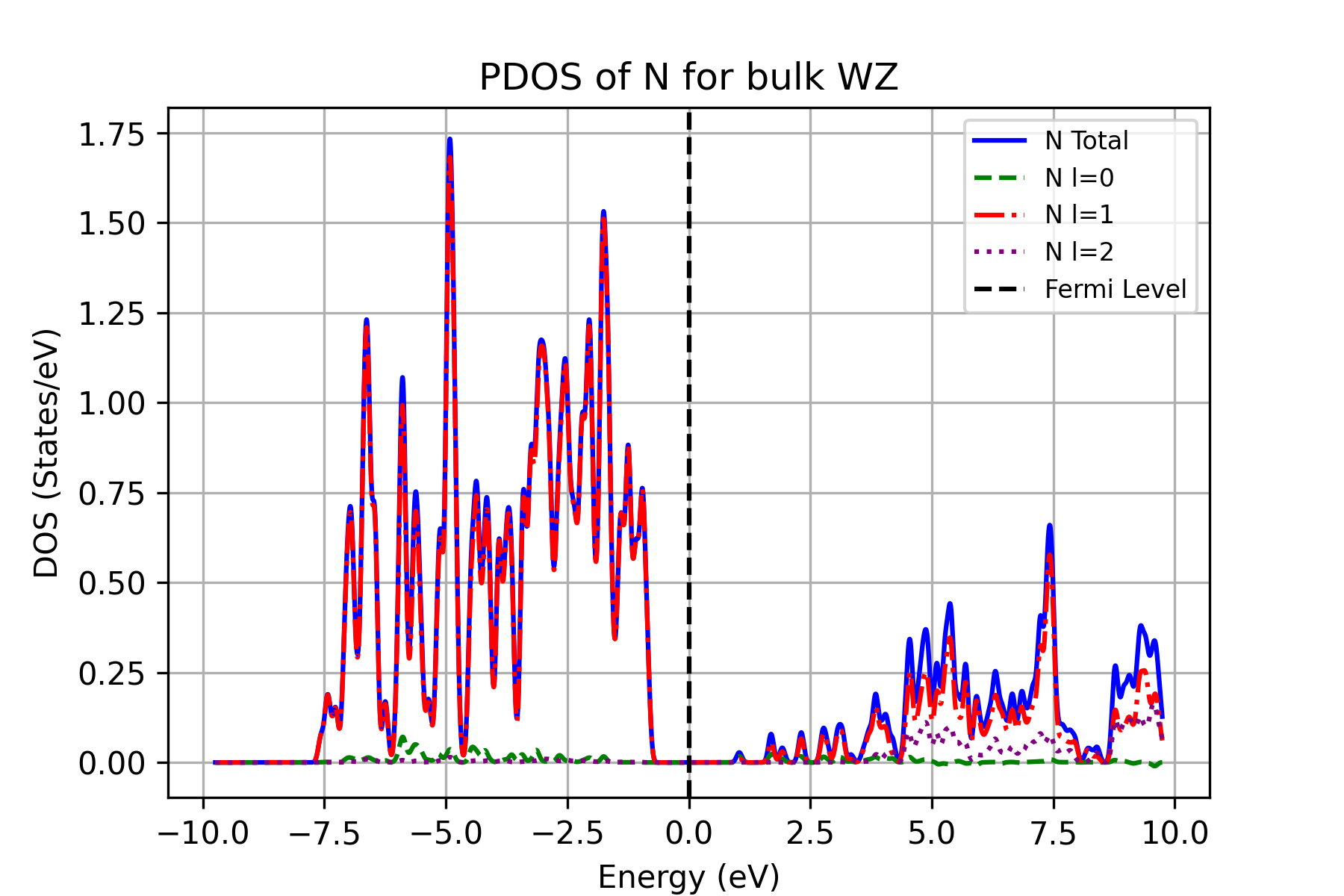}}\\
\subfloat[]{\includegraphics[width = 0.32\columnwidth]{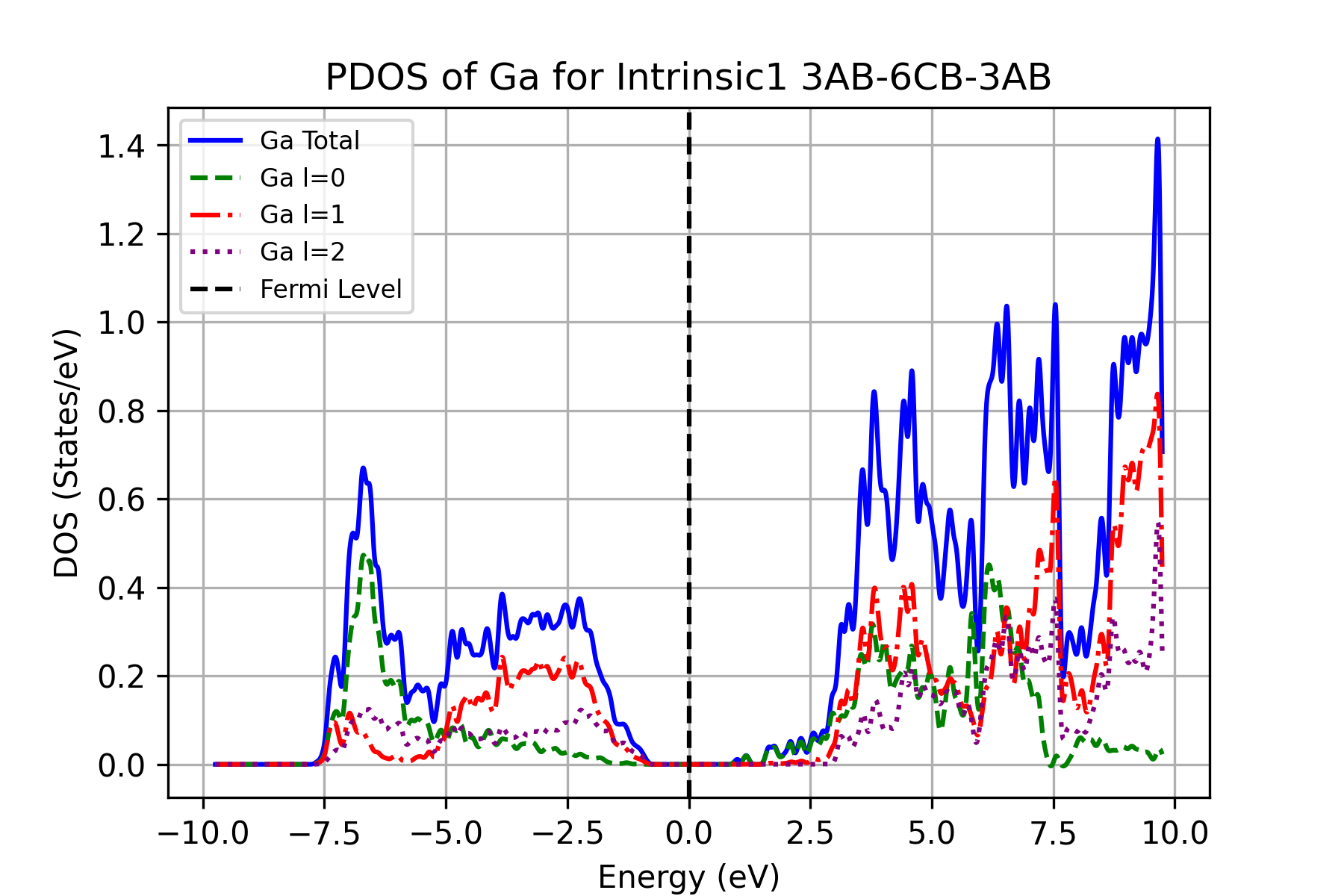}}
\subfloat[]{\includegraphics[width = 0.32\columnwidth]{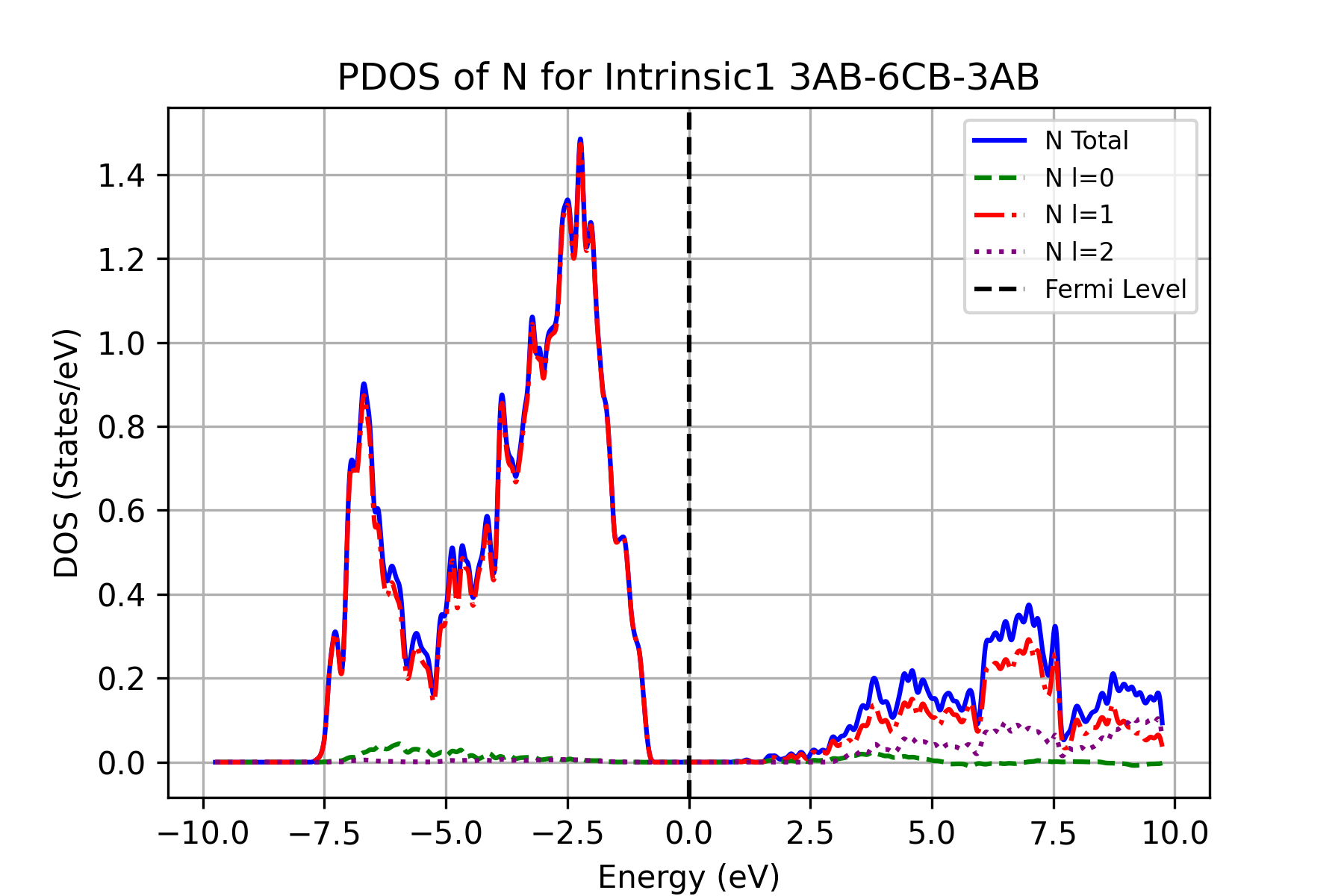}}\\
\subfloat[]{\includegraphics[width = 0.32\columnwidth]{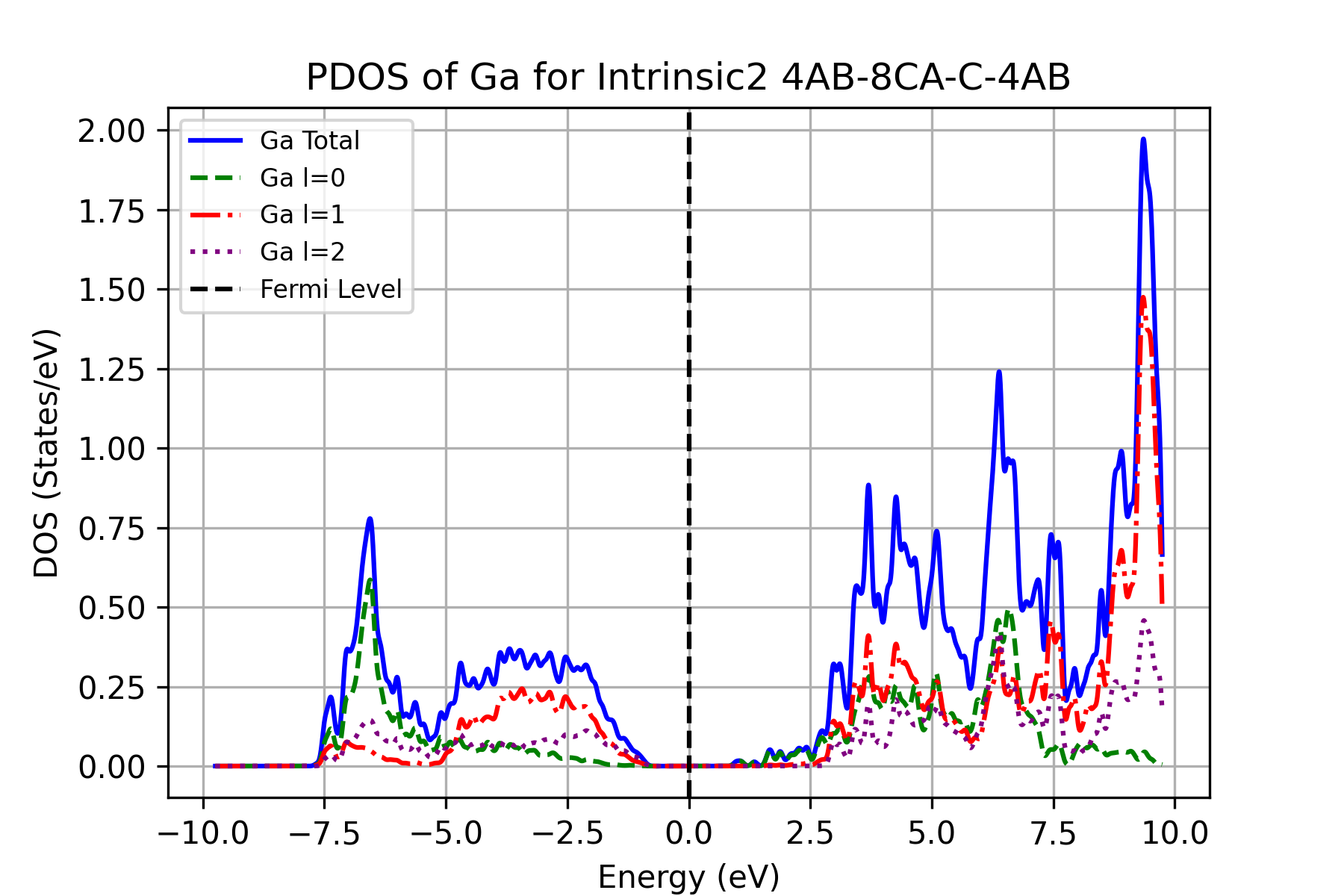}}
\subfloat[]{\includegraphics[width = 0.32\columnwidth]{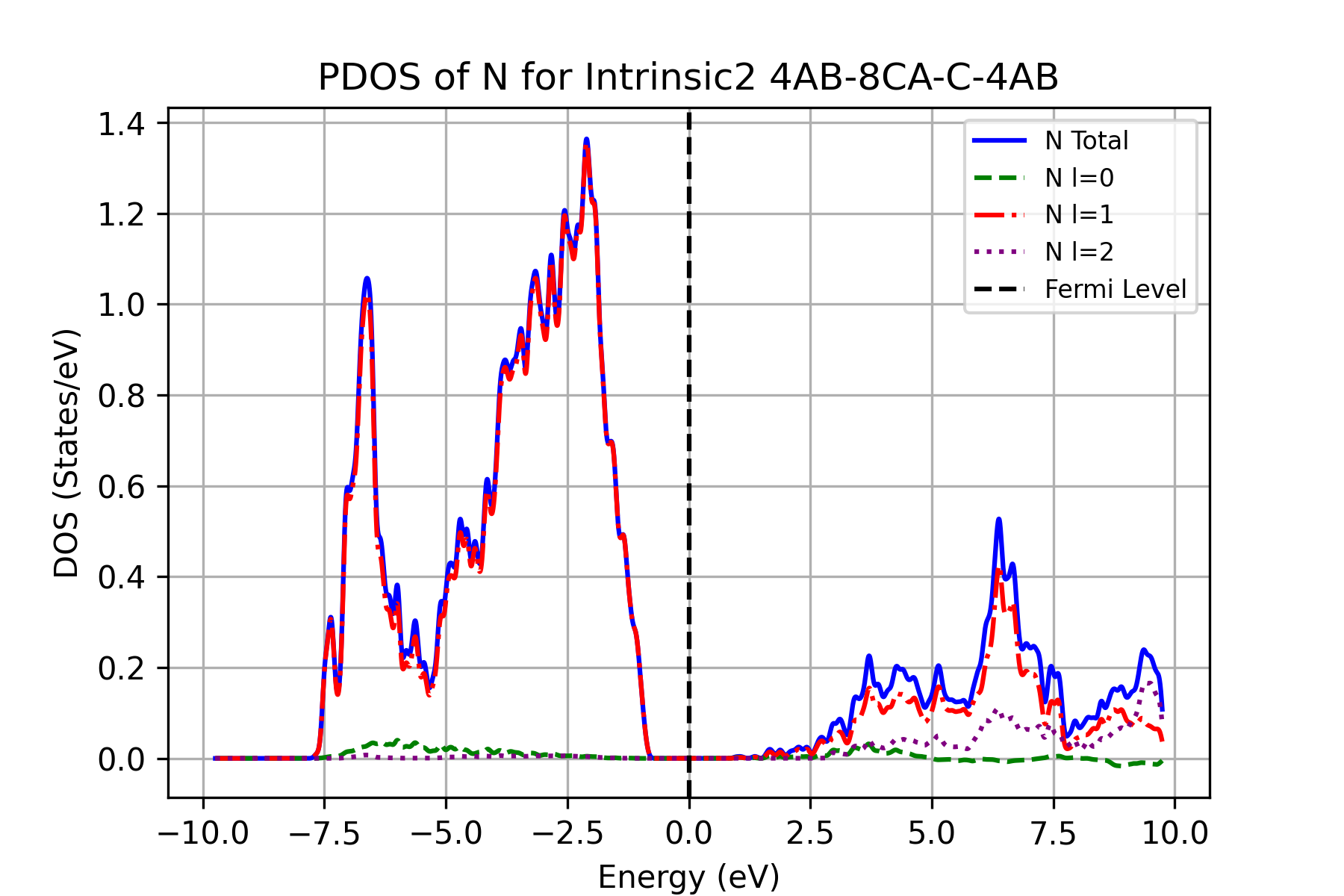}}\\
\subfloat[]{\includegraphics[width = 0.32\columnwidth]{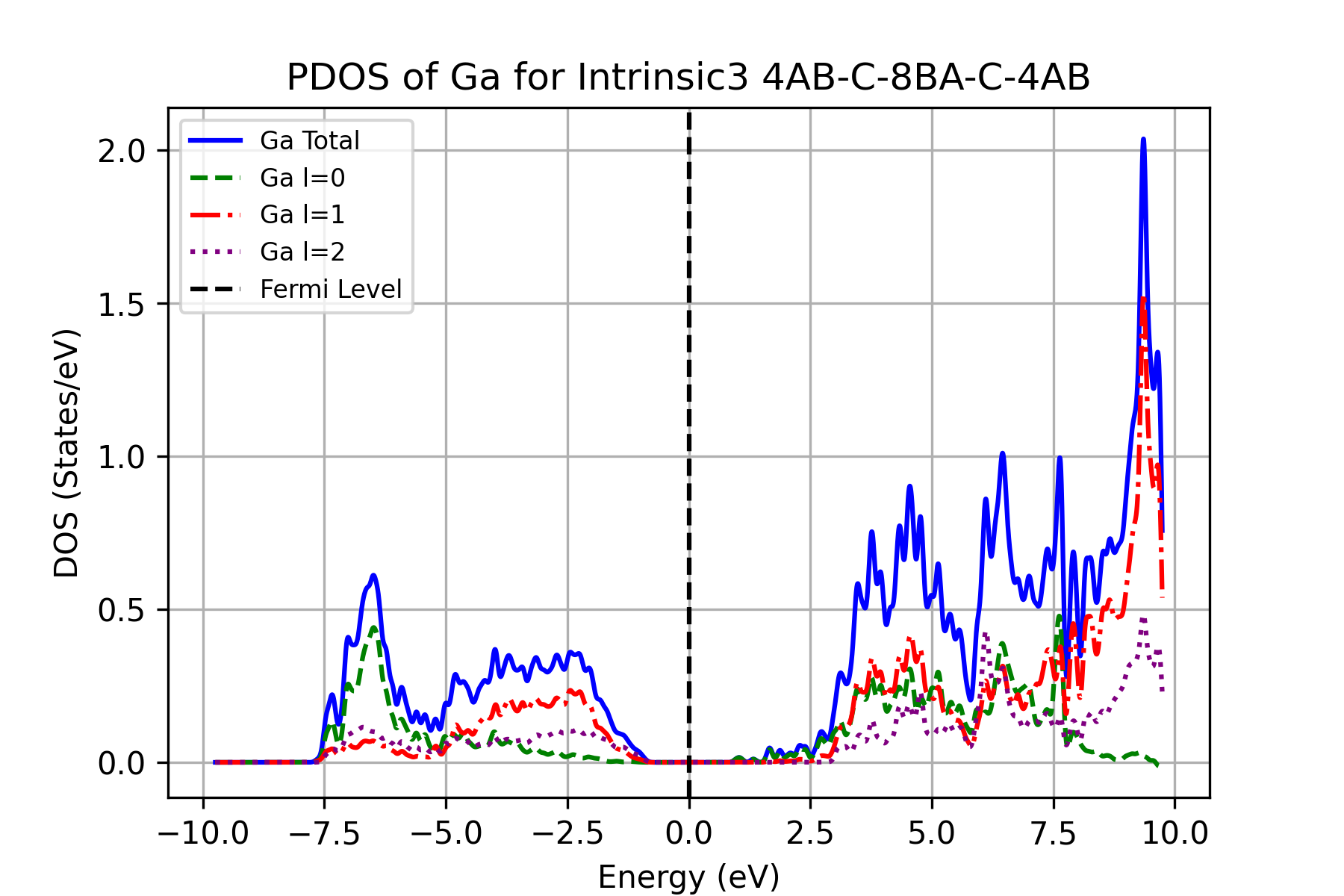}}
\subfloat[]{\includegraphics[width = 0.32\columnwidth]{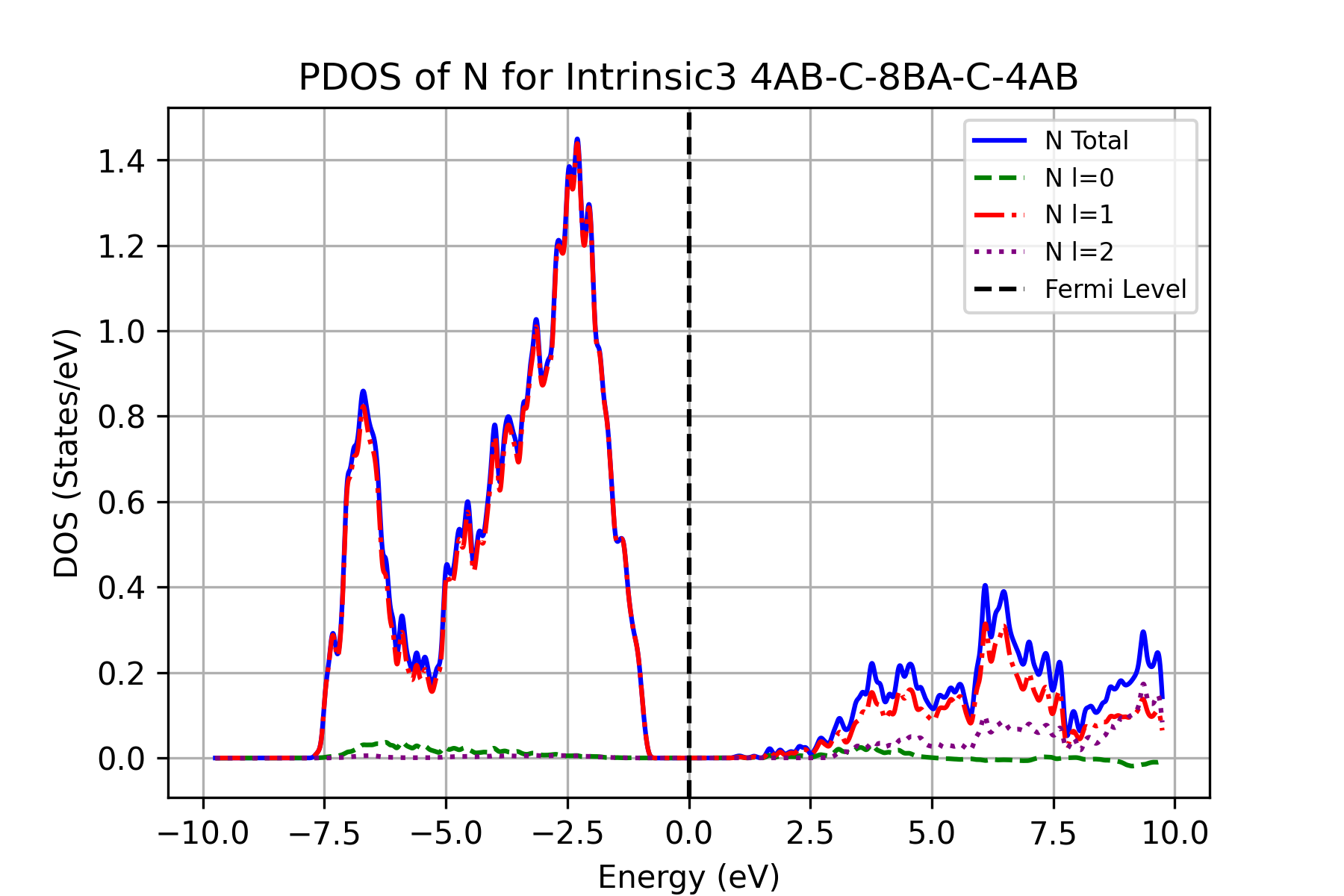}}\\
\subfloat[]{\includegraphics[width = 0.32\columnwidth]{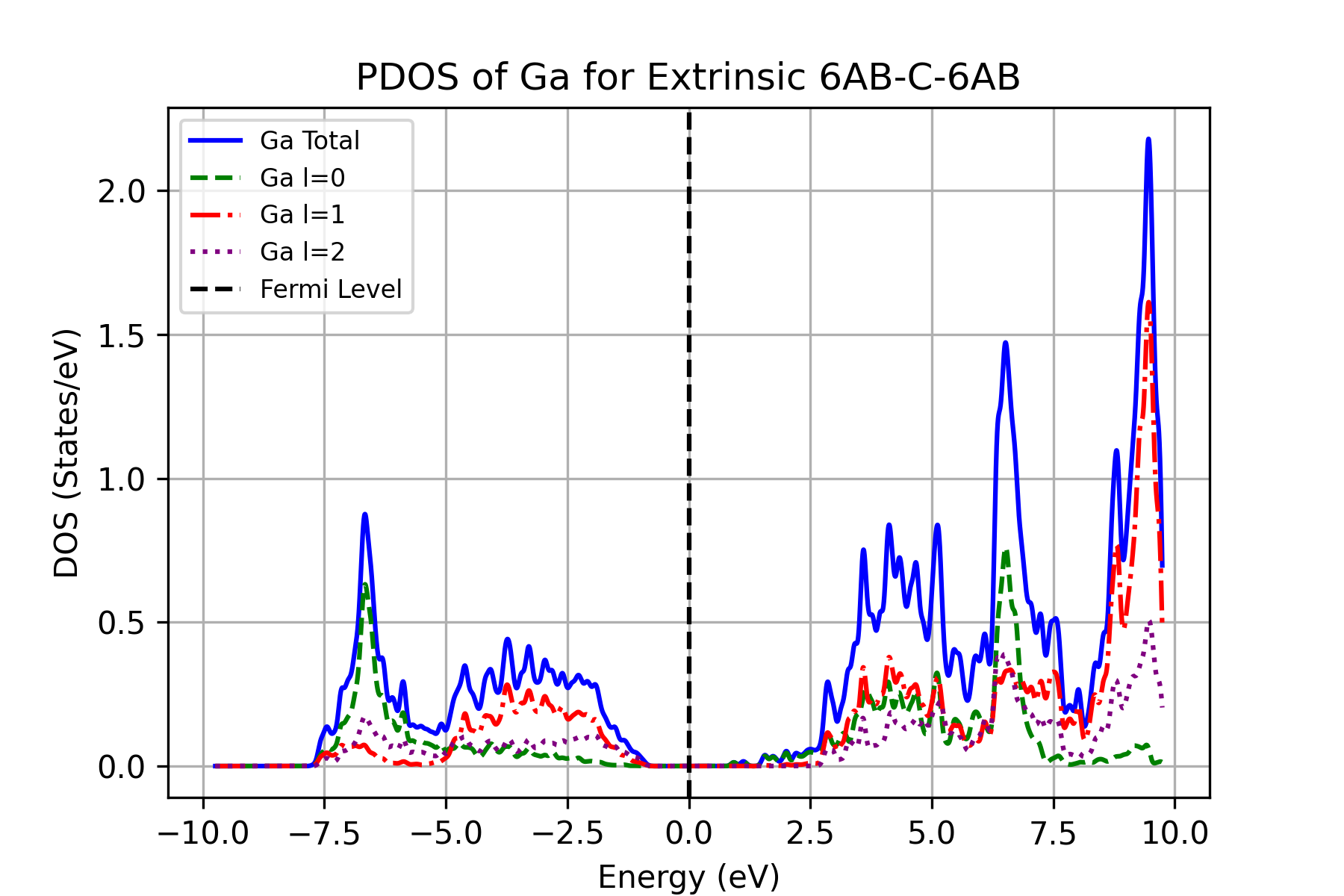}}
\subfloat[]{\includegraphics[width = 0.32\columnwidth]{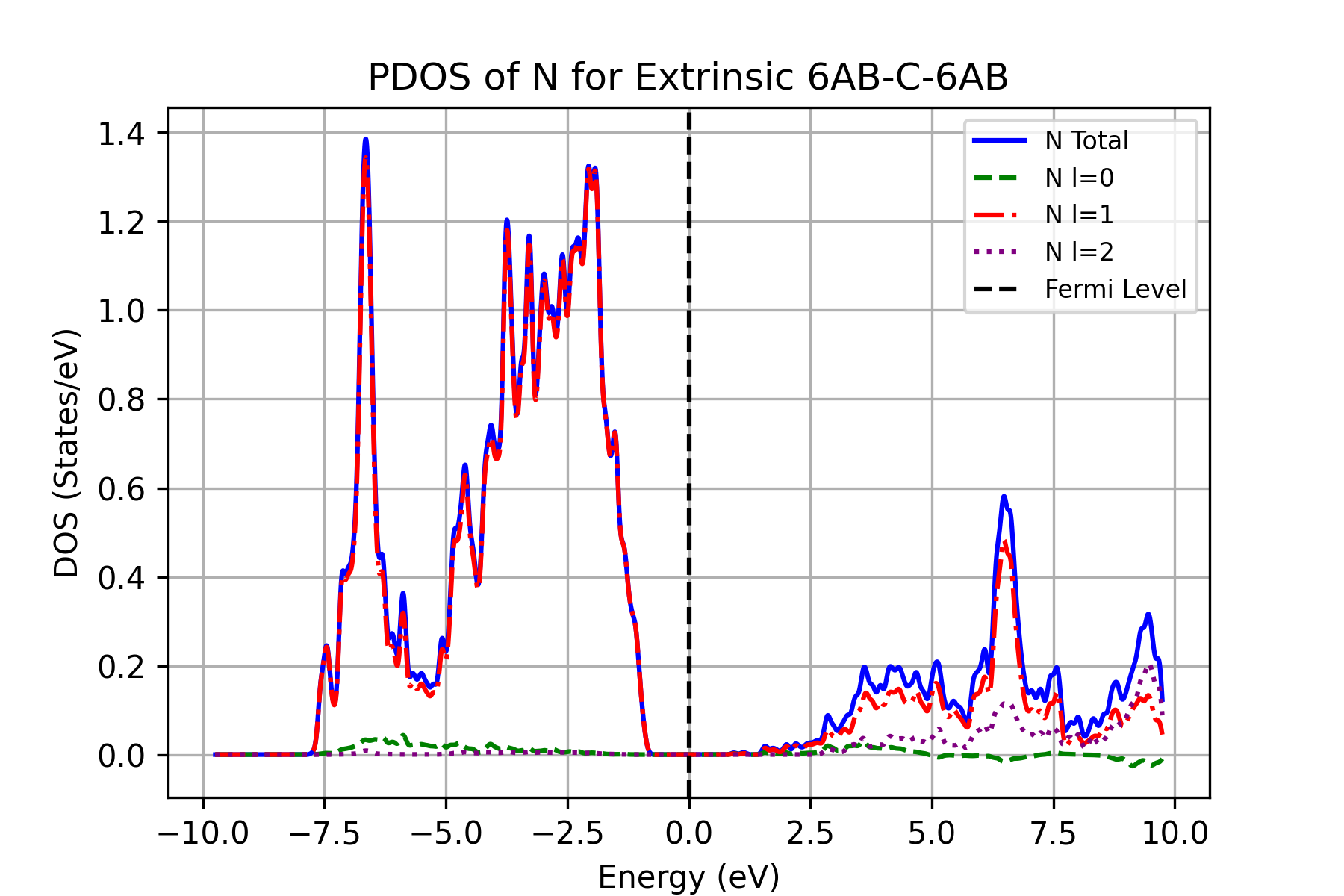}}
\caption{Projected density of states (pDOS) for wurtzite bulk and atoms located at the SFs, showing total (blue) and decomposed by angular momentum: (a) and (b) bulk (Ga and N respectively); (c) and (d) I$_{1}$ SF (Ga and N respectively); (e) and (f) I$_{2}$ SF (Ga and N respectively); (g) and (h) I$_{3}$ SF (Ga and N respectively); (i) and (j) Extrinsic SF (Ga and N respectively).}
\label{fgr:PDOS_WZ}
\end{figure*}

\begin{figure*}[ht]
\subfloat[]{\includegraphics[width = 0.32\columnwidth]{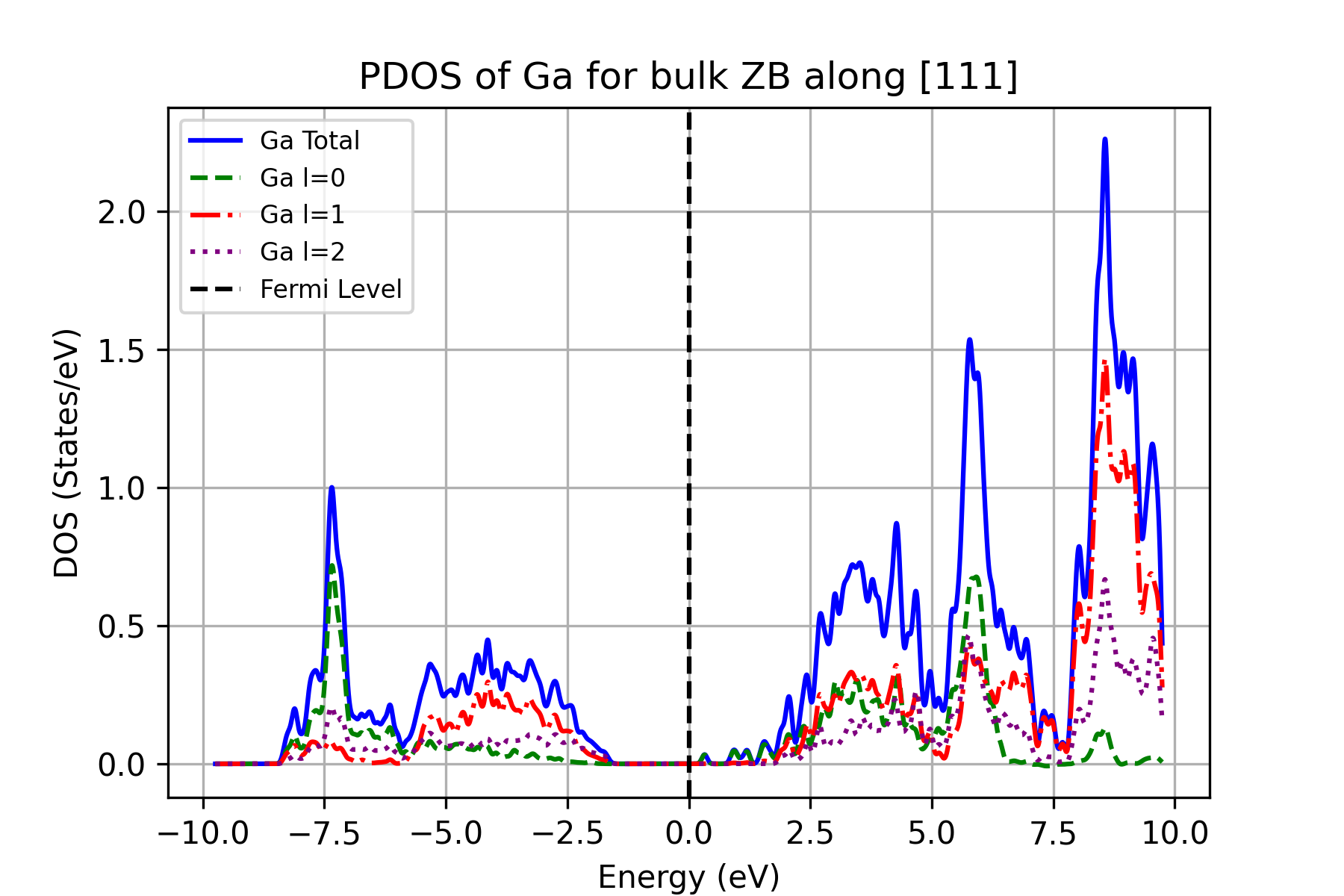}} 
\subfloat[]{\includegraphics[width = 0.32\columnwidth]{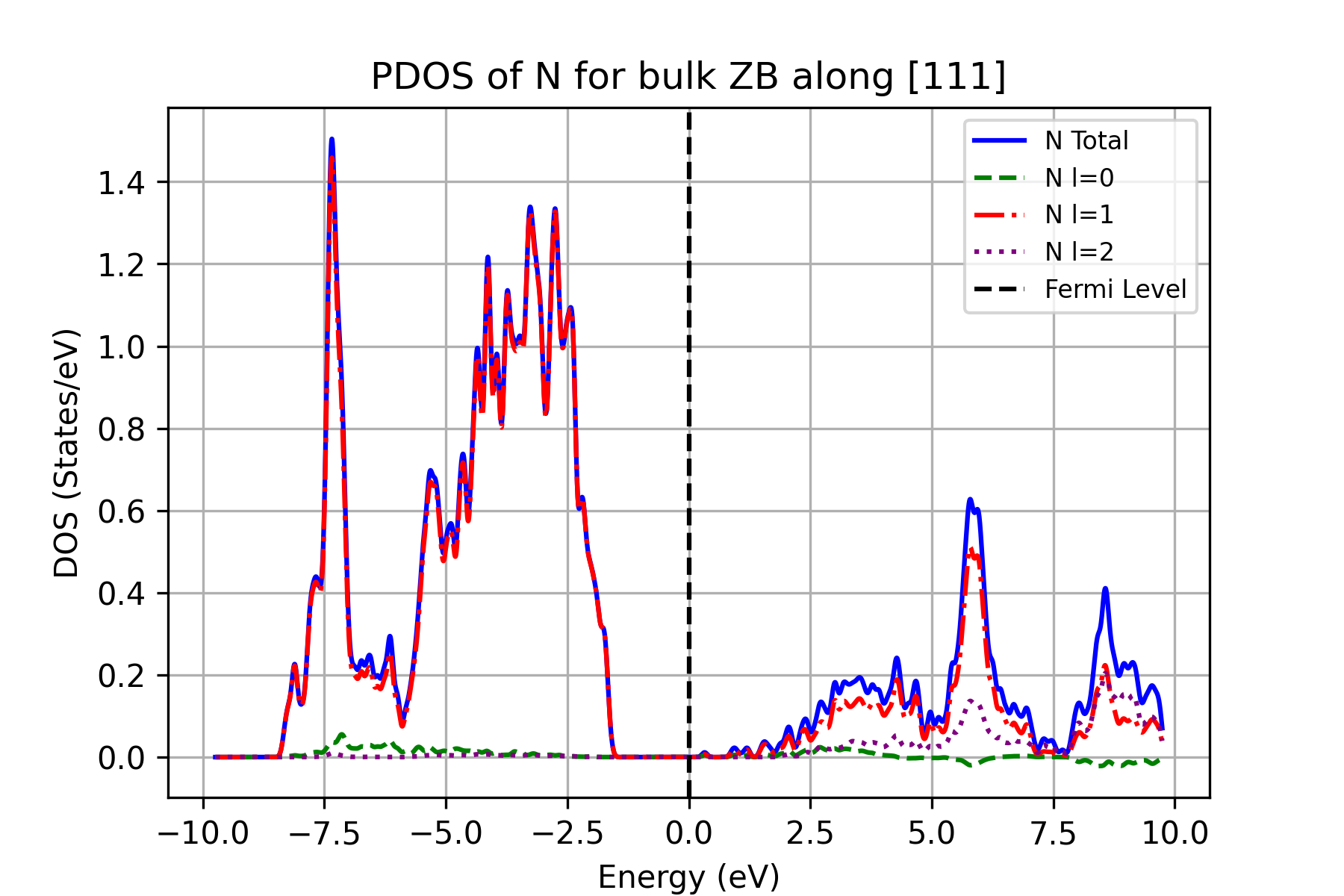}}\\
\subfloat[]{\includegraphics[width = 0.32\columnwidth]{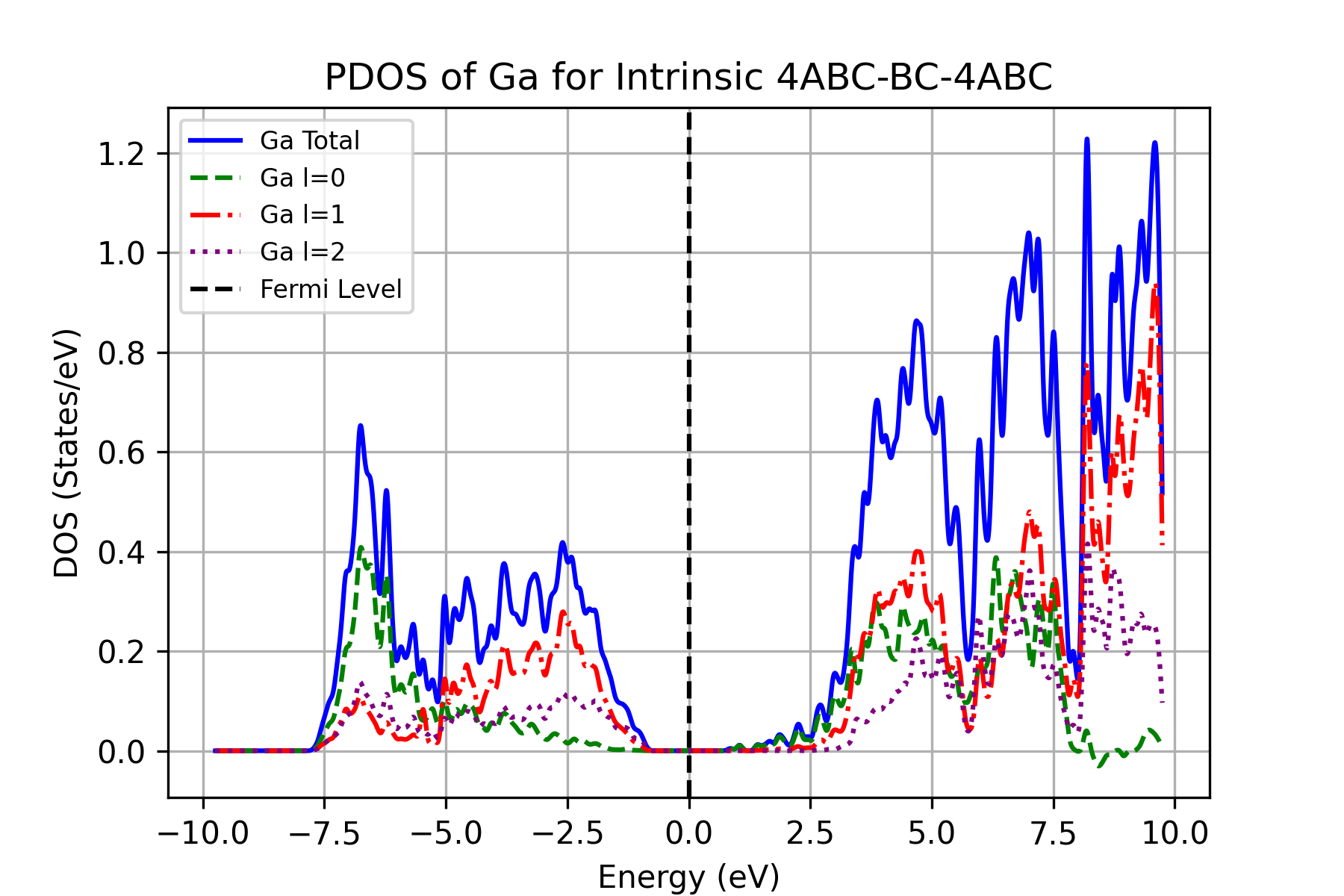}}
\subfloat[]{\includegraphics[width = 0.32\columnwidth]{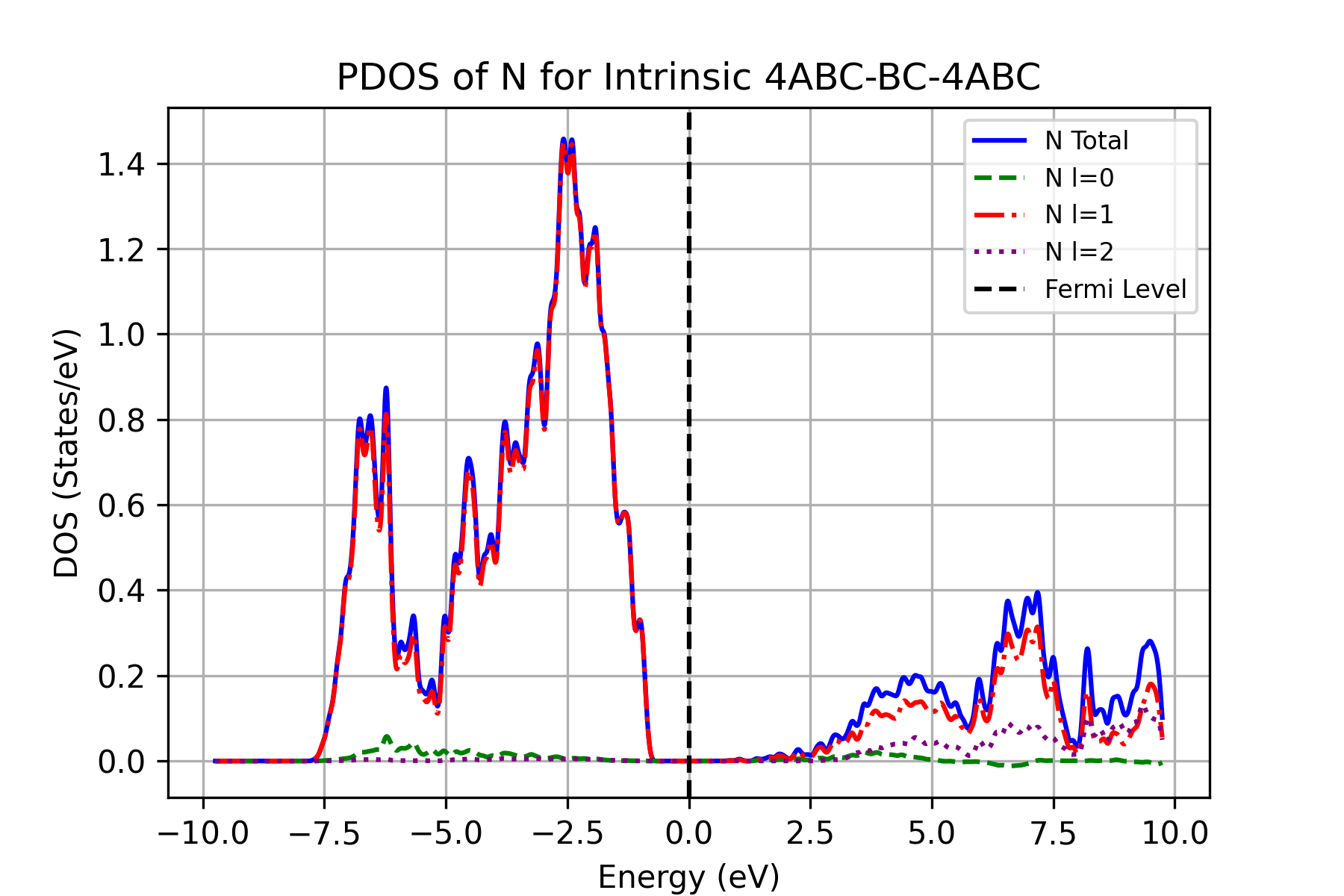}}\\
\subfloat[]{\includegraphics[width = 0.32\columnwidth]{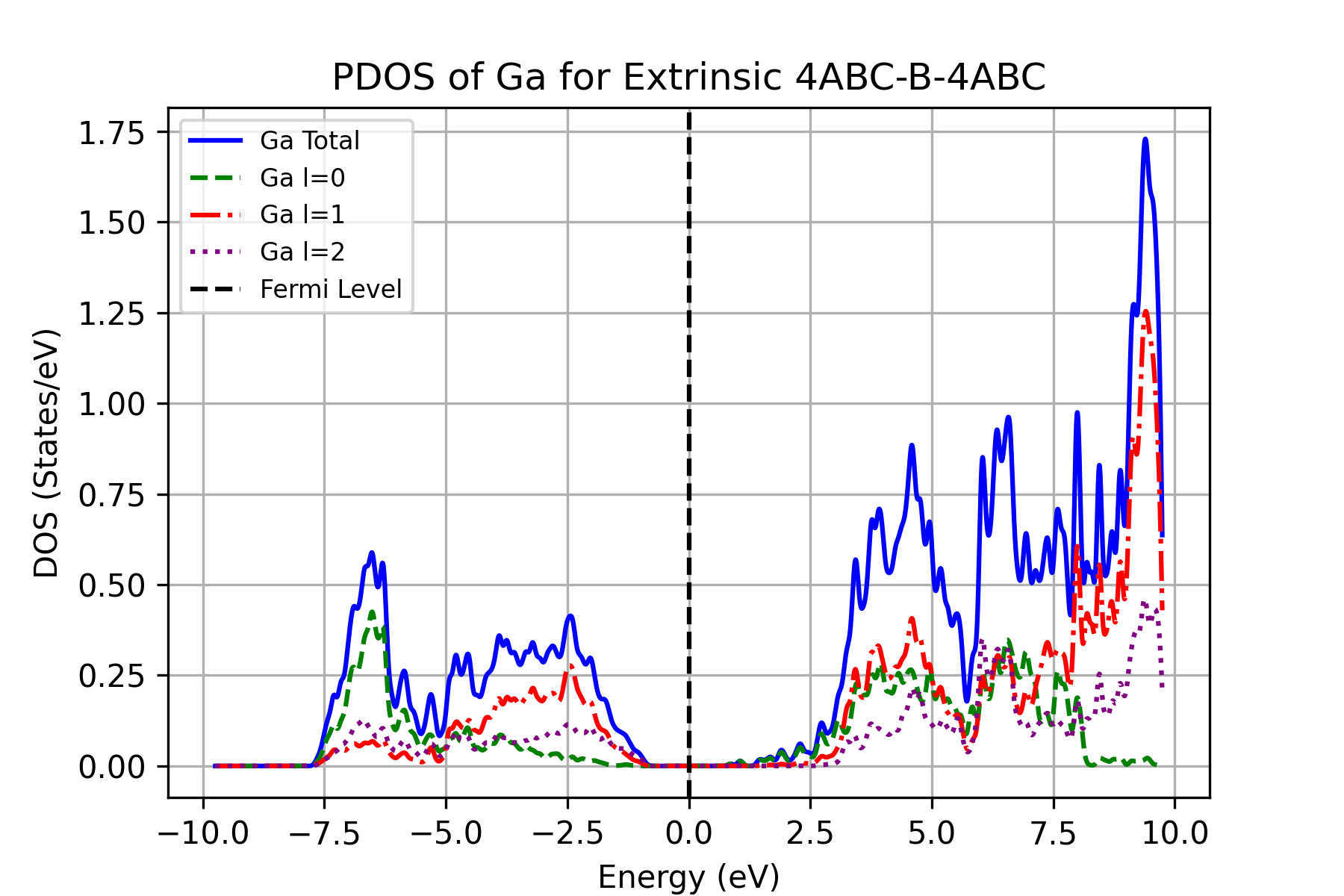}}
\subfloat[]{\includegraphics[width = 0.32\columnwidth]{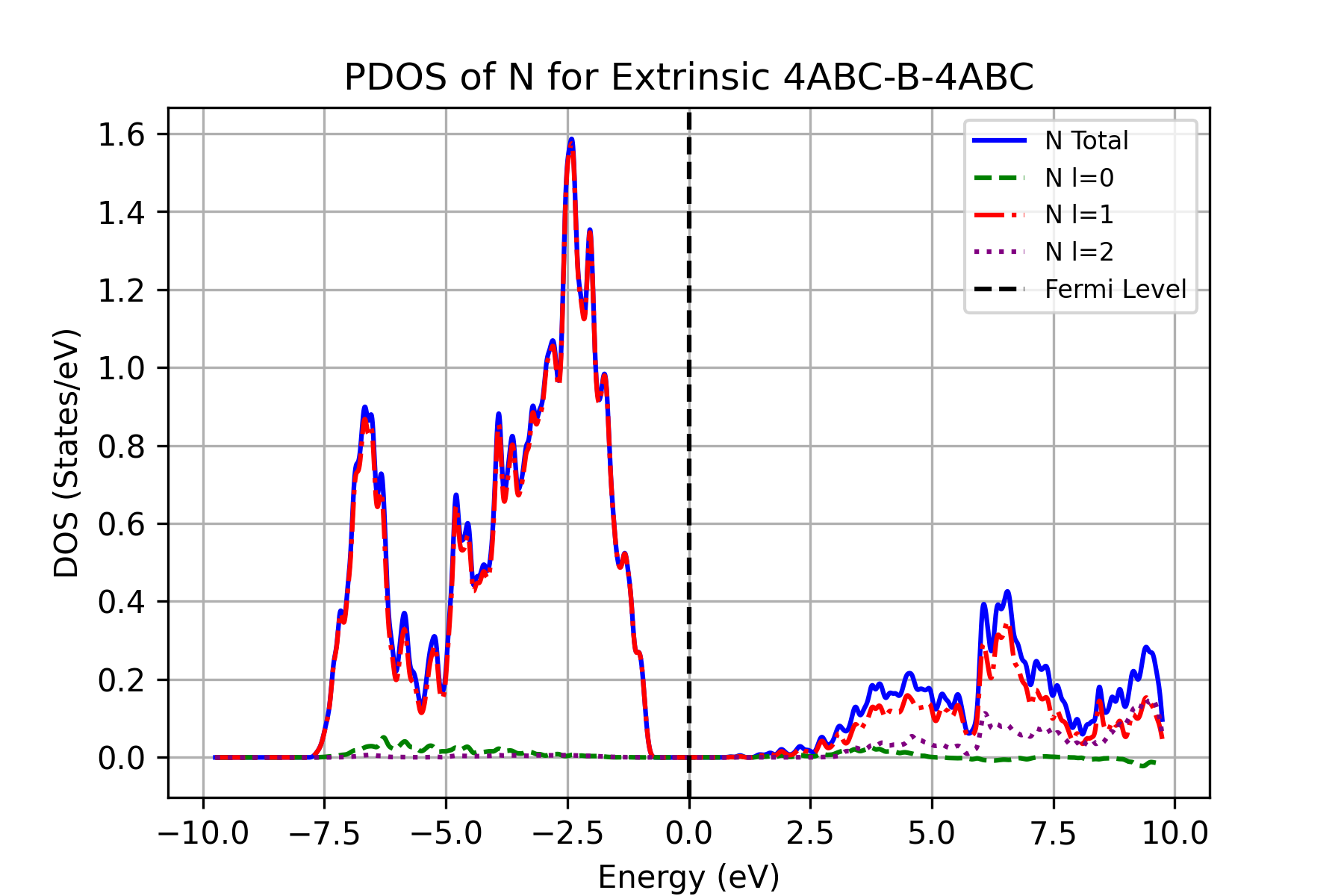}}\\
\subfloat[]{\includegraphics[width = 0.32\columnwidth]{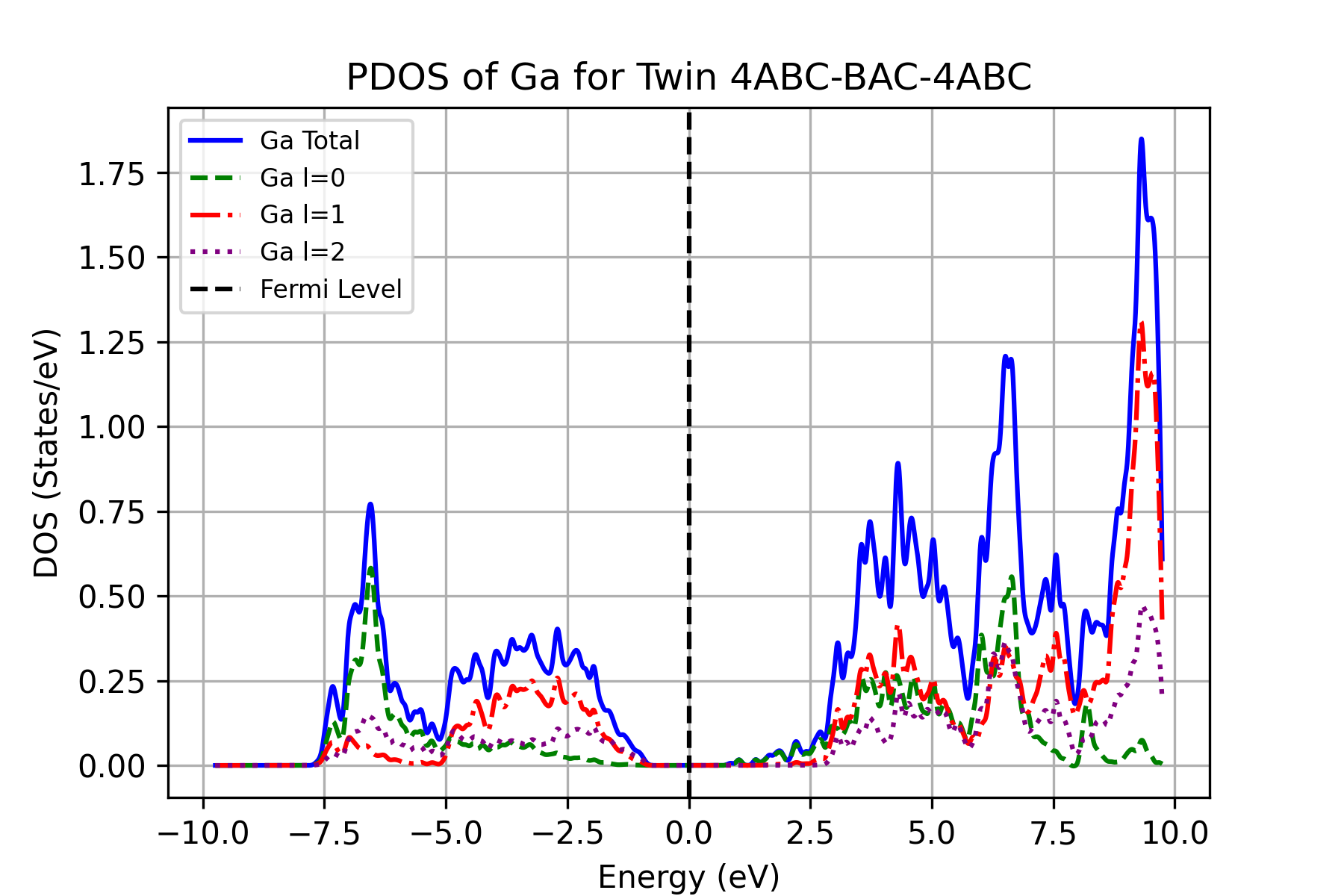}}
\subfloat[]{\includegraphics[width = 0.32\columnwidth]{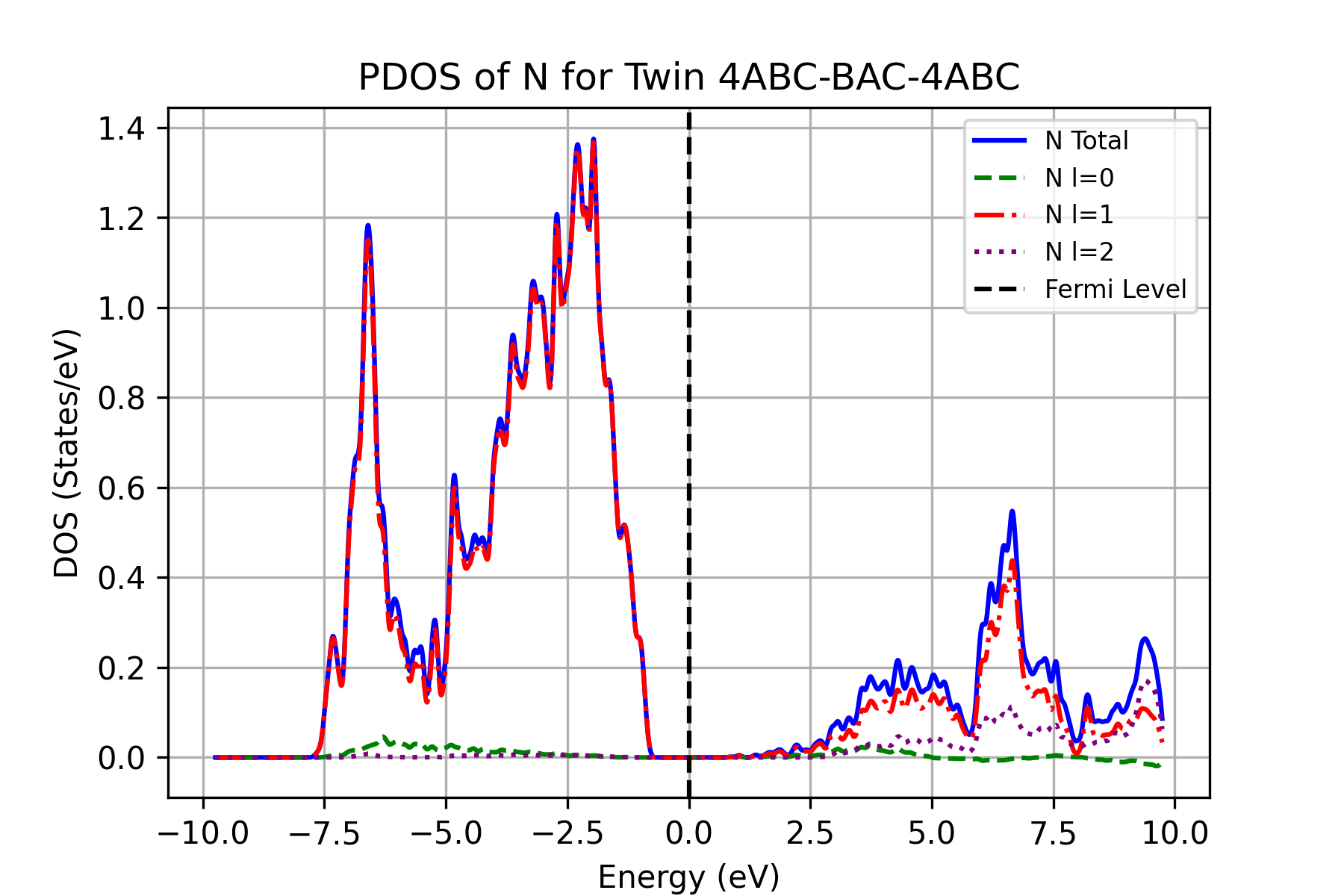}}
\caption{Projected density of states (pDOS) for zincblende bulk and atoms located at the SFs, showing total (blue) and decomposed by angular momentum: (a) and (b) bulk (Ga and N respectively); (c) and (d) intrinsic SF (Ga and N respectively); (e) and (f) extrinsic SF (Ga and N respectively); (g) and (h) twin SF (Ga and N respectively).}
\label{fgr:PDOS_zb}
\end{figure*}

\clearpage

\section{Band Gap Changes}
\label{sec:band-gap-changes}

\begin{figure*}[ht]
\subfloat[]{\includegraphics[width = 0.5\columnwidth]{BG_ZB_In.png}}
\subfloat[]{\includegraphics[width = 0.5\columnwidth]{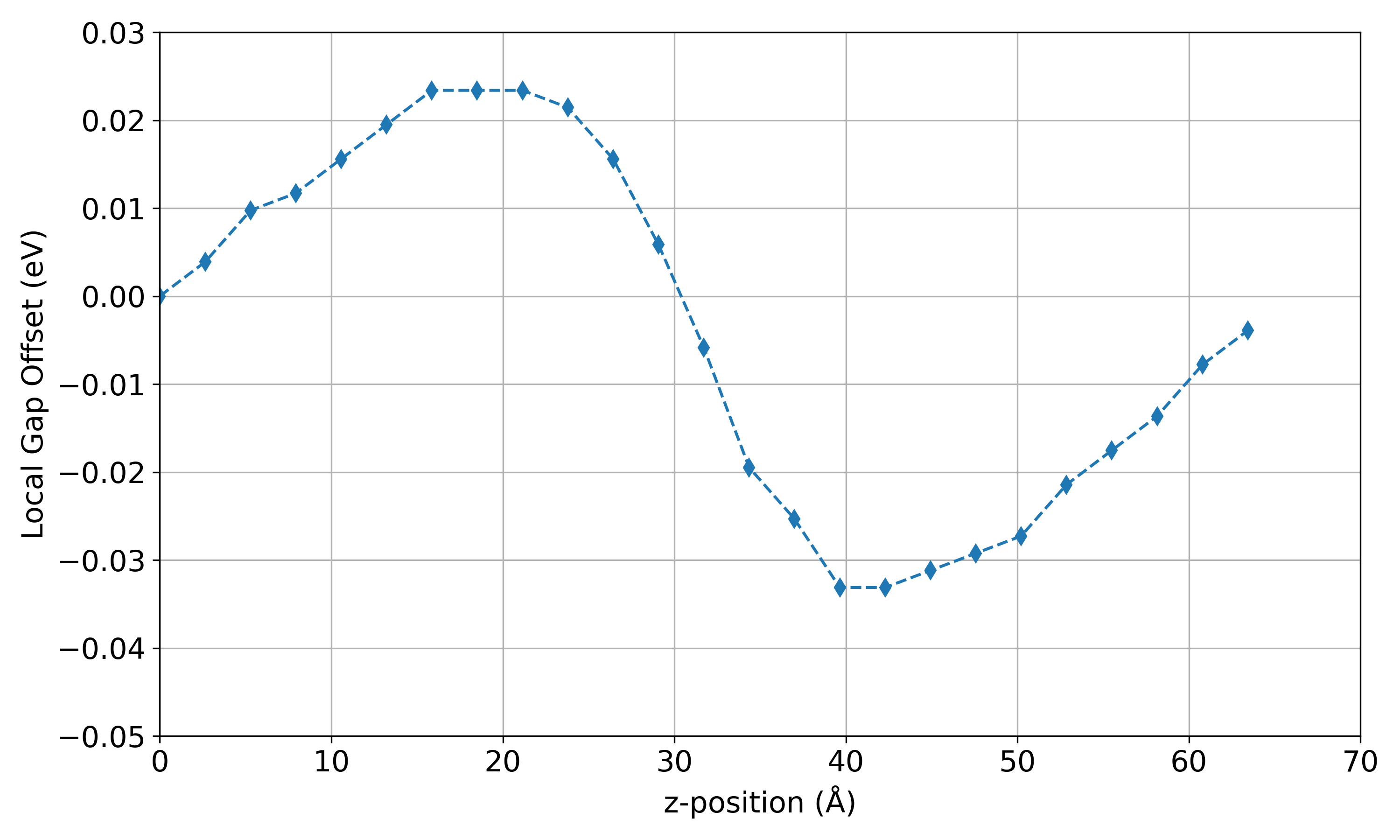}}\\
\subfloat[]{\includegraphics[width = 0.5\columnwidth]{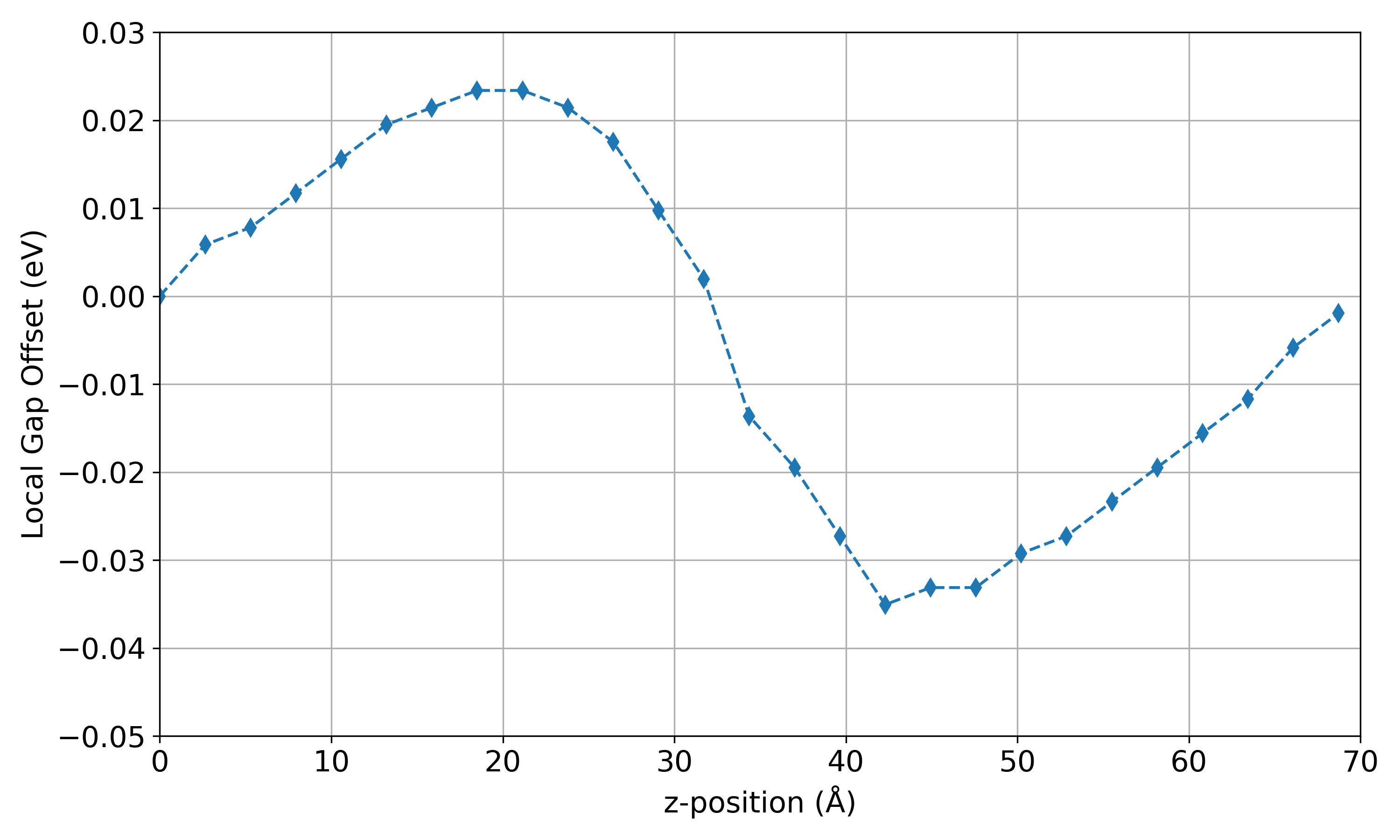}}
\caption{Band gap evolution along c-axis of simulation cell for stacking faults of zb GaN. Intrinsic (a), Extrinsic (b) and Twin (c).}
\label{fgr:Band_Gap_zb}
\end{figure*}

The local band gap offset is plotted below in Figure~\ref{fgr:Band_Gap_WZ} for wurtzite stacking faults and Figure~\ref{fgr:Band_Gap_zb} for zincblende, calculated from the projected DOS.

\begin{table*}[ht]
\begin{center}
  \caption{Calculated values of band gap difference $\Delta E_g$ (eV), band offsets $\Delta E_C$ (eV) and $\Delta E_V$ (eV) for the stacking faults with respect to ideal crystals of wz and zb GaN}
  \label{tbl:El-Prop}
  \begin{tabular}{*{10}{c}}
    \hline
    \hline
    \multicolumn{5}{c}{Stacking faults based on Wurtzite GaN}     \\
    \hline
    \hline
        & Intrinsic 1 (I$_{1}$)  &  Intrinsic 2 (I$_{2}$)  & Intrinsic 3 (I$_{3}$)  & Extrinsic  \\
     \hline         
    $\Delta E_g$        &   -0.015 (-0.050\cite{benbedra2025energetics})   &     -0.046 (-0.120\cite{benbedra2025energetics})    & -0.043   &  -0.095 (-0.140\cite{benbedra2025energetics})  \\
    $\Delta E_C$      &   0.005  (-0.045\cite{benbedra2025energetics})  &  0.039 (-0.096\cite{benbedra2025energetics})   &   0.028        &   0.054(-0.160\cite{benbedra2025energetics})\\
    $\Delta E_V$    &  0.021(0.011\cite{benbedra2025energetics})    & 0.085 (0.019\cite{benbedra2025energetics})  &     0.071      &   0.149(0.028\cite{benbedra2025energetics})  \\
    \hline
    \hline
    \multicolumn{4}{c}{Stacking faults based on Zincblende GaN}     \\
    \hline
    \hline
      &   Intrinsic    &  Extrinsic   &   Twin        \\
    $\Delta E_g$  &    -0.037       &     -0.032    &    -0.033      \\
    $\Delta E_C$  &     0.064      &   0.060      &    0.059     \\  
    $\Delta E_V$  &      0.101    &    0.092     &     0.092       \\
    \\
    \hline
     \hline
  \end{tabular}
\end{center}
\end{table*}

\begin{figure*}[h]
\subfloat[]{\includegraphics[width = 0.5\columnwidth]{BG_WZ_I1.png}} 
\subfloat[]{\includegraphics[width = 0.5\columnwidth]{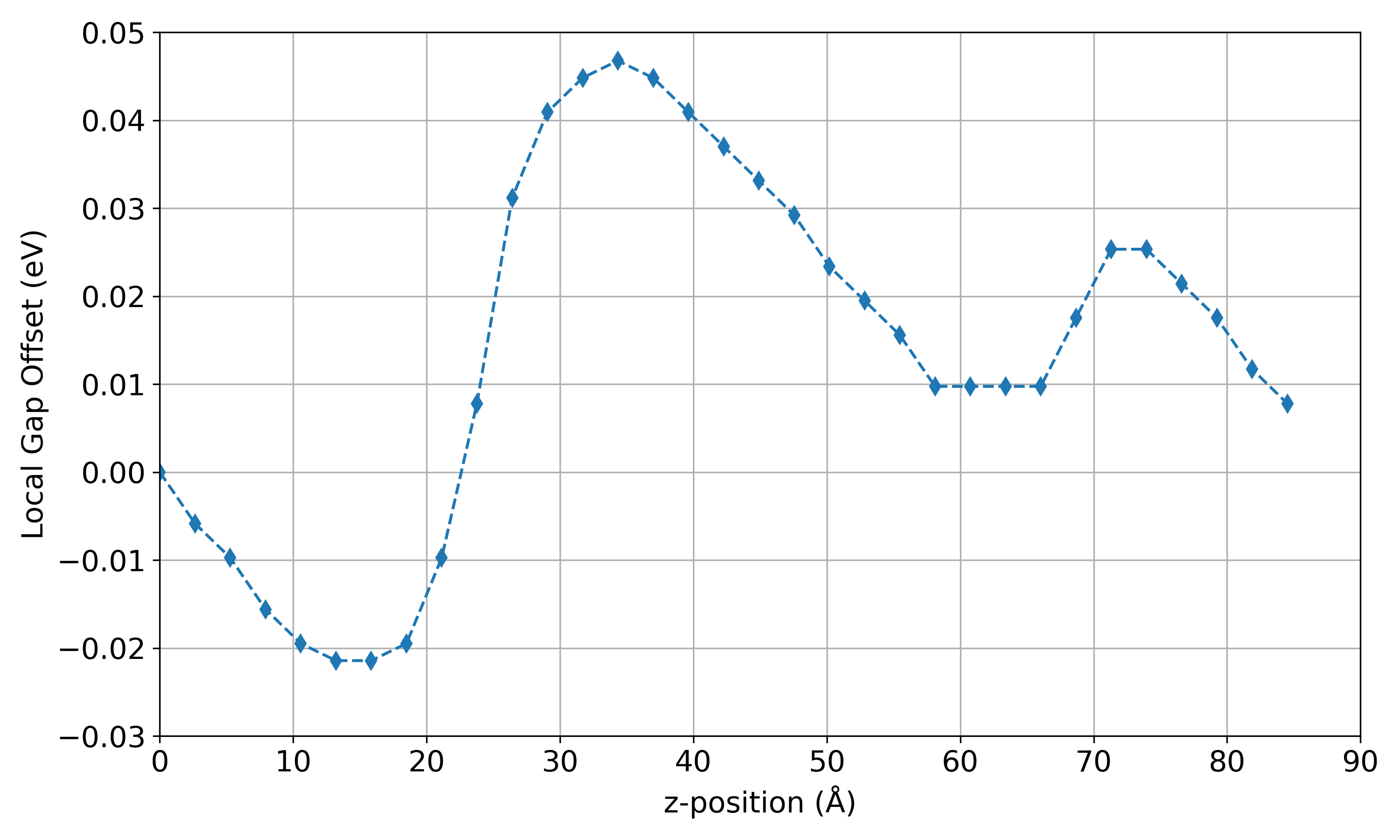}}\\
\subfloat[]{\includegraphics[width = 0.5\columnwidth]{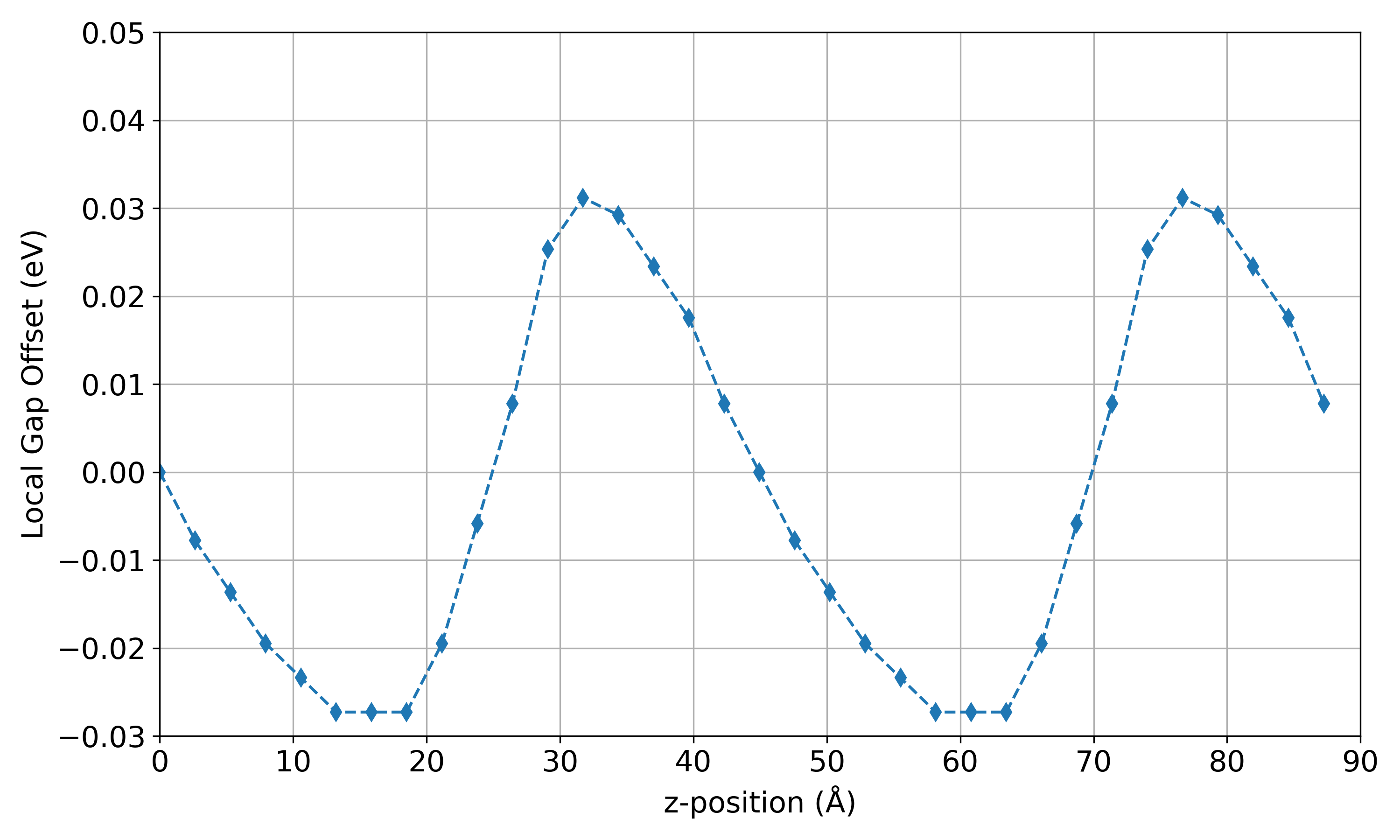}}
\subfloat[]{\includegraphics[width = 0.5\columnwidth]{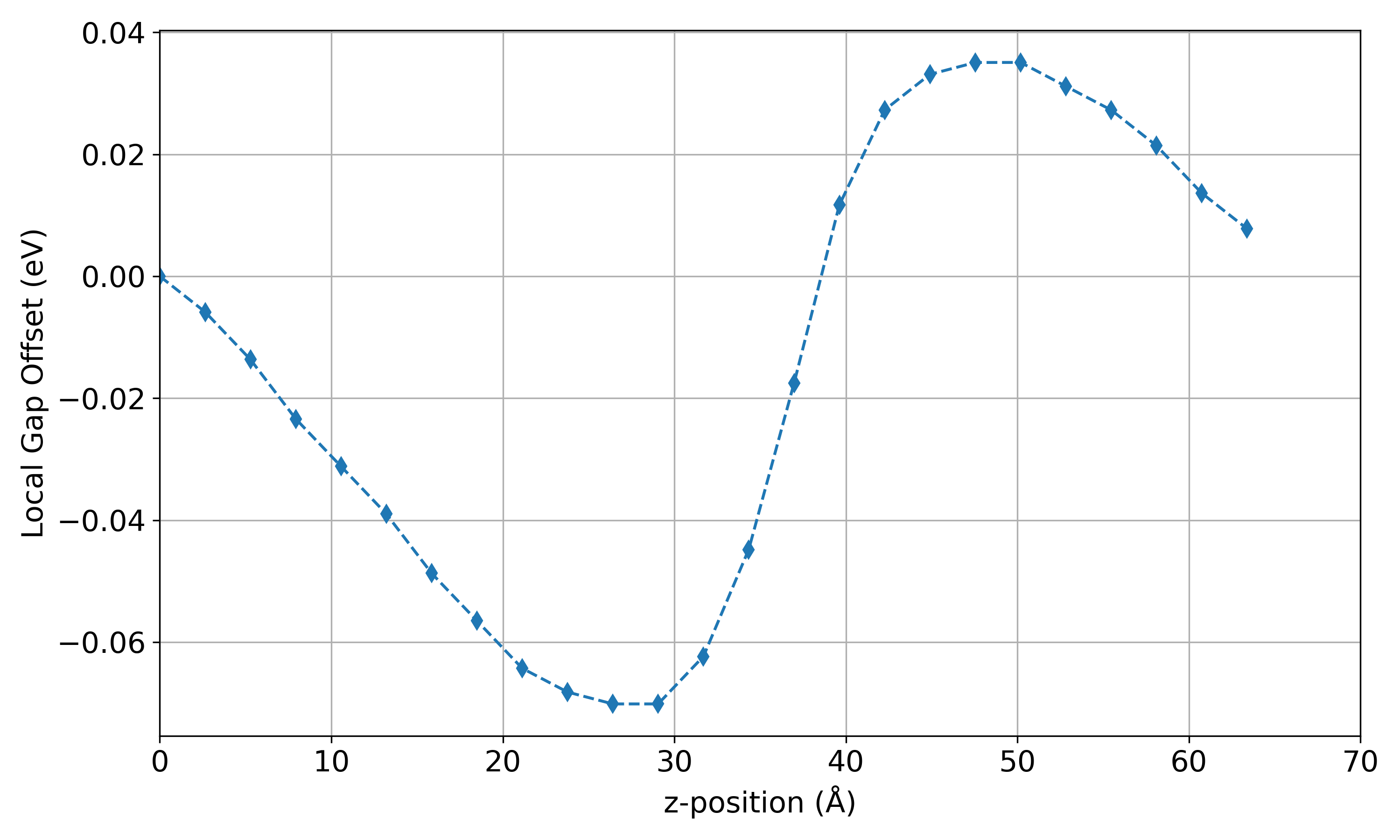}}
\caption{ Band gap evolution along c-axis of simulation cell for stacking faults in wz GaN. I$_{1}$ (a), I$_{2}$ (b), I$_{3}$ (c) and Extrinsic (d)}
\label{fgr:Band_Gap_WZ}
\end{figure*}

\clearpage
\bibliography{bibliography}
\end{document}